\documentclass [twoside,12pt]{report}
\usepackage{amsmath,latexsym,amsfonts,bbold}
\usepackage{units,nicefrac,amssymb,latexsym} \usepackage{graphicx}
 
\columnwidth\textwidth

\begin{document}

\newtheorem{df}{Definition} \newtheorem{thm}{Theorem} \newtheorem{lem}{Lemma}
\newtheorem{prop}{Proposition} \newtheorem{assump}{Assumption}
\newtheorem{rl}{Rule} \newtheorem{sh}{Trial Hypothesis}

\begin{titlepage}
 
\noindent

\vspace*{3cm}

\begin{center} {\LARGE Realism-Completeness-Universality interpretation of
quantum mechanics} \vspace{1cm}

P. H\'{a}j\'{\i}\v{c}ek \\ Institute for Theoretical Physics \\ University of
Berne \\ Sidlerstrasse 5, CH-3012 Bern, Switzerland \\

\vspace{.5cm}

Preliminary version \\ July 2016 \\ \vspace{1cm}

\end{center} \vspace*{2cm}

\noindent {\bf Abstract}: The paper reviews and discusses four ideas scattered
in previous papers of the author. First, objective properties of quantum
systems are not associated with observables but are defined by
preparations. Second, measurable results of classical theories are not
sharp. All such results can be obtained as high-entropy limits of quantum
mechanics. Third, a careful study of exchange symmetry requires a modification
of the theory of measurement. Fourth, screens and detectors are closely
associated with state reduction. A new, rather improved understanding of
quantum mechanics is achieved without any deep changes of the theory
itself. In the framework of a model approach to the philosophy of science, the
RCU interpretation is developed step by step by adding specific supplementary
hypotheses to the Minimum Interpretation of quantum mechanics.

\nopagebreak[4]

\end{titlepage}

\section*{Introduction} The non-relativistic quantum mechanics was borne
during 1920's, was described in the books by Heisenberg, Dirac, Pauli and von
Neumann (later editions of these books are \cite{heisenberg,dirac,pauli,JvN})
and has been in use unchanged in its main concepts and methods until now. It
is a very successful theory and the source of our present knowledge on the
structure of matter and its properties. The practical consequences of the
knowledge have completely changed the daily life of an average human.

Moreover, the theoretical structure of quantum mechanics is marvellous and
beautiful. The basic concepts and the associated mathematical entities can be
briefly listed as follows.
\begin{enumerate}
\item Each quantum phenomenon or experiment can be split into the so-called
preparation and the so-called registration. Preparations and registrations are
described by classical models.
\item Each quantum systems is associated with a copy of Hilbert space that
carries a unitary representation of a central extension of Galilean group.
\item A state of the system is described by a positive operator with trace 1
on the system Hilbert space.
\item An observable of the system is described by a self-adjoint operator on
the system Hilbert space.
\item Values of an observable obtained by a registration are eigenvalues of
the observable. The probability of registering a given eigenvalue of an
observable on a prepared state is calculated from the state and the observable
by the so-called Born rule.
\item The most important observables of the system are the generators of the
spacetime symmetry group.
\item The Hilbert space of the system composed of two system of different type
is the tensor product of the Hilbert spaces of the subsystems.
\item Subsystems of the same type cannot be distinguished from each
other. This leads to the symmetry under the permutation group of the
subsystems, the so-called exchange symmetry.
\end{enumerate}

In addition, quantum mechanics is a highly symmetric theory. One of its
symmetry groups is the infinitely dimensional transformation group of
(generalized) Hilbert-space bases. This is comparable to the symmetry of the
classical mechanics with respect to canonical transformations. The second
symmetry group is formed by the space-time transformations. Finally, the third
one is the exchange-symmetry group, unknown in classical theories.

Let us turn to interpretation of this apparatus. There is a popular belief
among physicists that quantum states refer to individual quantum systems
(Realism), that quantum mechanics gives a complete description of the
micro-world (Completeness) and that it is applicable to all physical objects
(Universality). This agrees more or less with the so-called Dirac-von-Neumann
version of the Copenhagen interpretation. Our aim is to reformulate this
interpretation more precisely by explicitly stating what assumptions it adds
to the so-called Minimum Interpretation, which could be roughly identified
with Bohr's version of the Copenhagen interpretation in the rigorous form
given to it by \cite{ludwig1}. Minimum Interpretation is also explained and
applied in textbooks \cite{peres} or \cite{ballent}.

We are going to study what can be added to the Minimum Interpretation and
concentrate all such supplementary assumptions into several specific
hypotheses, called Trial Hypotheses (TH). Some of such TH's are well known,
but we will also introduce new and sometimes rather heretic ideas. In each
case, we shall study the consequences of such changes, localize new problems
and try to remove them. The resulting understanding of quantum mechanics will
be called Realism-Completeness-Universality (RCU) interpretation.

The first bunch of TH's concerns the realism, Chapter 1. We shall first assume
that properties defined uniquely by preparation are objective properties of
the prepared system. This is a strong enhancement of the notion that states
refer to individual quantum systems. Second, the values of observables are not
objective but only created by registrations. This removes the contradictions
that result from the assumption that values of observables exist before, and
are only revealed by, registrations. Moreover, we are going to embed the
realist notion of quantum mechanics into a specific branch of philosophy of
science: the so-called Constructive Realism \cite{giere}, which views physical
theories as classes of models, each of which ought to describe some aspect of
reality in an approximate way. We shall find that the language of models
allows to formulate many difficult ideas in a precise and clear way.

As for the completeness, we shall assume that there are no unknown causes
beyond the probabilities given by the Born rule. At least, quantum mechanics
does not identify such causes. This does not lead to any internal logical
problem, it just contradicts the philosophy of determinism. We call this
assumption {\em Completeness Hypothesis}.

If we want quantum mechanics to be universally applicable, then it has to
explain the so-called classical properties. Newtonian mechanics, Maxwellian
electrodynamics, phenomenological thermodynamics and classical chemistry are
classical theories. They construct models of real objects and these models
describe classical properties, such as position and momentum, electric and
magnetic fields or volume, pressure and temperature as well as chemical
composition. Certain sets of such properties define (classical) {\em states}
and certain properties are values of quantities that can be called (classical)
{\em observables}.

The classical theories are not obsolete or made invalid by quantum mechanics,
but they describe certain aspects of certain physical objects in certain
approximation. Indeed, they are still in use and successful, if their accuracy
is sufficient for given aims, because the mathematics of the classical
theories is usually much simpler than the corresponding quantum calculation
would be. However, we have then to give a quantum-mechanical explanation of
the successful classical predictions.

We are going to give an explanation that is rather different from what is
usually assumed. We shall first postulate that a (real) physical object can
have both classical and quantum models. Second, that the states of the
classical model are associated with some high-entropy quantum states and
third, that classical properties are then some properties of these
high-entropy states. This must also hold for Newtonian mechanics and can be
achieved by requiring the mechanics to describe only measurable aspects of
``mechanical'' reality. To this aim, the mathematical theory of the so-called
maximum-entropy (ME) packets is introduced, Chapters 2 and 3. Chapter 2 shows
that Newtonian mechanics is then analogous to statistical thermodynamics. The
main result of Chapter 3 is that the classical limit is a well-defined
high-entropy limit.

Finally, a well-known paradox results from the assumption that states refer to
individual systems, namely the existence of two different kinds of dynamics:
the unitary evolution generated by the generator of the time translation of
Galilean group, and the creation of definite values of observables by a
registration process, the so-called state reduction. Thus, we have to deal
with the problem of the two dynamics. This will be done in Chapters 4 and
5. Chapter 4 shows that current quantum theory of measurement in all its
variants is deficient because it neglects the strong influence of the exchange
symmetry on measurement processes that necessarily follows from a consequent
application of quantum mechanical principles. A consequence of this influence
is that registration apparatuses can work only if they are {\em
incomplete}. That is, Born rule does not hold for registered values on all
states. Moreover, any preparation must elevate the prepared system from the
sea of identical particles. We say that such a system has a {\em separation
status}. The chapter gives a rigorous definition of the separation status and
introduces the mathematics needed to describe evolution of such systems.

Chapter 5 gives a reformulation of the quantum theory of measurement so that
it becomes compatible with Chapters 3 and 4. It analyzes the registration
processes and postulates that each registration apparatus contains a special
object called {\em detector}. It then shows that the state reduction occurs
inside the detector or inside the screen if they cause a loss of separation
status. Thus, the existence of the two dynamics is not removed but objective
conditions of when the state reduction occurs and where it happens, are
specified. This leads to observable phenomena, at least in principle. Finally,
Chapter 6 contains a review of all Trial Hypotheses and a concluding
discussion.

Most of these ideas have been published some time ago
(\cite{hajicek}--\cite{survey}). The present paper is written more carefully
than these papers, corrects some errors and fills some gaps. Moreover, the new
exposition is appreciably simplified and the focus is on examples and models
rather than on attempts to formulate the most general statements. This is in
agreement with the adopted kind of philosophy of science. Thus, the exposition
is better accessible and I believe that the paper can be read by students that
have finished an ordinary course of quantum mechanics. In any case, there are
excellent textbooks such as \cite{peres} and \cite{ballent} that can be
recommended as preparatory reading.

The present paper not only collects some ideas that are scattered in
literature but also answers the question whether the whole quantum mechanics
can be formulated in a coherent way so that it is compatible with these
ideas. The ultimate aim is to show that our understanding of quantum mechanics
can be rather improved without any deep changes of quantum mechanics itself.

During my work, I have profited from the enthusiasm and collaboration by
Ji\v{r}\'{\i} Tolar, who also made some calculations and helped with the
formulation of the first version \cite{survey} of this paper. I am thankful to
my colleagues at the Institute for Theoretical Physics in Berne, especially to
Heinrich Leutwyler and Uwe-Jens Wiese for their interest and
discussions. Special thanks are due to Stefan Janos for supervising my
excursions into experimental physics. A great and rather unexpected
encouragement has come from the organizers of the conference series
``DICE''. They have enabled me to present my three main ideas by three talks
during subsequent meetings in the years 2010, 2012 and 2014. I am very obliged
especially to Thomas Elze and Claus Kiefer for their interest and
encouragement.

\tableofcontents

{ \renewcommand{\theequation}{\arabic{chapter}.\arabic{equation}}
\renewcommand{\thethm}{\arabic{chapter}.\arabic{thm}}
\renewcommand{\theassump}{\arabic{chapter}.\arabic{assump}}
\renewcommand{\thedf}{\arabic{chapter}.\arabic{df}}
\renewcommand{\thelem}{\arabic{chapter}.\arabic{lem}}
\renewcommand{\thesh}{\arabic{chapter}.\arabic{sh}}

\chapter{Objective properties} \setcounter{equation}{0} \setcounter{thm}{0}
\setcounter{assump}{0} \setcounter{df}{0} \setcounter{sh}{0}

\section{Realism in philosophy of science} A realist interpretation of a
physical theory is a subtler and deeper problem than an answer to the question
of whether the world exists for itself rather than being just a construction
of our mind. This question can always be answered in positive without any
danger of falsification.

There is a more interesting question. Every physical theory introduces some
general, abstract concepts. For example, Newtonian mechanics works with mass
points, their coordinates, momenta and their dynamical trajectories. The
question is, whether such concepts possess any counterparts in the real
world. On the one hand, it seems very plausible today that mass points and
their sharp trajectories cannot exist and are at most some idealisations. On
the other hand, if we are going to understand a real object, such as a snooker
ball moving on a table, then we can work with a construction that uses these
concepts but is more closely related to the reality. For example, we choose a
system of infinitely many mass points forming an elastic body of a spherical
shape and calculate the motion of this composite system using Newton's laws
valid for its constituent points. Then, some calculated properties of such a
model can be compared with interesting observable properties of the real
object. Thus, even if the general concepts of the theory do not describe
directly anything existing, a suitable model constructed with the help of the
general concepts can account for some aspects of a real object.

Motivated by this observation, one can divide any physical theory into two
parts. First, there is a treasure of successful {\em models}. Each model gives
an {\em approximative} representation of some {\em aspects} of a real object
\cite{giere} (Giere uses words ``degree'' and ``respects''
instead). Historically, models form a primary but open part of the theory. For
example, in Newtonian mechanics, the Sun and its planets were carefully
observed by Tycho de Brahe and then some model of it was constructed by
Kepler. Apparently, Newton was able to calculate accelerations and doing so
for Kepler trajectories, he might discover that they pointed towards the
Sun. Perhaps this lead to the Second Law. The hydrogen atom had a similar role
in quantum mechanics.

Second, there is a general {\em language} part. If we restrict ourselves to
physics, it contains the mathematical structure of state space, conditions on
trajectories in the state space, their symmetries and the form of observables
\cite{fraassen}. It is obtained by generalisation from the study of models but
it is also an instrument of further model construction and a tool for
unification of the models. For example, in Newtonian mechanics, the phase
space is the state space, Newton's dynamical equations are the conditions on
trajectories, Galilean transformations are symmetries and real functions on
the phase space, such as a Hamiltonian, are observables.

A model is constructed as a particular subset of trajectories in a particular
state space as well as a choice of important observables. For example, to
describe the solar system, assumptions on the number of bodies, their
point-like form, their masses, the form of gravitational force and certain
class of their trajectories can be made if we want to construct a model. The
observed positions of the planets ought then to match the theoretical
trajectories of the model within certain accuracy. Thus, each model is
associated with a real object. The model always contains simplifying
assumptions, always holds only for some aspects of the object and only within
some approximation. The accuracy of the approximation that is referred to may
be unknown and is different from the accuracy of measurements that can be
performed at a given historical stage. This is measurable and can be expressed
numerically by statistical variances.

The models of a given theory are not predetermined by the language part but
obtained in the historical evolution and dependent on observation of real
objects. Indeed, on the one hand, the language part can also be used to
construct models that do not have any real counterparts. On the other, the
model part is steadily evolving and never closed. For example, a satisfactory
quantum model of high temperature superconductivity is not yet known. This is
why the treasure of successful models is an independent and, in fact, the
fundamental part of any theory.

Such an approach lies somewhat within the recent trend of the philosophy of
science that defines a theory as a class of models (see, e.g.,
\cite{suppes,sneed,fraassen,giere,cart}). It can be said that it combines
ideas of Constructive Realism by Ronald Giere with van Fraassen notions of
state space and symmetries \cite{fraassen} as a basis of the general language
part. It is important that Constructive Realism is immune to the usual
objections against naive realism. Naive realism is roughly characterised by
the statement: ``The world is as it is perceived'', which is obviously wrong.

To proceed any further, we have to clarify the relation between real objects
and their theoretical models. What is a real object?

An object (we shall leave out the adjective ``real'') is assumed to be the
cause of certain empirical experience we can have. These observations also
motivate assumptions about certain objective properties of the
object. Objective means here that the properties can be ascribed to the object
alone, each such ascription being a kind of {\em ontological hypothesis}. Lead
by such heuristic ideas, we shall later define objective properties of quantum
objects in a more precise way simply by listing the mathematical entities that
can describe them.

A few words have perhaps to be said on ontological hypotheses. As is
well-known, the objective existence of anything cannot be proved (even that of
the chair on which I am now sitting, see, e.g., Ref.\ \cite{d'Espagnat}, where
this old philosophical tenet is explained from the point of view of a
physicist). Thus, all such statements are only hypotheses. However, a
sufficiently specific ontological hypothesis may lead to contradictions with
some observations and some hypotheses which do not lead to contradictions may
be useful. For example, the objective existence of the chair nicely explains
why we all agree on its properties. Similarly, the assumption that quantum
systems possess certain objective properties might be useful for the quantum
theory of classical properties or for a solution of the problem of quantum
measurement.

We shall say ``Object $a$ has a property $A$'' if $A$ is an objective property
of $a$. An object must then satisfy the following requirements:
\begin{description}
\item[O1] An object has enough objective properties: all objective properties
of an object define the object uniquely in the sense that we can recognise the
same object at different times and different space position, as well as
distinguish it from other objects in the environment.
\item[O2] The proposition ``Object $a$ has a property $A$'' for a given object
$a$ and for all its objective properties $A$ is always either true or false.
\end{description} As an example, consider that dice I am holding in my
hand. The dice is defined by its geometry, chemical composition, and
colours. It can exist during several years and take part in different
processes at different positions. Its positions at different times are
objective properties that do not define it. If I toss the dice, some number
will fall.  Such number is an example of a property that is not an objective
property of the dice but it is an objective property of the toss.

Theoretical models of a given object ought to explain the observed properties
of the object. We stress again that the models do not describe the assumed
object exactly and completely. Such an exact and complete knowledge of any
given object will perhaps never be achieved. On the one hand, for a given
object, there can be several models that can differ in sophistication and
accuracy. On the other hand, one model can describe a whole class of objects.

\section{Application to quantum mechanics} However, the approach is not as
easily applied to quantum mechanics as it is to Newtonian mechanics. A
question looms large at the very start: What are the object, of which quantum
mechanics makes models? Could we leave the Minimum Interpretation and assume
that such objects are not just classical apparatuses but also some microscopic
objects met empirically in preparations and registrations? The character of
such assumptions has been analysed by Giere \cite{giere}, p.\ 115. Giere
distinguishes trial suppositions about reality, which are still the subject of
investigations and experiments, from assured knowledge about real objects,
which can be recognized by ability to control and manipulate the object and
thus to use it in investigations and experiments on other problems.

Our notion of properties allows us to express a similar idea. An existing
individual object must have a sufficient number of objective properties so
that they satisfy requirement O1. Then, we may know enough about the object in
order to be able to manipulate and control it. Another idea of a similar kind
has been introduced by Gisin \cite{gisin}:
\begin{quote} A theory is {\em realistic} if and only if, according to the
mathematical structure of the theory, the collection of all physical
quantities written in the system unambiguously determines the probabilities of
all possible measurement outcomes.
\end{quote} Here, ``physical quantities written in the system'' means what we
termed ``objective properties''.

In Newtonian mechanics, any mechanical object is modelled by a mechanical
system. The values of all point masses and their coordinates and momenta of
the system distinguish different objects. On the one hand, Newtonian
coordinates and momenta define a state, that is a point in the phase space of
the system, on the other, they are observables. In Newtonian mechanics, both
states and values of observables are directly associated with, or explain,
objective properties of a physical object without any danger of
contradictions.

The situation in quantum mechanics is more complicated. Let us first formulate
the ontological hypotheses that are compatible with the Minimum
Interpretation.
\begin{enumerate}
\item Specific processes running in an arrangement of classical apparatuses
are quantum measurements. A subset of the apparatuses are preparation, another
are registration ones. The apparatuses are objects, their classical properties
distinguishing them from any other such objects.
\item A system type and a prepared state can be considered as certain classes
of preparation apparatuses. An observable can be considered as a class of
registration apparatuses.
\item An {\em individual} measurement process is defined as such that results
in a unique definitive registration outcome: a unique value of a quantum
observable. Quantum mechanics is a set of rules allowing the computation of
probabilities for outcomes of individual registrations that follow specified
preparations \cite{peres}, p.\ 13.
\end{enumerate}

Actually, there can be different states associated with an individual
measurement process, each referring to a different time instant of the
process.

Now, as promised in the Introduction, we are going to augment the Minimum
Interpretation by further trial hypotheses.
\begin{df}\label{dfstructprop} A {\em structural} property of a quantum system
is a property that is common to all systems of the same type.
\end{df} For example, mass, spin and electric charge are structural properties
of particles while composition and Hamiltonian are those of composite systems.

The basis of the realist part of RCU interpretation is:
\begin{sh}\label{shobject} A quantum object is defined by a preparations. The
objects are thus distinguished from each other by the properties that are
determined by their preparations. These include the structural properties
describing a system type, the prepared state and properties that are uniquely
defined by the state. Objects of non-relativistic quantum mechanics can be
classified into electrons, neutrons, nuclei, atoms, molecules and their
composites.
\end{sh}

What we have added is that system types and prepared states are objective
properties of microscopic objects. Such objects can take part in measuring
processes but are then different from arrangements of classical apparatuses
and measurement processes. Further properties, different from structural ones
and states, will be added in Section 2.3 by TH \ref{shobject1}.

Sometimes, TH \ref{shobject} meets one of the following two questions. First,
how can the Hypothesis be applied to cosmology, when there was nobody there at
the Big Bang to perform any state preparation? Second, a state preparation is
an action of some human subject; how can its result be an objective property?
Both objections originate in a too narrow view of preparation: it is not
necessarily a human activity. Moreover, if we don't know the preparation, we
can still assume that the considered object is in some state, for example in
the case of cosmology. This is, in principle, a testable hypothesis. Actually,
the second objection is not much more than a pun. It is not logically
impossible that a human manipulation of a object results in an objective
property of the object. For example, pushing a snooker ball imparts it a
certain momentum and angular momentum that can then be assumed to be objective
properties of the pushed ball.

We distinguish quantum objects and quantum systems. Quantum object is an
object of which quantum model is constructed. Quantum system is a part of the
quantum model. A system is distinguished from other systems by some symbol and
is mathematically described by a Hilbert space that carries a representation
of a spacetime group. We shall denote objects by calligraphic capital letters
such as ${\mathcal S}$ and systems by capital letters such as $S$. Because of
the exchange symmetry, a quantum system is just an auxiliary mathematical
notion that has no really existing counterpart. This will be clarified in
Chapter 4. Moreover, there are also fully abstract quantum systems, such as
the centre of mass of an isolated composite system.

As a comment to requirement O2 applied to microscopic objects of quantum
mechanics (this idea is due to G\"{u}nter Ludwig \cite{ludwig1}), let us
clarify the relation to the quantum logic \cite{BvN}, which does not satisfy
requirement O2. The properties studied by the quantum logic are values of some
observables. But values of observables are analogous to values obtained by
tossing a dice: each such value is not a property of the dice alone but also
of a particular toss. They cannot be attributed to the object alone. The fact
that the mathematics of the quantum logic is still a beautiful kind of algebra
is due to von Neumann's smart choice of very special observables (projections)
that satisfy specific geometric relations. Thus, on the one hand, they are not
properties of one object, on the other hand, they are not properties of all
possible registrations on a given state.

According to TH \ref{shobject}, a sufficient condition for a property of a
quantum system to be objective is that its value can be uniquely determined by
a preparation according to the rules of standard quantum mechanics. The
``value'' is the value of the mathematical expression that describes the
property and it may be more general than just a real number. To observe an
objective property, many registrations of different kinds may be necessary.

We consider properties that are complex in the following sense \cite{survey}:
\begin{enumerate}
\item Their values may be arbitrary mathematical entities (sets, maps between
sets, etc.). For example, the Hamiltonian of a closed quantum system involves
a relation between energy and some other quantities of the system. This
relation is an example of such a complex property.
\item Their values need not be directly obtained by individual
registrations. For example, to measure a cross-section a whole series of
scattering experiments must be done. Thus, their values need not possess
probability distribution but may be equivalent to, or derivable from,
probability distributions.
\end{enumerate} Such complex properties are nothing new. For example, in
Newtonian mechanics, a value of any given observable $O$ for any mechanical
object is never ``known exactly'' but only as an expectation value with a
variance (i.e., mean quadratic deviation), and such a value is only obtained
by many registrations. The statement that there {\em is} an exact value and
the expectation value with the variance is only due to the inaccuracy of
measuring techniques is, in fact, one of the simplifying hypotheses of the
language part of the theory that need not have anything to do with
reality. This will be utilised in Chapter 2.

A value of an observable is an outcome of an individual registration performed
on an already existing quantum object. In this way, RCU interpretation leads
to a separation of the prepared object from its registration. This motivates
postulating more for registrations than is directly observed (such directly
observed properties are individual randomness and large-number regularity of
registered values) and than would be required by a strict adherence to the
Minimum Interpretation. Thus, we arrive at our second Trial Hypothesis.
\begin{sh}\label{OCR} (Outcomes Created by Registration) The outcome of an
individual registration performed on a quantum object ${\mathcal S}$ in state
${\mathsf T}$ is in general only created during the registration. It is an
objective property of the whole registration process.
\end{sh} The opposite conjecture, that every possible outcome is always
already determined before any registration, means that each quantum object has
some further objective properties that are not uniquely determined by its
preparations (the so-called ``hidden variables''). As quantum mechanics does
not determine these further properties, the conjecture about their existence
contradicts the Completeness Hypothesis (see the Introduction). Moreover, this
conjecture is also directly incompatible with quantum mechanics and a number
of fine experiments concerning Bell inequalities \cite{peres,ballent} and
contextuality \cite{peres,ballent,bub}. Of course, in some exceptional cases,
a value of an observable is determined before its registration (if the state
is an eigenstate of the observable).

To see the real meaning of TH \ref{OCR}, we just apply it to values of
position. Position of a given quantum system is an observable and its
eigenvalues can be measured. However, if a position value $\vec{x}$ is an
outcome of a registration on a state, this does not imply that some real part
of the registered object has been at the point $\vec{x}$ immediately before
the registration. If the possible position values for the state are
distributed over a finite region of space, we cannot think that the system in
this state describes an object that is extended over this region in a similar
way as a classical matter continuum would be extended. Indeed, the
corresponding ``real parts'' of such an object would then have to move with
superluminal velocity during some registrations.

Finally, we consider application of quantum mechanics to objects for which
classical models are good approximations. As the basis of our approach to the
issue of Universality, the third Trial Hypothesis reads:
\begin{sh}\label{rhmacro} Every object has a quantum model that accounts for
all its known physical properties.
\end{sh} This does not require that each quantum model of the object must
explain all its observed properties. We just assume that such models are in
principle possible.

A particular application of TH \ref{rhmacro} is that a physical object can
have both a classical model and a quantum model. This defines a relation of
the two models. Moreover, the classical model can be considered, in the sense
of Minimum Interpretation, as a preparation as well as a registration
apparatus for the quantum model. This is a very important observation that
will be used in the remaining sections.

The three Trial Hypotheses \ref{shobject}, \ref{OCR} and \ref{rhmacro} and
their consequences will strongly transform the Minimum Interpretation picture
of quantum mechanics although the resulting theory can still be used to
calculate the probabilities of outcomes of registrations following certain
preparation by the methods provided by the standard quantum mechanics. But now
it will be a theory that describes non-relativistic physical properties of
objects including microscopic ones and the calculation of probabilities will
be based on constructions of models of such objects.

Trial hypotheses \ref{shobject} and \ref{rhmacro} imply an atomistic picture
of the world, with only few types of ``atoms'' that are present in huge
numbers. The Hypotheses are, however, just starting points of a serious
work. We must remove various paradoxes and problems that might result from
them. This will be done in Chapters 2, 4 and 5.

\section{Spin systems} In this and the next sections, the heuristic idea that
properties are objective if they are uniquely defined by preparations will be
specified and a list of objective properties will be given. The space of
states will be described and the notions of proper mixture, of ontic and of
epistemic states introduced.

Consider first a spin system. This is described by the two-dimensional Hilbert
space ${\mathbf H}$ carrying a unitary representation of the central extension
$SU(2)$ of the proper rotation group $SO(3)$ (see e.g.\ \cite{ballent},
Section 7.4). The relation between the two groups is defined by the so-called
central homomorphism, the map $h_c : SU(2) \mapsto SO(3)$. The most important
observables of this system are generators of $SU(2)$ and they are called spin
components.

The space of all self-adjoint operators on ${\mathbf H}$ will be denoted by
${\mathbf L}_r({\mathbf H})$. It is a four-dimensional real linear space. If
we choose an orthonormal basis ${|1\rangle},|2\rangle$ of ${\mathbf H}$, then
any ${\mathsf A} \in {\mathbf L}_r({\mathbf H})$ is described by a $2\times2$
complex symmetric matrix. We can define coordinates $a_0,a_1,a_2$ and $a_3$ on
${\mathbf L}_r({\mathbf H})$ by writing ${\mathsf A}$ as follows:
$$
{\mathsf A} = a_0 {\mathsf 1} + \sum_k a_k {\mathsf \sigma}_k\ ,
$$
where
$$
{\mathsf 1} = \left(\begin{array}{ll} 1 & 0 \\ 0 & 1 \end{array}\right)\ ,
{\mathsf \sigma}_1 = \left(\begin{array}{ll} 0 & 1 \\ 1 &
0 \end{array}\right)\ ,\quad {\mathsf \sigma}_2 = \left(\begin{array}{ll} 0 &
-i \\ i & 0 \end{array}\right)\ ,\quad {\mathsf \sigma}_3 =
\left(\begin{array}{ll} 1 & 0 \\ 0 & -1 \end{array}\right)\ ,
$$
is the so-called Pauli basis.

The action ${\mathsf A} \mapsto {\mathsf U}{\mathsf A}{\mathsf U}^\dagger$ of
$SU(2)$ on ${\mathbf L}_r({\mathbf H})$ can then be expressed by the
coordinates as follows:
\begin{equation}\label{rotL} {\mathsf U}{\mathsf A}{\mathsf U}^\dagger =
a_0{\mathsf 1} + \sum_l\left(\sum_k {\mathsf O}_{lk}a_k\right)\sigma_l\ ,
\end{equation} the relation valid for any ${\mathsf A} \in {\mathbf
L}_r({\mathbf H})$, where ${\mathsf O} = h_c({\mathsf U})$ and
$a_0,a_1,a_2,a_3$ are the components of ${\mathsf A}$ in the Pauli
basis. Thus, the representation of $SO(3)$ on ${\mathbf L}_r({\mathbf H})$ is
not irreducible. It consists of a trivial representation on the subspace $a_1
= a_2 = a_3 = 0$ and the standard representation by the orthogonal $3\times
3$-matrices on the three-dimensional subspace $a_0 = 0$.

Each element of ${\mathbf L}_r({\mathbf H})$ can be considered as an
observable of the spin system. The positive elements with trace 1 can be
considered as states of the system. Let us denote by ${\mathbf T}({\mathbf
H})$ the set of states. Clearly, ${\mathbf T}({\mathbf H})$ is invariant with
respect to the action of $SU(2)$, so that $SU(2)$ also acts on it.

Let us determine which subset of ${\mathbf L}_r({\mathbf H})$ the space of
states is. The decomposition of ${\mathsf T}$ into Pauli basis defines its
components $t_j$, $j = 0,1,2,3$. Then the positivity conditions can be
expressed by:
$$
t_1^2 + t_2^2 + t_3^2 \leq 1/4\ ,
$$
and the trace condition by
$$
t_0 = 1/2\ .
$$
Thus, ${\mathbf T}({\mathbf H})$ can be identified with a three-dimensional
disk of radius $1/2$ with centre $(1/2,0,0,0)$ lying in the hyperplane $t_0 =
1/2$ in ${\mathbf L}_r({\mathbf H})$.

The most interesting states lie at the boundary of the disk and are called
{\em extremal} (more often ``pure''). Let us study these states. In terms of
the components in the Pauli basis, the matrix ${\mathsf M}_{\mathsf T}$ of
${\mathsf T}$ is
$$
{\mathsf M}_{\mathsf T} = \left(\begin{array}{ll}1/2 + t_3 & t_1 - it_2 \\ t_1
+ it_2 & 1/2 - t_3\end{array}\right)\ ,
$$
The determinant is easily calculated:
$$
\det({\mathsf T}) = 1/4 - (t_1^2 + t_2^2 + t_3^2)\ .
$$
Hence, the boundary points have zero determinant. As the determinant of matrix
${\mathsf M}_{\mathsf T}$ is the product of its eigenvalues and the trace
their sum, one of their eigenvalues is zero while the other is 1. Let us
denote the normalised eigenvector of ${\mathsf T}$ to the eigenvalue 1 by
$|T\rangle$. Then
$$
{\mathsf T} = |T\rangle \langle T|\ .
$$
It is an important result: every extremal state is a projection onto a
one-dimensional subspace of ${\mathbf H}$ and vice versa.

${\mathbf T}({\mathbf H})$ is not a linear space, but it is a convex subset of
${\mathbf L}_r({\mathbf H})$. If ${\mathsf T}_1,{\mathsf T}_2$ are two
operators, then their convex combination is defined as
\begin{equation}\label{defcc} c_1 {\mathsf T}_1 + c_2 {\mathsf T}_2
\end{equation} with $c_1 \geq 0$, $c_2 \geq 0$ and $c_1 + c_2 = 1$. From the
definition of ${\mathbf T}({\mathbf H})$ it follows that ${\mathsf T} = c_1
{\mathsf T}_1 + c_2 {\mathsf T}_2$ lies in ${\mathbf T}({\mathbf H})$ if
${\mathsf T}_1$ and ${\mathsf T}_2$ do.

Convex combination ${\mathsf T} = c_1 {\mathsf T}_1 + c_2 {\mathsf T}_2$ can
be represented by a line segment in the three-dimensional disk ${\mathbf
T}({\mathbf H})$ with end points ${\mathsf T}_1$ and ${\mathsf T}_2$. The
point corresponding to ${\mathsf T}$ lies then between the end points so that
it divides the line segment in the ratio $c_1:c_2$. This follows from the fact
that the linear structure of ${\mathbf L}_r({\mathbf H})$ is represented by
that of ${\mathbb R}^4$.

Using this geometric picture, we can see that the boundary points cannot be
written as convex combinations of other states. On the other hand, any
internal point ${\mathsf T}$ of ${\mathbf T}({\mathbf H})$ can be written as
convex combinations with any state ${\mathsf T}_1$ as the first
component. Indeed, just draw a line through ${\mathsf T}_1$ and ${\mathsf T}$
and choose any point ${\mathsf T}_2$ at this line that lies farther from
${\mathsf T}_1$ than ${\mathsf T}$.

In particular, if ${\mathbf T}_1$ is extremal (lies at the boundary), then
${\mathsf T}_2$ can also be chosen at the boundary, simply where the line
intersect the boundary for the second time. Thus, each state inside ${\mathbf
T}({\mathbf H})$ can be written in infinitely many ways as a convex
combination of extremal states.

However, from another point of view, any state ${\mathbf T}$ that is different
from the centre of the disk (proportional to the unit operator and called
``completely chaotic state'') defines a unique decomposition into a convex
combination of orthogonal extremal states. Indeed, it is just the spectral
decomposition of ${\mathbf T}$,
\begin{equation}\label{exampstatedec} {\mathbf T} = t_1 {\mathsf \Pi}_1 + t_2
{\mathsf \Pi}_2\ ,
\end{equation} where $t_1$ and $t_2$ are the eigenvalues of ${\mathbf T}$
while ${\mathsf \Pi}_1$ and ${\mathsf \Pi}_2$ are projections onto the
eigenspaces. The state at the centre of the sphere can be written as:
$$
\frac{1}{2}{\mathsf 1} = \frac{1}{2}|\psi_1\rangle \langle \psi_1| +
\frac{1}{2}|\psi_2\rangle \langle \psi_2|\ ,
$$
where $\{|\psi_1\rangle, |\psi_2\rangle\}$ is any basis of ${\mathbf H}$.

The interior of the disk ${\mathbf T}({\mathbf H})$ and each point of the the
boundary are examples of the so-called faces of the convex set ${\mathbf
T}({\mathbf H})$ (for a general definition of faces, see , \cite{ludwig1},
Chapter III, Section 6 (p.\ 75)). Roughly, a face of ${\mathbf T}({\mathbf
H})$ is an intersection of ${\mathbf T}({\mathbf H})$ with a hyperplane
${\mathbf V} \in {\mathbf L}_r({\mathbf H})$ that satisfied the condition: if
a point ${\mathsf T}$ of ${\mathbf T}({\mathbf H}) \cap {\mathbf V}$ can be
written as a convex combination of ${\mathsf T}_1,{\mathsf T}_2 \in {\mathbf
T}({\mathbf H})$ then ${\mathsf T}_1,{\mathsf T}_2 \in {\mathbf T}({\mathbf
H}) \cap {\mathbf V}$. Thus, extremal points form zero-dimensional faces of
${\mathbf T}({\mathbf H})$ and the interior of ${\mathbf T}({\mathbf H})$
itself is a three-dimensional face. In our case, the hyperplanes can be only
three- and zero-dimensional.

Observe that the geometry of the faces reflects the geometry of the Hilbert
space as follows. The three-dimensional face contains states associated with
the total Hilbert space while the zero-dimensional faces contain states
associated with the one-dimensional subspaces of the Hilbert space. This will
be a general picture. The association is defined as follows. Let ${\mathbf
H}_1$ be a subspace of ${\mathbf H}$ and ${\mathsf \Pi}_1$ be the projection
onto ${\mathbf H}_1$. State ${\mathsf T}$ is associated with ${\mathbf H}_1$
if
$$
{\mathsf T} = {\mathsf \Pi}_1{\mathsf T}{\mathsf \Pi}_1\ .
$$
It follows that ${\mathsf T} \in {\mathbf T}({\mathbf H}_1)$.

Let us now turn to the question of how observables can be measured and state
prepared. Stern and Gerlach \cite{SG} have measured the spin components by
observing the deflection of a neutral beam of silver atoms in an inhomogeneous
magnetic field in 1922. We shall use the account of an idealised experiment of
this kind given in \cite{peres}.

Silver atoms evaporate in an oven and pass through a velocity selector. Let us
call this preparation ${\mathcal P}$. To be registered, the resulting beam
crosses an inhomogeneous magnetic field in a magnet apparatus ${\mathcal M}$
and, finally, strikes a photo plate perpendicular to the direction of the
beam. All the impacts are then found in two narrow equally dense strips. Let
$\vec{n}$ be one of the two unit vectors parallel to the plate and
perpendicular to the strips. Vector $\vec{n}$ depends on the orientation of
the magnet. The two strips would be rotated by the angle by which the magnet
${\mathcal M}$ in the plane perpendicular to the beam were. In this way,
vector $\vec{n}$ also indicates the orientation of the magnet, that is the
orientation of the Stern-gerlach apparatus. Then, a registration by
Stern-Gerlach apparatus with orientation $\vec{n}$ can have only two outcomes:
the atom hits either the upper or the lower strip, where ``upper'' and
``lower'' are defined by $\vec{n}$ so that the upper strip lies in the
direction of $\vec{n}$ from the lower one.

Let us choose a Cartesian coordinate frame $(x_1,x_2,x_3)$ and describe the
orientation of the magnet by the components of a unit vector $\vec{n}$ in this
frame. Then the observable that is registered by the apparatus with
orientation $\vec{n}$ is the spin in direction $\vec{n}$. It is described by
the operator
$$
{\mathsf S}(\vec{n}) = \frac{1}{2}\hbar (n_1\sigma_1 + n_2\sigma_2 +
n_3\sigma_3)\ ,
$$
where $\sigma_k$ is the $k$-th Pauli matrix. The spin with orientation of the
third axis has eigenvalues $\pm \hbar/2$ and eigenvectors
$$
|\uparrow\rangle = (1,0)\,\quad |\downarrow\rangle = (0,1)\ .
$$
In this way, the basis of ${\mathbf H}$ in which Pauli matrices are defined is
associated with the frame $(x_1,x_2,x_3)$ and relation (\ref{rotL}) implies
that rotation of a Stern-Gerlach apparatus by ${\mathsf O}$ corresponds to the
$SU(2)$ transformation of the registered observable.

A Stern-Gerlach apparatus can also be used for preparations. Let preparation
${\mathcal P}(\vec{n})$ be defined as follows. The first part of it is as
above with the magnet ${\mathcal M}$ being set in direction $\vec{n}$. Then,
instead of a detector, a screen is placed in the way of the beams coming from
${\mathcal M}$ so that the lower beam is blocked off and the upper beam is let
through.

Now, we can embark on the discussion of states. RCU interpretation of states
as objective properties of individual systems enables us to introduce the
concept of state statistics that is different and independent of the
statistics of registration outcomes. To explain this point, we can use
Newtonian mechanics. A probability distribution on the phase space of a
Newtonian object can be called an {\em epistemic} state if the notion of
Newtonian mechanics is adopted that a system always {\em is} at a definite
point of the phase space. Then, the epistemic state expresses our incomplete
knowledge of the system. A point of the phase space is an example of a state
that provides a maximal possible information on the Newtonian object. States
of this kind can be called {\em ontic}.

Let us now study the epistemic-or-ontic question for quantum states. Consider
a preparation that uses two Stern-Gerlach apparatuses in a parallel
arrangement. That is, the first of them performs a preparation ${\mathcal
P}(\vec{n}_1)$ while the second does ${\mathcal P}(\vec{n}_2)$, and they are
oriented in such a way that the beams created by them are approximately
parallel but cross at some distance from the apparatuses. Let further the
intensity of the beams differ so that they have rates, ${\mathrm P}_1$ and
${\mathrm P}_2$, respectively, ${\mathrm P}_1 \neq {\mathrm P}_2$ and
${\mathrm P}_1 + {\mathrm P}_2 = 1$. The preparation of the state that results
at the intersection of the beams is called {\em statistical mixture} ${\mathrm
P}_1{\mathcal P}(\vec{n}_1) + {\mathrm P}_2{\mathcal P}(\vec{n}_2)$ of
preparations ${\mathcal P}(\vec{n}_1)$ and ${\mathcal P}(\vec{n}_2)$.

A general definition of statistical mixture of preparations is as follows:
\begin{df}\label{dfstatprep} Let ${\mathcal P}_1$ and ${\mathcal P}_2$ be two
preparation of $S$ and ${\mathrm P} \in [0,1]$.  {\em Statistical mixture}
\begin{equation}\label{statmixP} {\mathcal P} = {\mathrm P}{\mathcal P}_1 + (1
- {\mathrm P}){\mathcal P}_2
\end{equation} of the two preparations is a new preparation constructed as
follows. Let object ${\mathcal S}$ be prepared either by ${\mathcal P}_1$ or
by ${\mathcal P}_2$ in a random way so that ${\mathcal P}_1$ is applied with
probability ${\mathrm P}_1$ and ${\mathcal P}_2$ with probability $1 -
{\mathrm P}$.
\end{df} This definition can easily be extended to any number of
preparations. The preparation of the state at the intersection of the beams is
not completely known, but some knowledge about it is available (namely that it
prepares two known states, each with a known rate).

Now, we can formulate a standard assumption of quantum mechanics.
\begin{assump}\label{asconvcomb} Let preparations ${\mathcal P}_1$ and
${\mathcal P}_2$ preparing states ${\mathsf T}_1$ and ${\mathsf T}_1$ of
object ${\mathcal S}$ are {\em mixed with rates} ${\mathrm P}_1$ and ${\mathrm
P}_2$. The state prepared in this way is
\begin{equation}\label{2spinsprop} {\mathrm P}_1{\mathsf T}_1 + {\mathrm
P}_2{\mathsf T}_2\ .
\end{equation} It is called {\em proper mixture} of states ${\mathsf T}_1$ and
${\mathsf T}_2$.
\end{assump} The generalisation of this assumption to any quantum system is
easy. In literature, there are various names proposed for such a state: {\em
direct mixture} \cite{ludwig1}, {\em proper mixture} \cite{d'Espagnat} or {\em
Gemenge} \cite{BLM}.

It follows that a convex combination (\ref{defcc}) can also have a physical
meaning: it can be a proper mixture of states ${\mathsf T}_1$ and ${\mathsf
T}_2$. Suppose that state ${\mathsf T}$ is a such a proper mixture. Then it is
not an extremal state and hence decomposable into an infinite number of
non-equivalent convex combination. Only one of them has then a physical
meaning of proper mixture. The others are just mathematical possibilities of
how ${\mathsf T}$ can be written. The proper mixture does not differ from all
other convex combination by any mathematical property, only by the method of
preparation. The expression (\ref{defcc}) itself does not contain any
information on this preparation. Thus, we find it useful, to introduce a
special notation for those convex combination that are proper mixtures:
$$
{\mathrm P}_1{\mathsf T}_1 +_s {\mathrm P}_2{\mathsf T}_2\ .
$$

Proper mixtures are still states uniquely determined by their preparations and
hence they are objective properties of the prepared systems. This is similar
as in Newtonian mechanics. One can also understand it as follows. A proper
mixture gives a direct information on an ensemble of equally prepared states
and surely is an objective property of this ensemble. However, to be an
element of a particular ensemble is then also an objective property for an
individual subsystem of this ensemble.

For a state to be a proper mixture, there must be some observational reasons,
such as special preparations or other observations of macroscopic
objects. Even if, mathematically, there is a unique convex decomposition it
would be wrong to consider such a convex combination as a proper mixture. An
example of such a unique decomposition is the spectral decomposition (see,
e.g., \cite{ballent}, p.\ 20) of a state operator that has a non-degenerate
spectrum. Indeed, state (\ref{2spinsprop}) can also have a unique spectral
decomposition into extremal states but this decomposition does not correspond
to the way the state has been prepared.

The possibility that a convex combination can represent a proper mixture can
lure one into a conjecture that there is some deeper physical difference
between extremal states, which cannot be written as a non-trivial convex
compositions, and non-extremal states, which always can. Thus, extremal states
are often called ``pure'' and non-extremal ones ``mixtures''. However, a
non-extremal state lies inside a three-dimensional face and can be written as
convex combination in an infinity of different ways. If there is no further
reason to choose a particular combination, there is no reason either to
consider the state as a particular proper mixture. Such states are often
called ``improper mixtures''.

Quantum states are thus usually classified into extremal states, improper and
proper mixtures. We assume now:
\begin{sh}\label{shontic} Extremal states and improper mixtures are ontic,
proper mixtures are epistemic, quantum states.
\end{sh} TH \ref{shontic} refines Trial Hypothesis \ref{shobject} by stating
that the quantum states that are not proper mixtures give a maximum
information on quantum systems.

Observe that the notions of ontic and epistemic states we have introduced are
different from how such notions are understood within Minimum
Interpretation. Minimum Interpretation considers a whole measurement process,
including its preparation and its registration, as a real entity. The prepared
state and the registered value of an observable are objective properties of
each individual measurement process. However, the state does not determine
which value really occurs. In this sense, a state is always an epistemic
description of a measuring process. This is of course also true for RCU
interpretation but the state is then also an objective property of the
prepared system, for which it can be an ontic description. The registered
values become real only in registration. It is important to realise this
difference to prevent misunderstanding.

Examples of three different preparations that illuminate the ideas about
proper mixtures can be constructed for the system of two spins. Consider the
system composite from an electron (index $(2)$) and a positron (index $(1)$)
prepared in extremal state ${\mathsf T}_- = |\psi_-\rangle\langle \psi_-|$,
where
\begin{equation}\label{EPRBs} |\psi_-\rangle =
\frac{1}{\sqrt{2}}(|\uparrow\rangle^{(1)} \otimes |\downarrow\rangle^{(2)} -
|\downarrow\rangle^{(1)} \otimes |\uparrow\rangle^{(2)})\ .
\end{equation} Subsystem states
$|\uparrow\rangle^{(1)},|\downarrow\rangle^{(1)}$ and
$|\uparrow\rangle^{(2)},|\downarrow\rangle^{(2)}$ define basis
$|\uparrow\rangle^{(1)} \otimes |\uparrow\rangle^{(2)},
|\downarrow\rangle^{(1)} \otimes |\uparrow\rangle^{(1)},
|\downarrow\rangle^{(1)} \otimes |\downarrow\rangle^{(2)})$ of ${\mathbf H}
\otimes {\mathbf H}$.

Let Alice is going to register spins of the positron and Bob those of the
electron by Stern-Gerlach meters. Let the first experiment be such that Alice
does not do anything just leaving the positron to pass without
registration. Then the state prepared in this way for Bob is the partial trace
${\mathsf T}^{(2)} = tr^{(1)}({\mathsf T}_-)$, see \cite{ballent}, Section
8.3. An easy calculation gives:
$$
{\mathsf T}^{(2)} = \frac{1}{2}|\uparrow\rangle^{(2)}\langle \uparrow|^{(2)} +
\frac{1}{2}|\downarrow\rangle^{(2)} \langle \downarrow|^{(2)}\ ,
$$
which has the following components in the basis of ${\mathbf H}$:
$$
{\mathsf T}^{(2)} = \left(\begin{array}{ll}1/2 & 0 \\ 0 &
1/2 \end{array}\right)\ .
$$
This is a completely random state so that the probability of obtaining value
$+\hbar/2$ by registration of observable ${\mathsf S}^{(2)}(\vec{n})$ is $1/2$
independently of $\vec{n}$. Hence, ${\mathsf T}^{(2)}$ contains zero
information on the spin of the electron. Still, it is an ontic state in the
sense that it is not a proper mixture of states each of which contains any
non-trivial information about the spin. If this were the case, then each
individual electron arriving at Bob's detector would be objectively in some of
these more specific states. But this would contradict the rule that objective
states must be prepared. The studied example is thus an interesting case of
ontic state.

In the second experiment, let Alice register ${\mathsf S}^{(1)}_3$ on each
positron but do not record the outcomes. Each such registration creates one of
the states
$$
{\mathsf T}_\uparrow^{(2)} = |\uparrow\rangle^{(2)} \langle \uparrow|^{(2)}\
,\quad {\mathsf T}_\downarrow^{(2)} = |\downarrow\rangle^{(2)} \langle
\downarrow|^{(2)}
$$
with probability $1/2$ at Bob's laboratory. That is, the prepared state of the
electron is now a proper mixture, $1/2{\mathsf T}_\uparrow^{(2)} +_s
1/2{\mathsf T}_\downarrow^{(2)}$. This preparation of state ${\mathsf
T}^{(2)}$ is incompletely known by both Alice and Bob.

State ${\mathsf T}_-$ implies strong anticorrelations of spins
$S^{(1)}(\vec{n})$ and $S^{(2)}(\vec{n})$ for any unit vector $\vec{n}$. The
intuitive idea that a relation of $S^{(2)}$ to another system can be
considered as a property of $S^{(2)}$ motivates the following claim: The above
example shows that general rules of quantum mechanics imply the existence of
properties of system $S^{(2)}$ which, on the one hand, are prepared by the
above preparation but, on the other, are not encoded in its state ${\mathsf
T}^{(2)}$. The correlations between the spins of the electron and positron are
due to their entanglement and are described e.g.\ in Section 20.2 of
\cite{ballent}. Observe that the correlations {\em are} determined by a state,
namely ${\mathsf T}_-$ but it is a state of a composite system containing
$S^{(2)}$.

Hence, the first example suggests:
\begin{sh}\label{shobject1} Objective properties of a quantum system $S$ can
be divided into three classes: 1. structural properties of $S$, 2. a state of
$S$ and the properties determined uniquely by the state (such as expectation
values of a fixed observable), 3. the properties of the state of a system that
has been prepared so that it contains $S$ as a subsystem if such properties
concern $S$ but are not determined by the state of $S$ itself (such as the way
$S$ is entangled with other systems).
\end{sh} This will be a general hypothesis of RCU interpretation. It can also
be considered as a rigorous definition of objective properties of quantum
systems.

The fact that Bob's electron is in an ontic state in spite that it is totally
random must be interpreted as mathematical expression of the inefficiency of
the observables pertaining to an individual electron. If one registered
simultaneously also an observable of the positron, some non-trivial
information about the correlations between observables of the two systems can
be obtained.

The correlations between the spins of the positron and the electron encoded in
state ${\mathsf T}_-$ is e.g.\ a strong anticorrelation between the third
components of the spins. Thus, if Alice sees $+\hbar/2$, Bob must see
$-\hbar/2$. It is well-known that some kind of nonlocality is implied by
that. The registration apparatuses can be arbitrarily far away from each other
at the time of their simultaneous registrations. The electron spin value is
determined when the positron spin is registered, but not earlier. But the
positron spin is also only determined when it is registered. To be correlated
with the outcome for the electron, it must ``know'' the outcome for the
electron. This holds independently of whether we limit ourselves to the
Minimum Interpretation or add any assumption about reality of states, because
the correlations concern primarily outcomes of registrations, that is, states
of classical apparatuses. Within the Minimum Interpretation, the nonlocality
concerns spooky communication between remote meters, while within our
interpretation, the state of the composed object is a real ``thing'' extended
over two regions of the space that are far away from each other and it must
change abruptly in both regions if a registration is made in just one of
them. But the extension is ``abstract'' in the sense that it is not due to
different real (material) local subsystems as the extension of a classical
continuum is.

\section{Particles} Particles are quantum systems associated with the Hilbert
space ${\mathbf H}_s$ constructed from the complex linear space of functions
$\psi(\vec{x},m)$, $\vec{x} \in {\mathbb R}^3$, $m = -s,\ldots,s$ and $s =
0,1/2,1,\ldots$ The Hilbert space caries an irreducible unitary representation
of central extension $\bar{G}^+_\mu$ of proper Galilean group $G^+$. The
central extension depends on the mass $\mu$ of the particle (for details, see
Chapter XII, Section 8 of \cite{varad}).

The most important difference between the spin and particle systems is that
the space and time aspects of quantum mechanics are missing for the former and
fully accounted by the latter. This allows a further development of the ideas
introduced in the previous section.

In Section 1.3, a special sign, $+_s$, for proper mixing of states has been
introduced. Now, we can discuss the proper mixing in more detail. First, as a
mathematical operation, $+_s$ is commutative and associative:
$$
{\mathrm P}_1{\mathsf T}_1\ +_s\ {\mathrm P}_2{\mathsf T}_2 = {\mathrm
P}_2{\mathsf T}_2\ +_s\ {\mathrm P}_1{\mathsf T}_1
$$
for any two states ${\mathsf T}_1$, ${\mathsf T}_2$ and rates ${\mathrm P}_1$,
${\mathrm P}_2$, and
$$
({\mathrm P}_1{\mathsf T}_1\ +_s\ {\mathrm P}_2{\mathsf T}_2)\ +_s\ {\mathrm
P}_3{\mathsf T}_3 = {\mathrm P}_1{\mathsf T}_1\ +_s\ ({\mathrm P}_2{\mathsf
T}_2\ +_s\ {\mathrm P}_3{\mathsf T}_3)
$$
for any three states ${\mathsf T}_1$, ${\mathsf T}_2$, ${\mathsf T}_3$ and
rates ${\mathrm P}_1$, ${\mathrm P}_2$, ${\mathrm P}_3$. This follows directly
from the definition.

Second, the definition also directly implies that proper mixing is independent
of representation and invariant with respect to transformation by Galilean
group,
$$
{\mathsf U}({\mathrm P}_1{\mathsf T}_1\ +_s\ {\mathrm P}_2{\mathsf
T}_2){\mathsf U}^\dagger = {\mathrm P}_1{\mathsf U}{\mathsf T}_1{\mathsf
U}^\dagger\ +_s\ {\mathrm P}_2{\mathsf U}{\mathsf T}_2{\mathsf U}^\dagger\ ,
$$
where ${\mathsf U}$ may be both the unitary transformation between
representations and transformation by a representative of an element of
Galilean group. In particular, a proper mixture remains a proper mixture
during time evolution.

Finally, the invariance of the proper mixture with respect to system
composition follows also from the definition of $+_s$ \cite{survey}:

\vspace{0.5cm}

\noindent {\bf Composition Invariance of Proper Mixture} {\it Let the state of
composed system $S^{(1)} + S^{(1)}$ be ${\mathsf T}$. The necessary and
sufficient condition for state $tr^{(2)}({\mathsf T})$ of $S^{(1)}$ to be a
proper mixture,
$$
tr^{(2)}({\mathsf T}) = {\mathrm P}_1{\mathsf T}^{(1)}_1\ +_s\ {\mathrm
P}_2{\mathsf T}^{(1)}_2\ ,
$$
where ${\mathsf T}^{(1)}_1$ and ${\mathsf T}^{(1)}_2$ are some states of
$S^{(1)}$, is that ${\mathsf T}$ itself is a proper mixture of the form
$$
{\mathsf T} = {\mathrm P}_1{\mathsf T}^{(1)}_1 \otimes {\mathsf T}^{(2)}_1\
+_s\ {\mathrm P}_2{\mathsf T}^{(1)}_2 \otimes {\mathsf T}^{(2)}_2\ ,
$$
where ${\mathsf T}^{(2)}_1$ and ${\mathsf T}^{(2)}_2$ are some states of
$S^{(2)}$.}
\par\vspace{0.5cm}

The question of whether the states $w{\mathsf T}_1\ +_s\ (1-w){\mathsf T}_2$
and $w{\mathsf T}_1 + (1-w){\mathsf T}_2$ of object ${\mathcal S}$ can be
distinguished by registrations is interesting and important. State operator
${\mathsf T}$ does not, by itself, determine the statistical decomposition of
a prepared state described by it, unless ${\mathsf T}$ is extremal so that its
every convex decomposition is trivial. Registrations that are limited to
observables of $S$ cannot distinguish proper and improper mixtures because the
registration probabilities depend only on the state operator.

However, if observables of composite systems containing $S$ are also admitted,
then the difference of the two states can be found by measurements, for
instance if the state of the composite system is extremal and Composition
Invariance of Proper Mixture holds. Another aspect of the distinction is the
following. If one part of the studied system is a macroscopic object with both
classical and quantum models such as a registration apparatus, and if the
state of the whole system is a convex combination of two quantum states each
of them associated with a different classical state of the macroscopic object,
then such a convex combination must be a proper mixture because the
macroscopic object is always only in one of the two classical states. We shall
return to this in Chapter 5.

Extremal states allow another mathematical operation, a linear superposition,
which is different from a convex combination. If
$$
|\psi\rangle = c_1|\psi_1\rangle + c_2|\psi_2\rangle\ ,
$$
the corresponding state operator is
\begin{equation}\label{pures} |\psi\rangle \langle \psi| =
|c_1|^2|\psi_1\rangle\langle \psi_1| + c_1c_2^*|\psi_1\rangle\langle \psi_2| +
c_1^*c_2|\psi_2\rangle\langle \psi_1| + |c_2|^2|\psi_2\rangle\langle \psi_2|\
.
\end{equation} This differs from the convex combination of the two states,
$$
{\mathsf T} = |c_1|^2|\psi_1\rangle\langle \psi_1| +
|c_2|^2|\psi_2\rangle\langle \psi_2|\ ,
$$
by the non-diagonal (cross) terms. The difference can be revealed by the
registration of suitable observables. For example, expectation values of
observable ${\mathsf O} = |\psi_1\rangle \langle \psi_2| + |\psi_2\rangle
\langle \psi_1|$, for orthogonal states $|\psi_1\rangle$ and $|\psi_2\rangle$,
are: $\langle \psi |{\mathsf O}|\psi\rangle = c^*_1c_2 + c^*_2c_1$ and
$tr({\mathsf O}{\mathsf T}) = 0$. The non-diagonal terms also describe
correlations that can be revealed by registrations. The cross terms lead also
to interference phenomena (such as the electron interference in interference
experiments, e.g.\ \cite{tono}).

If a preparation is not completely known, we can still assume that it prepares
a state described by some state operator of which we do not know whether it is
a proper or improper mixture. In many cases, the structure of a possible
proper mixture is not important because sufficiently many properties of the
state are independent of it and everything one needs can be obtained from the
state operator.

It may be helpful to compare quantum states of our interpretation with states
of Newtonian mechanics. Let us define a state of an arbitrary Newtonian system
as a point of the (many-dimensional) phase space ${\mathbf \Gamma}$ of the
system. Newtonian state defined in this way is generally assumed to satisfy:
\begin{enumerate}
\item {\em objectivity}: a state of a system is an objective property of the
system,
\item {\em generality}: any system is always in some state,
\item {\em exclusivity}: a system cannot be in two different states
simultaneously,
\item {\em completeness}: any state of a system determines the values of all
observables that can be measured on the system,
\item{separability}: the state of a composite system is determined by the
states of its subsystems,
\item {\em locality} the state of a system determines the space position of
the system, that is positions of all its subsystems.
\end{enumerate} An incomplete information about the state of a system can be
described by a probability distributions on ${\mathbf \Gamma}$. Indeed,
because of the generality, the system always is at a particular point of
${\mathbf \Gamma}$, but we do not know at which. Such a distribution is
sometimes called {\em statistical state}. In any case, we distinguish a state
from a statistical state.

As was mentioned in Section 1.3, statistical states can be called epistemic
and points of ${\mathbf \Gamma}$ can be called ontic. This distinction depends
clearly on hypotheses about what can and what cannot be known\footnote{Of
course, this is based on the hypotheses that point-like states really
exist. The assumption of real existence of phase space points is, however,
surely incorrect, if quantum mechanics is valid. We shall study these
questions in more detail in Chapter 2.}.

In quantum mechanics, we have the following picture: The space of quantum
ontic states is ${\mathbf T}({\mathbf H}_s)$ (more details about the structure
of this space can be found in Chapter III, Section 6 of \cite{ludwig1}; an
example of it for the two-dimensional Hilbert space is described in the
section on the spin system). Epistemic quantum states---proper mixtures---can
be described by probability distributions on ${\mathbf T}({\mathbf H}_s)$.

Here, a ``state of a system'' is interpreted as a state that is prepared for
the system. Then, point 1., 2. and 3. are also valid for quantum
mechanics. Point 4. does not hold in quantum mechanics because each ontic
state of a system determines only probability distributions of values of its
observables. Still, the information given by an ontic state is maximal in the
following sense. If the same ontic states ${\mathsf T}$ is repeatedly prepared
then each element of the ensemble of systems obtained in this way cannot be
considered to be in another state than ${\mathsf T}$. That an improper mixture
can give a maximum information exactly as a pure state does is also a
consequence of Composition Invariance of Proper Mixture. Indeed, let, in
Composition Invariance of Proper Mixture, ${\mathsf T}$ be extremal and
${\mathsf T}^{(1)}$ be not. If there is more information to have on system
$S^{(1)}$ than that contained state ${\mathsf T}^{(1)}$ then there is also
more information on the composed system $S$ than state ${\mathsf T}$
provides. Indeed, state ${\mathsf T}$ is to give a maximum information on $S$
and thus on $S^{(1)}$ and ${\mathsf T}^{(1)}$ gives as much information on
$S^{(1)}$ as ${\mathsf T}$ does.

As we have seen in the previous section, Point 5 does not hold in quantum
mechanics. Finally, the position of a subsystem is an observable and it has,
therefore, no predetermined value if the ontic state of a composite system is
known. Hence, Point 6. is also wrong in quantum mechanics.

The properties that distinguish quantum objects from each other have been
classified and listed by TH \ref{shobject1}. The arguments and motivation for
this TH have been given there for a two-spin system. Analogous arguments and
motivations are valid for general systems defined in the present section.

Another difference between quantum and Newtonian mechanics is that a proper
mixture can be mathematically represented by the same state operator as an
ontic state while Newtonian ontic and epistemic states are always represented
by different mathematical entities.

\chapter{Maximum-entropy packets} \setcounter{equation}{0} \setcounter{thm}{0}
\setcounter{assump}{0} \setcounter{df}{0} \setcounter{sh}{0} According to
Trial Hypothesis \ref{rhmacro}, the physical objects that have classical
models also admit quantum models. The subject of the present chapter is the
questions of what the relation between the classical and the quantum model of
an object is, how such quantum models can be constructed and how the classical
properties can be derived from them.

A well-known problem of quantum theory of classical properties is that some
obvious features of the classical models seem incompatible with quantum
mechanics. Let us list the most important features of this kind.
\begin{enumerate}
\item Every classical system is always only in one of its possible classical
states independently of whether it is observed or not.
\item Every classical system possesses all its properties independently of
whether it is observed or not. Classical properties are objective.
\item Classical systems are durable, that is they do not suddenly appear or
disappear except in very special cases.
\item Classical properties (including states) are robust, that is, a classical
measurement can be done in such a way that the state of the measured object is
arbitrarily weakly disturbed.
\end{enumerate}

One important consequence of Point 2 and some properties of classical models
is the existence of {\em sharp trajectories}: sharp values of classical
observables, such as position, momentum, field strengths, charge current, as
well as temperature and internal energy of phenomenological thermodynamics,
can be ascribed to the objects independently of whether they are observed or
not. Another one is that of {\em no superpositions}: Nobody has ever seen a
chair, say, to be in a linear superposition of being simultaneously in the
kitchen as well as in the bedroom.

Hence, we shall have at least the following difficulties.
\begin{description}
\item[A] According to Trial Hypothesis \ref{OCR}, values of quantum
observables are only created by registrations. Then Point 2 would lead to a
problem, if classical observables were too closely related to quantum ones.
\item[B] The basic classical assumption of sharp trajectories do not seem
easily compatible with the Heisenberg uncertainty relations.
\item[C] Attempts to get quantum trajectories as sharp as possible lead to
states with minimum uncertainty. But the quantum states of minimum uncertainty
are extremal and such states can be linearly superposed.
\item[D] Quantum states are not disturbed only by measurements of very few
very special observables (for more discussion, see Refs.\
\cite{leggett,BLM}). Especially, the extremal states are rather fragile. Then
Point 4 would lead to a problem if classical states are too closely related to
quantum ones.
\end{description}

There are various proposals of how at least some of these difficulties could
be dealt with. For example, one assumes that some phenomena exist at the
macroscopic level which are not compatible with standard quantum
mechanics. Such phenomena may prevent linear superpositions of extremal states
(see, e.g., \cite{leggett} and the references therein). Or they can lead to a
spontaneous dissipation of extremal states. The dynamical collapse theory
(see, e.g., \cite{ghirSEP}) is of this kind. The second example are theories
based on the idea that certain kind of coarse-grained operators
\cite{kampen,poulin,Kofler} associated with macroscopic systems are measurable
but fine grained are not. The third example are the Coleman-Hepp theory
\cite{Hepp,Bell3,Bona} and its modifications \cite{Sewell,Primas,wanb}: they
are based on some specific theorems that hold only for infinite quantum
systems (see the analysis in \cite{Bell3}) or for asymptotic regions
\cite{wanb}. Other examples assume that the macroscopic realism is only
apparent in the sense that there {\em are} linear superpositions of
macroscopic states but the corresponding correlations are difficult or
impossible to observe. For example, the quantum decoherence theory
\cite{Zeh,schloss,Zurek} works only if certain observables concerning both the
environment and the quantum system cannot be measured (see the analysis in
\cite{d'Espagnat,bub}). Here, also the so-called modular interpretations
belong (\cite{bub}).

Admittedly, the list is too concise and rather incomplete. Its only purpose is
to suggest that there {\em are} problems and there is a vast literature on
them. However, the aim of this chapter is to focus on our approach, which is
new and very different. It rejects sharp trajectories as an idealisation and
is limited to looking for quantum derivation of only those classical
properties that are themselves fuzzy. This opens a way to an application of
statistical methods. A motivation of such an approach is that some classical
properties have already been successfully derived from quantum mechanics: the
properties are the thermodynamic ones and the method is that of quantum
statistical thermodynamics. We are going to generalise statistical methods to
Newtonian mechanics.

\section{Hypothesis of High-Entropy States} To see how thermodynamic
properties are derived, look at a vessel of gas in a laboratory. Let us denote
the gas object by ${\mathcal S}$. First, we describe a classical model $S_c$
of it. Let the gas be in thermodynamic equilibrium, the volume of the vessel
be $\Omega$, the gas pressure be $P$, its temperature be $T$ and its mass
$M$. As for the chemical composition, let the gas be the monoatomic helium. We
can calculate various thermodynamic quantities using formulas of
phenomenological thermodynamics. For example, the number of molecules $N$ is
given by $MN_A/M_{\text{mol}}$, where $N_A$ is the Avogadro number and
$M_{\text{mol}}$ is the molecular weight, while the internal energy $E$ is
given by
$$
E = \frac{3}{2}kNT\ ,
$$
where $k$ is the Boltzmann constant. In this way, the classical model $S_c$ of
${\mathcal S}$ is defined.

Second, let us construct a quantum model $S_q$ of ${\mathcal S}$: the system
of $N$ spin-zero point particles, each with mass $\mu = M_{\text{mol}}$, in a
deep potential well of volume $\Omega$ with Hamiltonian
$$
{\mathsf H} = \sum_{k=1}^N\frac{|\vec{\mathsf p}^{(k)}|^2}{2\mu}\ ,
$$
where $\vec{\mathsf p}^{(k)}$ is the momentum of $k$-th particle in the rest
system of $\Omega$. ${\mathsf H}$ is simultaneously the operator of the
internal energy of $S_q$ (total energy in a rest frame).

An important assumption of the quantum model is the choice of state. It is the
state that maximises the von Neumann entropy\footnote{For definition, see
e.g.\ Section 9-1 of \cite{peres}.} for fixed value $E$ of the expectation
value of the internal energy. Such state is called {\em Gibbs state}. All
properties of $S_c$ can then be calculated from $S_q$ as properties of the
Gibbs states.

We can of course see that both models $S_c$ and $S_q$ are incomplete pictures
of object ${\mathcal S}$. This is a general property of models. Still, we have
a classical system and a quantum system and a definite relation between the
two: they ought to refer to one and the same object. This relation can be made
deeper if we realise that the classical system $S_c$ can play the role of a
preparation apparatus for $S_q$ or, at least, an essential part of such an
apparatus. Then, the system $S_q$ and its quantum state are determined by
$S_c$ and its classical state. Moreover, $S_c$ can also play the role of a
meter for $S_q$. Indeed, by observing a classical property of $S_c$, we obtain
an information on $S_q$. For instance, if we (macroscopically) isolate the
vessel and keep it so for some time the gas settles in a state of
thermodynamic equilibrium. This is achieved with the help of some further
(macroscopic) tools different from the system $S_c$ alone. Similarly, if we
measure the temperature of the $S_c$, we use a classical system, the
thermometer, that is different from $S_c$. However, $S_c$ itself must take
part in these procedures and as it is ``at the classical side'' of the
experiments, it is an important part of the preparation and registration of
$S_q$.

We also know that many classical properties of classical systems can be
observed directly via our senses, that is without mediation of any other
registration devices and that such observations do not disturb the observed
classical system. As mentioned at the beginning of the present chapter, this
is a feature of classical systems and we shall try to explain it by the
quantum mechanics of the corresponding quantum model.

The thermodynamic example motivates a general trial hypothesis about a
relation between classical and quantum models of the same object. The
assumption works at least in the case of thermodynamics and it can be
formulated as follows.
\begin{sh}\label{asobjectcq} Let $S_c$ be a classical model and $S_q$ a
quantum model of object ${\mathcal S}$. Then $S_c$ can be considered as an
essential part of the preparation device and, simultaneously, as an essential
part of a meter, for the quantum model $S_q$. The meter in question registers
values distinguishing the quantum states of $S_q$ that are associated with
different classical states of $S_c$.
\end{sh}

The main principle of our theory is a generalisation of the Gibbs-state idea
to all classical properties, including the mechanical ones. Thus, we
supplement TH \ref{asobjectcq} by the following trial typothesis:
\begin{sh}\label{ashighS} Let a real object ${\mathcal S}$ has a classical
model $S_c$ and a quantum model $S_q$. Then all properties of $S_c$ are
selected properties of some high-entropy states of $S_q$.
\end{sh} This hypothesis is only possible if some of already stated trial
hypotheses hold true, such as TH \ref{rhmacro} and
\ref{asobjectcq}. Hypothesis \ref{ashighS} is a heuristic one and it is
therefore formulated a little vaguely. It will be made clearer by examples of
its use studied in this chapter. But some examples can make it clearer already
now.

Consider first states of macroscopic systems that are at or close to absolute
zero of temperature. These are approximately or exactly extremal and maximize
entropy at the same time but the entropy, though it could be maximal, is not
high. We are not going to consider these objects as classical. Second,
consider states of macroscopic systems at room temperature that are not at
their thermodynamic equilibrium but are close to it. There are many such
states and they and the systems can be described by classical physics to a
good approximation. They are not in maximum- but in high-entropy states. As
the final example, consider a mechanical watch such as an old reliable Swiss
chronometer. It is composed of a great number of small mechanical subsystems,
cogs, shafts, bearings, etc., that are ordered according to an ingenious plan,
so that the state of the system composed of such subsystems as units can be
considered to have zero entropy. However, each such subsystem is macroscopic
and its internal state is close to thermal equilibrium that is to the state of
maximum entropy. The loss of entropy due to the order of the mechanical
subsystems is negligible with respect to the sum of their internal
entropies. Hence, the watch can also be considered as being in a high-entropy
state.

An important advantage of TH's \ref{asobjectcq} and \ref{ashighS} is that they
suggest ways in which the problems mentioned at the start of this chapter can
be solved. Problem A could be approached as follows. Our theory of objective
properties of quantum systems in Sections 1.2 (TH \ref{shobject}), 1.3 and 1.4
justifies the assumption that quantum states, even the high-entropy ones, are
objective. If classical properties are properties of some states of the
quantum model, they will also be objective. For example, the classical
internal energy and pressure can be assumed to be the expectation values of
some quantum observables in a high-entropy state. Also, there are many
classical properties of real objects that have been successfully modelled by
quantum mechanics, such as temperature, entropy, electrical conductivity or
specific heats that are not expectation values of quantum observables. They
are still properties of some high-entropy states.

Problem C is connected with the assumption that classical systems are modelled
on coherent states of the corresponding quantum model and with the fact that
the coherent states are extremal. If we however look at any object of our
everyday experience, such as a chair, we can immediately see that, as a
quantum system, it cannot be in an extremal state: it is near its
thermodynamic equilibrium at the room temperature. In principle, it might be
possible to prepare a macroscopic quantum system in an extremal state, but it
is fiendishly difficult. One had to bring its entropy to zero for that! Hence,
not only is its state a high-entropy one, but it is also no proper mixture of
extremal states. According to Assumption 2.3.5, existence of such a proper
mixture could only be justified if each of the extremal states were extra
prepared. For the above reasons, such a preparation is practically impossible.

As for Problem D, we first observe that any classical state in our sense is a
state of a classical model. For example, in the above example of the ideal
gas, the classical state is determined by values of three quantities: the
volume, the particle number and the internal energy. From the quantum point of
view, the volume is an external field while the energy and particle number are
expectation values of the internal energy and the particle number
operators. Although the three values determine the classical state uniquely,
many different quantum states are compatible with the same three values. Then,
even if quantum states may be disturbed by an observation, the corresponding
classical states need not be. This idea of a conceptual difference between
quantum and classical state is essential for our theory. For example, in the
theory of Leggett-Garg inequality \cite{LG}, such a difference is ignored.

Finally, Problem B is solved in phenomenological thermodynamics by accepting
that the sharpness is only an idealisation or approximation and the ``sharp''
values are in fact fuzzy.

Of course, even if such project of constructing quantum models corresponding
to classical ones worked nicely the question would remain open of what is the
origin of all the ontic high-entropy states that are observed in such a great
abundance around us. The physical foundations of thermodynamics are not yet
completely understood but there are many ideas around about the origin of
high-entropy states. Their existence might follow partially from logic
(Bayesian approach, \cite{Jaynes}) and partially from quantum mechanics
(thermodynamic limit, \cite{thirring}, Vol.\ 4). Some very interesting models
of how maximum entropy quantum states come into being are based on
entanglement \cite{GMM, short1,short2,goldstein}). In particular, in
\cite{GMM} a spontaneous evolution to such a state of a system $S_q$ is
proved. The main premises of the proof are that $S_q$ is a subsystems of
$S_q'$ that $S_q'$ is in an extremal state and that there is a weak
interaction between $S_q$ and the rest of $S_q'$. It then would follow from
our analysis in Section 1.3 that the state of $S_q$ could not be a proper
mixture at any time.  However, even if we do not know the physical cause of
hight-entropy states, we can just try to construct quantum models of classical
properties and the high-entropy states can be used as one of the assumptions
without really understanding their origin.

An understanding of classical macroscopic world mediated by quantum mechanics
would not be complete without a quantum theory of classical measuring
apparatuses. It seems that the final stage of many classical measurements is a
sight by a human eye. The pictures that are formed with the help of eyes are
relatively smooth functions on a two-dimensional plane representing the
intensities of light of different small frequency intervals.

From the quantum point of view, they are apparently results of a huge number
of photons registered simultaneously and not decomposed into the single photon
events. The retina contains millions of cone cells and each such cell can give
its signal only if a number of photons, as a rule tens to hundreds,
simultaneously or at least in short intervals after each other, hit the cell
so that a sufficiently high potential is built in it. Thus, the eye
accumulates and synthesise the individual photon registrations while an ideal
quantum measurement apparatus must be able to detect and distinguish
individual photons.

Similarly, a photo-emulsion works. An emulsion consists of millions of
silver-halogen corns, tiny crystals of diameter about 100 nm. Each corn must
be hit by at least four photons so that pure silver impurities form in the
crystal. These impurities are condensation centres around which the developer
causes a change in the structure of the corn from silver-halogen to pure
silver. Again, a continuous field of intensities of different colours is the
result, which is not decomposed into individual photon registrations. If we
know the bulk intensities around each point and the energies of the photons of
the corresponding colour, we could do the analysis and find approximately the
intensity and colour distribution over the picture from which the quantum
probabilities of the captured photons could be established.

The light that is registered by these apparatuses in the described way is what
is called in optics ``incoherent light'': clouds of many photons in a state of
high entropy. It seems therefore, that classical measuring apparatuses are
quantum measuring apparatuses that are deficient in the sense that they only
register if clouds of a large number of particles arrives at them. The
apparatuses are constructed in such a way that the results of their
registrations are associated with probability distributions of quantum
values. On the one hand, the probability distributions can be considered as
properties of the registered quantum states. On the other, the apparatus can
then be considered as registering directly a classical property. We shall call
such a measurement a {\em cumulative measurement}. These ideas are not new. We
find, e.g., in \cite{JvN}, p.\ 4:
\begin{quote} \ldots the idea that the principle of continuity (``natura non
facit saltus''), prevailing in the perceived macroscopic world, is merely
simulated by an averaging process in a world which in truth is discontinuous
by its very nature. This simulation is such that man generally perceives the
sum of many billions of elementary processes simultaneously, so that the
levelling law of large numbers completely obscures the real nature of the
individual processes.
\end{quote}

The registrations of the eye or the camera are relatively simply related to
the similar quantum registrations. However, many classical measurements are
less directly related to registrations of quantum observables. Examples are
measurements of the temperature, of the heat capacities or of the
pressure. Still, we assume that even such classical measurements ultimately
give information about some quantum observables.

\section{High-entropy states in Newtonian mechanics} There seems to be a
difficulty with TH \ref{ashighS}. For a body such as the chair, the Galilean
invariance of quantum theory leads to the separation of the bulk motion from
all other degrees of freedom. The motion of mass centre and of the total
angular momentum with respect to the mass centre is described by Newtonian
mechanics. It comprises only six degrees of freedom while statistical methods
seem to show their full power for systems with a huge number of weakly
coupled degrees of freedom.

Let us look more closely at Newtonian mechanics. One of its basic hypotheses
is that any system at any time is objectively at some point of its phase
space. The time dependence of this point forms a trajectory, a curve in the
phase space. Let us call this {\em Sharp Trajectory Hypothesis} (STH). In
Newtonian mechanics, states may be more general than points of phase space:
probability distributions on the phase space are also viewed as states. The
points have been called ontic, the non-trivial distributions epistemic states
in Section 1.3. This is justified by STH: a non-trivial probability
distribution describes the (incomplete) state of our knowledge on the
system. As already observed in Section 1.3 this distinction between ontic and
epistemic states is well-defined only if there is some assumption on what can
in principle be known, represented here by the STH.

If we ask what is the evidence supporting the STH, the problem emerges that
any measurement of the position and momentum of a classical body is afflicted
with an uncertainty. What we really know from any carefully done experiment
are expectation values $Q$ and $P$ and variances $\Delta Q$ and $\Delta P$ of
positions and momenta. In fact, for a macroscopic body, we always have
$$
2\Delta Q\Delta P \gg \hbar\ ,
$$
where ``$\gg$'' represents many orders of magnitude (for the definition of
variance, see e.g.\ \cite{ballent}, p.\ 223).

There can be different attitudes concerning this fact.
\begin{enumerate}
\item With improving techniques, the expression on the left-hand side will
decrease eventually approaching zero. This is wrong if quantum mechanics is
valid.
\item A sharp trajectory is just a model of a real motion. Thus, it is
approximative and describes only some aspects of the motion. The model is
considered as valid if the sharp trajectory lies within the ``tube'' in the
phase space that is defined by the measured expectation values and
variances. This corresponds well with the common experimental practice, as
well as with Constructive Realism.
\item Another model of a real motion is a time-dependent probability
distribution on the phase space that have suitable expectation values and
variances. This seems to be a more accurate model of what is really observed.
\end{enumerate}

If we assume that quantum mechanics is true then the sharp trajectories used
in Points 2 and 3 do not exist in real world and are only a simplifying
assumptions allowing constructions of nice models. Actually, the probability
distribution of Point 3 is not measurable because the pointlike states do not
exist. It cannot therefore be considered as a purely epistemic state because
no knowledge of the sharp trajectories is possible.

These considerations motivate an understanding of Newtonian mechanics that is
different from the common one. Such an understanding is not new: the point of
view that the statistical character of classical observational results must
not only be due to inaccuracy of observational methods but also to genuine
uncertainty of quantum origin is due to Exner \cite{Exner}, (``physical laws
are only average laws'') p.~669, and Born \cite{Born} (the title: ``Is
classical mechanics really deterministic?''). It can be formulated as follows:

\vspace{0.5cm}

\noindent {\bf Exner-Born Conjecture} {\it States of Newtonian systems that
are described by sharp points of the phase space do not exist. Newtonian
models that can approach the reality better are non-trivial probability
distribution function on the phase space.}

\vspace{0.5cm}

\noindent The Conjecture suggests a change of interpretation of classical
theories and, with it, a change of expectation of what is to be approximately
obtained from quantum mechanics in the classical limit.

However, most physicists take the existence of sharp trajectories seriously
and try to obtain them from quantum mechanics as exactly as possible. Hence,
they focus at quantum states the phase-space picture of which is as sharp as
possible. That are states with minimum uncertainty allowed by quantum
mechanics. For one degree of freedom, described by coordinate ${\mathsf q}$
and momentum ${\mathsf p}$, the uncertainty is given by the quantity
\begin{equation}\label{uncertainty} \nu = \frac{2\Delta {\mathsf q}\Delta
{\mathsf p}}{\hbar}\ .
\end{equation}

The states with minimum uncertainty $\nu = 1$ are, however, very special
extremal states. Such states do exist for macroscopic quantum systems but are
very difficult to prepare unlike the usual states of macroscopic systems that
we observe around us. As explained at the beginning of the chapter, they also
have properties that are strange from the point of view of classical theories
and they are therefore not what can be successfully used for modelling of
classical systems. Thus, there are some reasons to abandon STH. The Exner-Born
interpretation is not only more realistic but it also makes TH \ref{ashighS}
applicable to Newtonian mechanics.

Let us discuss the objectivity of fuzzy states of classical objects. In
quantum mechanics, the basis of objectivity of dynamical properties is the
objectivity of the conditions that define preparation procedures. In other
words, if a property is uniquely determined by a preparation, then it is an
objective property. If we look closely, one hindrance to try the same idea in
classical theories is the custom to speak always about initial data instead of
preparations. An initial datum can be and mostly is a sharp state. The
question of exactly how a sharp state can come into being is ignored. This in
turn seems justified by the hypothesis that sharp states are objective, that
is, they just exist by themselves. It seems however also possible to accept
the idea that preparation procedures play the same basic role in the classical
as in quantum physics. Then, the nature and form of necessary preparation
procedures must be specified and the corresponding states described. Let us
give an example.

Consider a gun in a position that is mechanically fixed and that shoots
bullets using cartridges of a given provenance. All shots made under these
conditions form an ensemble with expected trajectory
$(\vec{Q}_{\text{gun}}(t),\vec{P}_{\text{gun}}(t))$ and the trajectory
variance $(\Delta \vec{Q}_{\text{gun}}(t),\Delta \vec{P}_{\text{gun}}(t))$
that describe objective properties of the ensemble. The Newtonian model of
this ensemble is the evolution $\rho_{\text{gun}}(\vec{Q},\vec{P};t)$ of a
suitable distribution function on the phase space.

The simplest construction of a fuzzy model is to fix initial expectation
values and variances of coordinates and momenta, $Q^k$, $\Delta Q^k$, $P^k$,
$\Delta P^k$, and consider everything else as unknown. This opens the problem
to application of Bayesian methods, see, e.g., \cite{Jaynes}, which recommend
maximising entropy in the cases of missing knowledge (see Section E.2 and
\cite{Jaynes}, Chapter 11). Let us define a fuzzy state called {\em
maximum-entropy packet} (ME packet) as the phase-space distribution maximising
entropy for given expectation values and variances of mechanical state
coordinates. Of course, this is a particular choice that represents only one
in a large number of possibilities.

For example, the above definition depends on the coordinates $q$ and $p$, and
it can be shown that it is not canonically invariant. Choose, e.g., $q'$ and
$p'$ defined by the following canonical transformation:
\begin{eqnarray*} q' &=& \frac{1}{\sqrt{2}}q - \frac{1}{\sqrt{2}}p\ , \\ p'
&=& \frac{1}{\sqrt{2}}q + \frac{1}{\sqrt{2}}p\ .
\end{eqnarray*} Then,
\begin{multline*} \Delta Q^{\prime 2} = \left\langle \left[\frac{1}{\sqrt{2}}q
- \frac{1}{\sqrt{2}}p - \frac{1}{\sqrt{2}}Q +
\frac{1}{\sqrt{2}}P\right]^2\right\rangle \\ = \left\langle \frac{1}{2}(q -
Q)^2 + \frac{1}{2}(p - P)^2 - (q - Q)(p - P)\right\rangle = \frac{1}{2}\Delta
Q^2 + \frac{1}{2}\Delta P^2 - \langle(q - Q)(p - P)\rangle\ ,
\end{multline*} where $\langle x \rangle$ represents the expectation value of
quantity $x$ in the state that is under consideration.

But $\langle(q - Q)(p - P)\rangle$ is the correlation function of the
variables $q$ and $p$ and it is not determined by $Q$, $P$, $\Delta Q$ and
$\Delta P$. Hence, the condition that $Q$, $P$, $\Delta Q$ and $\Delta P$ are
fixed is in general not equivalent to $Q'$, $P'$, $\Delta Q'$ and $\Delta P'$
being fixed. One consequence of the definition of an ME packet not being
canonically invariant is that the property of maximum entropy is not preserved
by the dynamical evolution of the state.

The fact that our definition of the maximum entropy packet depends on the
coordinates chosen for the description of the system and on the time instant
when it is applied is not necessarily a serious hindrance for our project: we
are just going to construct a model of a sufficiently fuzzy state. It is clear
that such construction is inherently arbitrary. Any such model will do and it
may be even advantageous to have some freedom. It seems plausible that
relevant properties of the fuzzy states are independent of the details of
their definition in some reasonable extent. Some suitable formulation of such
assumptions must yet be found and their validity must be studied.

\section{Classical ME packets} Let us first consider the sharp trajectories of
a classical mechanical system $S_c$ and their approximation by a quantum
system $S_q$. For any comparison of Newtonian and quantum mechanics, it is
necessary that the Newtonian canonical coordinates are chosen in such a way
that there are reasonable quantum observables corresponding to them. For
example, we ought to assume that the space coordinates are Cartesian. Let
$S_c$ have just one degree of freedom, canonical coordinates $q$ and $p$ and
Hamiltonian
\begin{equation}\label{comhamc} H = \frac{p^2}{2\mu} + V(q)\ ,
\end{equation} where $\mu$ is a mass and $V(q)$ a potential function. The
classical equations of motion are
\begin{equation}\label{motionc} \dot{q} = \frac{p}{\mu}\ ,\quad \dot{p} =
-\frac{dV}{dq}\ .
\end{equation} Their solution is a sharp trajectory,
$$
q = q(t)\ ,\quad p = p(t)\ ,
$$
for every initial values $q(0)$ and $p(0)$.

Let us choose the corresponding quantum model $S_q$ to be a system of one
degree of freedom with position operator ${\mathsf q}$, momentum operator
${\mathsf p}$ and spin 0. Operators ${\mathsf q}$ and ${\mathsf p}$ are
related to the chosen coordinates $q$ and $p$, e.g., through their
spectra. Let the Hamiltonian be
\begin{equation}\label{comham} {\mathsf H} = \frac{{\mathsf p}^2}{2\mu} +
V({\mathsf q})\ .
\end{equation} The Heisenberg equations of motion are
\begin{equation}\label{motionq} \dot{\mathsf q} = \frac{{\mathsf p}}{\mu}\
,\quad \dot{\mathsf p} = -\frac{dV}{d{\mathsf q}}\ .
\end{equation} Then the time dependence of position and momentum expectation
values $Q = \langle {\mathsf q} \rangle$ and $P = \langle {\mathsf p} \rangle$
in a state $|\psi\rangle$ is
$$
\dot{Q} = \frac{P}{\mu}\ ,\quad \dot{P} = -\left\langle \frac{dV}{d{\mathsf
q}}\right\rangle\ .
$$
To evaluate the right-hand side of the second equation, let us expand the
potential function in powers of ${\mathsf q} - Q$:
$$
V({\mathsf q}) = V(Q) + ({\mathsf q} - Q)\frac{dV}{dQ} + \frac{1}{2}({\mathsf
q} - Q)^2 \frac{d^2V}{dQ^2} + \ldots
$$
so that
$$
\frac{dV}{d{\mathsf q}} = \frac{dV}{dQ} + ({\mathsf q} - Q)\frac{d^2V}{dQ^2} +
\frac{1}{2}({\mathsf q} - Q)^2 \frac{d^3V}{dQ^3} + \ldots\ .
$$
If we take the expectation value of the last equation and use relations
$\langle ({\mathsf q} - Q)\rangle = 0$ and $\langle ({\mathsf q} - Q)^2\rangle
= \Delta Q^2$, where $\Delta Q$ is the variance of ${\mathsf q}$ in state
$|\psi\rangle$, we obtain
$$
\left\langle \frac{dV}{d{\mathsf q}}\right\rangle = \frac{dV}{dQ} +
\frac{1}{2}\Delta Q^2 \frac{d^3V}{dQ^3} + \ldots\ .
$$

Let us assume that coordinate $q$ and momentum $p$ of $S_c$ are obtained from
the quantum model by formulas
$$
q = Q\ ,\quad p = P\ .
$$
(This assumption is only natural if we worked with fuzzy classical models or
if we are going to compare a fuzzy model with a sharp trajectory one.) Then,
already for potentials of the third order, the quantum equations of motion for
expectation values deviate from classical equation of motion for sharp
trajectories. The quantum correction is proportional to the variation $\Delta
Q$. There are two important observations about this quantum correction. First,
the correction is not proportional to $\hbar$. Second, the correction would be
negligible for small $\Delta Q$. Hence, the difference to the classical
trajectory is smaller if the spread of the wave packet $|\psi\rangle$ over the
space is smaller. This implies that the minimum-uncertainty wave packets give
the best approximation to classical sharp trajectories.

However, for small $\Delta Q$ the variance $\Delta P$ is large, and $\Delta Q$
will quickly increase with time. Moreover, as already discussed, the minimum
uncertainty packets have some further disadvantages.

Next, we turn to that maximum-entropy packets. Let us start the theory of such
packets with the above system $S_c$ of one degree of freedom and then
generalise it to any number of degrees. A fuzzy state is a distribution
function $\rho(q,p)$ on the phase space spanned by Cartesian coordinates $q$
and $p$. The function $\rho(q,p)$ is dimensionless and normalized by
$$
\int\frac{dq\,dp}{v}\,\rho = 1\ ,
$$
where $v$ is an auxiliary phase-space volume to make $\rho$ dimensionless. The
entropy of $\rho(q,p)$ can be defined by
$$
\Sigma := -\int\frac{dq\,dp}{v}\,\rho \ln\rho\ .
$$
The value of entropy will depend on $v$ but most other results will
not. Classical mechanics does not offer any idea of how to fix the value of
$v$. We shall get a hint from quantum mechanics.

If we have chosen a different set of canonical coordinates, $q'$ and $p'$,
say, then the transformation
$$
q = q(q',p')\ ,\quad p = p(q',p')\ ,
$$
being canonical, satisfies
$$
\frac{\partial(q,p)}{\partial(q',p')} = 1\ ,
$$
where the left-hand side is the Jacobian of the transformation (see, e.g.,
\cite{LL1}, Section 46, p.\ 146). For the transformed distribution function,
$$
\rho'(q',p') = \rho(q(q',p'),p(q',p'))\ ,
$$
then holds that
$$
\int\frac{dq\,dp}{v}\,\rho = \int\frac{dq'\,dp'}{v}\,\rho'\ ,\quad
\int\frac{dq\,dp}{v}\,\rho \ln\rho = \int\frac{dq'\,dp'}{v}\,\rho' \ln\rho'\ .
$$
Hence, the normalisation condition and the value of entropy are invariant with
respect to a general canonical transformation in Newtonian mechanics.

Let us now give a rigorous definition of the classical ME packets following
\cite{hajicek,hajicekC3}.
\begin{df}\label{dfold21} ME packet is the distribution function
$\rho[Q,P,\Delta Q,\Delta P]$ that maximizes the entropy subject to the
conditions:
\begin{equation}\label{21.4} \langle q\rangle = Q\ ,\quad \langle q^2\rangle =
\Delta Q^2 + Q^2\ ,
\end{equation} and
\begin{equation}\label{21.5} \langle p\rangle = P\ ,\quad \langle p^2\rangle =
\Delta P^2 + P^2\ ,
\end{equation} where the values of $Q$, $P$, $\Delta Q$ and $\Delta P$ are
given.
\end{df} We have used the abbreviation
$$
\langle x\rangle = \int\frac{dq\,dp}{v}\,x\rho
$$
for any function $x(q,p)$.

The explicit form of ME packets can be found using the partition-function
method as it is derived, e.g., in Ref.\ \cite{Jaynes}, Chapter 11. The
variational principle,
$$
\delta \int\frac{dq\,dp}{v}\,(\rho \ln\rho + \lambda_0 \rho + \lambda_1\rho q
+ \lambda_2\rho p + \lambda_3\rho q^2 + \lambda_4\rho p^2) = 0\ ,
$$
where $\lambda_0$, $\lambda_1$, $\lambda_2$, $\lambda_3$ and $\lambda_4$ are
the five Lagrange multipliers corresponding to the normalisation condition and
to the four conditions (\ref{21.4}) and (\ref{21.5}), yields
\begin{equation}\label{H5} \rho =
\frac{1}{Z(\lambda_1,\lambda_2,\lambda_3,\lambda_4)} \exp(-\lambda_1q -
\lambda_2p - \lambda_3q^2 - \lambda_4p^2)\ ,
\end{equation} so that the normalisation condition for $\rho$ gives
$\exp(1+\lambda_0) = Z$, where
\begin{equation}\label{partfunc} Z = \int \frac{dq\,dp}{v}\exp(-\lambda_1q -
\lambda_2p - \lambda_3q^2 - \lambda_4p^2)
\end{equation} is the {\em partition function}. The integral is easy to
calculate:
\begin{equation}\label{22.1} Z=
\frac{\pi}{v}\frac{1}{\sqrt{\lambda_3\lambda_4}}
\exp\left(\frac{\lambda_1^2}{4\lambda_3} +
\frac{\lambda_2^2}{4\lambda_4}\right)\ .
\end{equation}

From the definition (\ref{partfunc}) of partition function, it follows that
the expectation value of any monomial of the form $q^k p^l q^{2m} p^{2n}$ can
be calculated with the help of partition-function method as follows:
\begin{equation}\label{43.1} \langle q^k p^l q^{2m} p^{2n}\rangle =
\frac{(-1)^{\mathbf N}}{Z}\ \frac{\partial^{\mathbf N}
Z}{\partial\lambda_1^k \partial\lambda_2^l \partial\lambda_3^m \partial\lambda_4^n
}\ ,
\end{equation} where ${\mathbf N} = k+l+2m+2n$ and $Z$ is given by Eq.\
(\ref{22.1}).

Observe that this allowes to calculate the expectation value of a monomial in
several different ways. Each of these ways, however, leads to the same result
due the identities
$$
\frac{\partial^2 Z}{\partial \lambda_1^2} = -\frac{\partial Z}{\partial
\lambda_3}\ ,\quad \frac{\partial^2 Z}{\partial \lambda_2^2} = -\frac{\partial
Z}{\partial \lambda_4}\ ,
$$
which are satisfied by the partition function.

In particular, the expressions for $\lambda_1$, $\lambda_2$, $\lambda_3$ and
$\lambda_4$ in terms of $Q$, $P$, $\Delta Q$ and $\Delta P$ can be obtained by
solving the equations
$$
\frac{\partial\ln Z}{\partial\lambda_1} = -Q\ , \quad \frac{\partial\ln
Z}{\partial\lambda_3} = -\Delta Q^2 - Q^2\ ,
$$
and
$$
\frac{\partial\ln Z}{\partial\lambda_2} = -P\ , \quad \frac{\partial\ln
Z}{\partial\lambda_4} = -\Delta P^2 - P^2\ .
$$
The result is:
\begin{equation}\label{lagrancl1} \lambda_1 = -\frac{Q}{\Delta Q^2}\ ,\quad
\lambda_3 = \frac{1}{2\Delta Q^2}\ ,
\end{equation} and
\begin{equation}\label{lagrancl2} \lambda_2 = -\frac{P}{\Delta P^2}\ ,\quad
\lambda_4 = \frac{1}{2\Delta P^2}\ .
\end{equation} Substituting Eqs.\ (\ref{lagrancl1}) and (\ref{lagrancl2}) into
Eq.\ (\ref{H5}), we obtain the distribution function of a one-dimensional ME
packet. The generalization to any number of dimensions is easy:
\begin{thm}\label{propold19} The distribution function of the ME packet for a
system $S_c$ of $n$ degrees of freedom with given expectation values and
variances $Q_1,\cdots,Q_n$, $\Delta Q_1,\cdots$, $\Delta Q_n$ of coordinates
and $P_1,\cdots,P_n$, $\Delta P_1,\cdots,\Delta P_n$ of momenta, is
\begin{multline}\label{23.1} \rho[Q,P,\Delta Q,\Delta P](q,p) \\ =
\left(\frac{v}{2\pi}\right)^n\prod_{k=1}^n\left(\frac{1}{\Delta Q_k\Delta
P_k}\exp\left[-\frac{(q_k-Q_k)^2}{2\Delta Q_k^2} -\frac{(p_k-P_k)^2}{2\Delta
P_k^2}\right]\right)\ ,
\end{multline} where $Q$, $P$, $\Delta Q$, $\Delta P$ $q$ and $p$ stand for
$n$-tuples of values, e.g., $Q \equiv Q_1,\ldots, Q_n$, etc.
\end{thm} Formula (\ref{23.1}) holds for general canonical coordinates, not
only for the Cartesian ones.

The state of system $S_c$ is described by distribution function
$\rho[Q,P,\Delta Q,\Delta P](q,p)$ which is determined by $4n$ values. In this
way, to describe the mechanical degrees of freedom, we need twice as many
variables as the standard mechanics. The doubling of state coordinates is due
to the necessity to define a fuzzy distribution rather than a sharp
trajectory.

We observe that all expectation values obtained from $\rho$ are independent of
$v$ and that the right-hand side of equation (\ref{23.1}) is a Gaussian
distribution in agreement with Jaynes' conjecture that the maximum entropy
principle gives the Gaussian distribution if the only conditions are fixed
values of the first two moments.

As $\Delta Q$ and $\Delta P$ approach zero, $\rho$ becomes a $\delta$-function
and the state becomes sharp. For some quantities this limit is sensible for
others it is not. In particular, the entropy, which can easily be calculated,
$$
\Sigma = 1 + \ln\frac{2\pi\Delta Q\Delta P}{v}\ ,
$$
diverges to $-\infty$. This is due to a general difficulty in giving a
definition of entropy for a continuous system that would be satisfactory in
every respect (see \cite{Jaynes}, Section 12.3). What one could do is to
divide the phase space into cells of volume $v$ so that $\Delta Q\Delta P$
could not be chosen smaller than $v$. Then, the limit $\Delta Q\Delta P
\rightarrow v$ of entropy would make more sense.

The importance of the ME packets for our theory is expressed by:

\vspace{0.5cm}

\noindent {\bf ME-Packet Conjecture} {\it For most mechanical objects
${\mathcal S}$, all measurable predictions of Newtonian mechanics can be
obtained from a classical model $S_c$ described by Theorem \ref{propold19}.}

\vspace{0.5cm}

\noindent This can be considered as a more specific form of Exner-Born
Conjecture.

\section{Polynomial potential function} Here, an account of some aspects of
Newtonian dynamics of ME packets is given with the aim to compare them with,
or obtain them in some approximation from, the dynamics of corresponding
quantum systems later. This comparison is easy for polynomial potential
functions.

Let the Hamiltonian of $S_c$ has the form (\ref{comhamc}) so that the
equations of motion are (\ref{motionc}). The general solution to these
equations can be written in the form
\begin{equation}\label{36.65} q(t) = \bar{q}(t;q,p)\ ,\quad p(t) =
\bar{p}(t;q,p)\ ,
\end{equation} where
\begin{equation}\label{36.3} \bar{q}(0;q,p) = q\ ,\quad \bar{p}(0;q,p) = p\ ,
\end{equation} $q$ and $p$ being arbitrary initial values. This implies for
the time dependence of the expectation values and variances, if the initial
state is an ME packet:
\begin{equation}\label{36.7} \bar{Q}(t) = \langle \bar{q}(t; q,p)\rangle\
,\quad \Delta \bar{Q}(t) = \sqrt{\langle \bar{q}^2(t;q,p)\rangle- \langle
\bar{q}(t;q,p)\rangle^2}
\end{equation} and
\begin{equation}\label{36.8} \bar{P}(t) = \langle \bar{p}(t; q,p)\rangle\
,\quad \Delta \bar{P}(t) = \sqrt{\langle \bar{p}^2(t;q,p)\rangle- \langle
\bar{p}(t;q,p)\rangle^2}\ .
\end{equation} We introduce the notation $\bar{Q}(t)$, $\bar{P}(t)$, $\Delta
\bar{Q}(t)$ and $\Delta \bar{P}(t)$ to distinguish the expectation values and
variances of time-dependent coordinates and momenta from the values $Q$, $P$,
$\Delta Q$ and $\Delta P$ that define the EM packet.

Let us first consider the special case of at most quadratic potential:
\begin{equation}\label{36.1} V(q) = V_0 + V_1 q + \frac{1}{2} V_2 q^2\ ,
\end{equation} where $V_k$ are constants with suitable dimensions. If $V_1 =
V_2 =0$, we have a free particle, if $V_2 = 0$, it is a particle in a
homogeneous force field and if $V_2 \neq 0$, it is an harmonic or
anti-harmonic oscillator.

For potential (\ref{36.1}), the dynamical equations are linear and their
general solution (\ref{36.65}) has the form
\begin{eqnarray}\label{37.1} \bar{q}(t) &=& f_0(t) + q f_1(t) + p f_2(t)\ , \\
\label{37.2} \bar{p}(t) &=& g_0(t) + q g_1(t) + p g_2(t)\ ,
\end{eqnarray} where $f_0(0) = f_2(0) = g_0(0) = g_1(0) = 0$ and $f_1(0) =
g_2(0) = 1$. If $V_2 \neq 0$, the functions are
\begin{equation}\label{37.4} f_0(t) = -\frac{V_1}{V_2}(1-\cos\omega t)\ ,\quad
f_1(t) = \cos \omega t\ ,\quad f_2(t) = \frac{1}{\xi}\sin\omega t\ ,
\end{equation}
\begin{equation}\label{37.5} g_0(t) = -\xi\frac{V_1}{V_2}\sin\omega t\ ,\quad
g_1(t) = -\xi\sin \omega t\ ,\quad g_2(t) = \cos\omega t\ ,
\end{equation} where
$$
\xi = \sqrt{\mu V_2}\ ,\quad \omega = \sqrt{\frac{V_2}{\mu}}\ .
$$
Only for $V_2 > 0$, the functions remain bounded. If $V_2 = 0$, we obtain
\begin{equation}\label{37.9a} f_0(t) = -\frac{V_1}{2\mu}t^2\ ,\quad f_1(t) =
1\ ,\quad f_2(t) = \frac{t}{\mu}\ ,
\end{equation}
\begin{equation}\label{37.9b} g_0(t) = -V_1t\ ,\quad g_1(t) = 0\ ,\quad g_2(t)
= 1\ .
\end{equation}

The time dependence of expectation values and variances resulting from Eqs.\
(\ref{36.65}), (\ref{21.4}) and (\ref{21.5}) are
\begin{equation}\label{38.3} \bar{Q}(t) = f_0(t) + Q f_1(t) + P f_2(t)
\end{equation} and
\begin{multline}\label{38.4} \Delta \bar{Q}^2(t) + \bar{Q}^2(t) = f_0^2(t) +
(\Delta Q^2 + Q^2) f_1^2(t) + (\Delta P^2 + P^2) f_2^2(t)\\ + 2Qf_0(t)f_1(t) +
2Pf_0(t)f_2(t) + 2\langle qp\rangle f_1(t)f_2(t)\ .
\end{multline} For the last term, we have from Eq.\ (\ref{43.1})
$$
\langle qp\rangle = \frac{1}{Z}\frac{\partial^2
Z}{\partial\lambda_1\partial\lambda_2}\ .
$$
Using Eqs.\ (\ref{22.1}), (\ref{lagrancl1}) and (\ref{lagrancl2}), we obtain
from Eq.\ (\ref{38.4})
\begin{equation}\label{39.1} \Delta \bar{Q}(t) = \sqrt{f_1^2(t)\Delta Q^2 +
f_2^2(t)\Delta P^2}\ .
\end{equation} Similarly,
\begin{eqnarray}\label{39.2} \bar{P}(t) &=& g_0(t) + Q g_1(t) + P g_2(t)\ ,\\
\label{39.3} \Delta \bar{P}(t) &=& \sqrt{g_1^2(t)\Delta Q^2 + g_2^2(t)\Delta
P^2}\ .
\end{eqnarray} We observe: if functions $f_1(t)$, $f_2(t)$, $g_1(t)$ and
$g_2(t)$ remain bounded, the variances also remain bounded and the predictions
are possible in arbitrary long intervals of time. Otherwise, there will always
be only limited time intervals in which the theory can make reasonable
predictions. We can also see that the evolution $\bar{Q}(t)$ and $\bar{P}(t)$
coincides with the sharp trajectories (\ref{37.1}) and (\ref{37.2}). In
particular, it is independent of $\Delta Q$ and $\Delta P$, which is a
well-known property of potential (\ref{36.1}). In general, $\bar{Q}(t)$ and
$\bar{P}(t)$ will depend not only on initial $Q$ and $P$, but also on $\Delta
Q$ and $\Delta P$.

From formulas (\ref{39.1}) and (\ref{39.3}) we can also see that the ME packet
form is not preserved by the evolution (the entropy ceases to be
maximal). First, both variances must increase near $t=0$. Second, the entropy
must stay constant because it is preserved by the dynamics. Third, the
relation between entropy and $\nu$ is fixed for ME packets.

After these preparation remarks, we turn to a general polynomial potential of
degree $N$,
\begin{equation}\label{50.2} V(q) = \sum_{k=0}^N \frac{1}{k!}V_k q^k\ ,
\end{equation} and study the general time derivatives of $\bar{Q}$, $\bar{P}$,
$\Delta \bar{Q}$, $\Delta \bar{P}$. We shall need some results of this study
for the proof of Theorem \ref{thmclaslim} on the classical limit. The
right-hand sides of dynamical equation (\ref{motionc}) for the potential
function (\ref{50.2}) are polynomials in $\bar{q}$ and $\bar{p}$:
$$
\frac{\partial \bar{q}}{\partial t} = \frac{1}{\mu}\bar{p}\ ,\quad
\frac{\partial \bar{p}}{\partial t} = -\sum_{k=0}^{N-1} \frac{1}{k!}V_{k+1}
\bar{q}^k\ .
$$
In general, for the $K$-th time derivatives, we obtain
$$
\frac{\partial^K\bar{q}}{\partial t^K} = A_K(\bar{q},\bar{p})\ ,
$$
$$
\frac{\partial^K\bar{p}}{\partial t^K} = B_K(\bar{q},\bar{p})\ ,
$$
where $A_K$ and $B_K$ are polynomials in $\bar{q}$ and $\bar{p}$. The proof is
by mathematical induction: For the first derivative, the claim is true. If it
is true for the $K$-th derivative, then we have for the $K+1$-th one:
$$
\frac{d^{K+1}\bar{q}}{\partial t^{K+1}} = \frac{\bar{p}}{\mu}\frac{\partial
A_K\bigl(\bar{q},\bar{p}\bigr)}{\partial \bar{q}} - \sum_{k=0}^{N-1}
\frac{1}{k!}V_{k+1} \bar{q}^k\frac{\partial
A_K\bigl(\bar{q},\bar{p}\bigr)}{\partial \bar{p}}\ ,
$$
$$
\frac{\partial^{K+1}\bar{p}}{\partial t^{K+1}} =
\frac{\bar{p}}{\mu}\frac{\partial B_K\bigl(\bar{q},\bar{p}\bigr)}{\partial
\bar{q}} - \sum_{k=0}^{N-1} \frac{1}{k!}V_{k+1} \bar{q}^k\frac{\partial
B_K\bigl(\bar{q},\bar{p}\bigr)}{\partial \bar{p}}\ ,
$$
which are again polynomials in $\bar{q}$ and $\bar{p}$. The equations for the
time derivatives of expectation values are
$$
\left(\frac{d^K\bar{Q}}{dt^K}\right)_0 = \langle A_K(q,p)\rangle\ ,\quad
\left(\frac{d^K\bar{P}}{dt^K}\right)_0 = \langle B_K(q,p)\rangle\ ,
$$
and the calculation of all expectation values can be reduced to that of
$q^kp^l$-products.

Slightly more complicated equations hold for the time derivatives of
variances. For the first derivatives, we obtain
\begin{equation}\label{dtdeltaq} \frac{d\Delta \bar{Q}(t)}{dt} =
\frac{1}{\mu\Delta \bar{Q}(t)}\bigl(\langle\bar{q}\bar{p}\rangle -
\langle\bar{q}\rangle\langle\bar{p}\rangle\bigr)\ ,
\end{equation}
\begin{equation}\label{dtdeltap} \frac{d\Delta \bar{P}(t)}{dt} = -
\frac{1}{\Delta \bar{P}(t)}\left(\left\langle\bar{p}\sum_{k=0}^{N-1}
\frac{1}{k!}V_{k+1} \bar{q}^k\right\rangle -
\langle\bar{p}\rangle\left\langle\sum_{k=0}^{N-1} \frac{1}{k!}V_{k+1}
\bar{q}^k\right\rangle\right)\ .
\end{equation} To get any further, we need the following property:
$$
\langle q^kp^l\rangle = \langle q^k\rangle\langle p^l\rangle \ .
$$
Indeed, we obtain easily from Eq.\ (\ref{43.1}) that
\begin{multline}\label{qpcl} \langle q^kp^l\rangle =
\left[(-1)^k\exp\left(-\frac{\lambda_1^2}{4\lambda_3}\right)\frac{\partial^k}{\partial\lambda_1^k}\exp\left(\frac{\lambda_1^2}{4\lambda_3}\right)\right]
\\ \times
\left[(-1)^l\exp\left(-\frac{\lambda_2^2}{4\lambda_4}\right)\frac{\partial^l}{\partial\lambda_2^l}\exp\left(\frac{\lambda_2^2}{4\lambda_4}\right)\right]\
.
\end{multline} From Eqs.\ (\ref{dtdeltaq}) and (\ref{dtdeltap}), it then
follows immediately that
$$
\left(\frac{d\Delta \bar{Q}}{dt}\right)_0 = 0\ ,\quad \left(\frac{d\Delta
\bar{P}}{dt}\right)_0 = 0\ ,
$$
This implies that, in calculating higher time derivatives of the right-hand
sides of Eqs.\ (\ref{dtdeltaq}) and (\ref{dtdeltap}), we can ignore the
variances in the denominator, so that we obtain
\begin{equation}\label{dtdeltaqK} \left(\frac{d^K\Delta
\bar{Q}}{dt^K}\right)_0 = \frac{1}{\mu\Delta
Q}\left[\frac{\partial^{K-1}}{\partial
t^{K-1}}\bigl(\langle\bar{q}\bar{p}\rangle -
\langle\bar{q}\rangle\langle\bar{p}\rangle\bigr)\right]_0\ ,
\end{equation}
\begin{multline}\label{dtdeltapK} \left(\frac{d^K\Delta
\bar{P}}{dt^K}\right)_0 \\ = - \frac{1}{\Delta
P}\left[\frac{\partial^{K-1}}{\partial
t^{K-1}}\left(\left\langle\bar{p}\sum_{k=0}^{N-1} \frac{1}{k!}V_{k+1}
\bar{q}^k\right\rangle - \langle\bar{p}\rangle\left\langle\sum_{k=0}^{N-1}
\frac{1}{k!}V_{k+1} \bar{q}^k\right\rangle\right)\right]_0\ .
\end{multline} Now, it is easy to show that the right-hand sides will be
expressions constructed from expectation values of polynomials in $q$ and $p$
by the same argument as that used for expectation values.

It follows that all time derivatives of $Q(t)$, $P(t)$, $\Delta Q(t)$ and
$\Delta P(t)$ at $t = 0$ can be calculated by an iterative application of
Eqs.\ (\ref{motionc}) and then using Formula (\ref{43.1}) (in \cite{hajicek},
the first four time derivatives for a fourth-degree potential function have
been calculated).

Finally, we have the following Lemma.
\begin{lem}\label{lemhopolyncl} The expectation value $\langle q^xp^y
\rangle_c$ for any non-negative integers $x$ and $y$ is a polynomial,
$$
\langle q^xp^y \rangle_{\text{cl}} = X(Q,P,\Delta Q,\Delta P)\ ,
$$
with integer coefficients. The term of the highest order of $\Delta Q$ and
$\Delta P$ in $X$ has the form,
$$
A_mB_nQ^{x - 2m}P^{y - 2n}\Delta Q^{2m}\Delta P^{2n}\ ,
$$
where $A_m$ and $B_n$ are positive integers,
$$
m = \left[\frac{x}{2}\right]\ ,\quad n = \left[\frac{y}{2}\right]
$$
and $[a]$ is the integer part of a real number $a$, i.e., the largest integer
not larger than $a$.
\end{lem} {\bf Proof} Consider first the case of $x = 0,1$ so that $m =0$, and
similarly for $y$ and $n$. An easy calculation using Eq.\ (\ref{qpcl}) gives:
$$
A_0 = 1\ , \quad B_0 = 1\ .
$$

For $m \neq 0$ and $n \neq 0$, Eq.\ (\ref{qpcl}) implies that $X$ is a product
of two functions,
$$
X(Q,P,\Delta Q,\Delta P) = Y_x(\lambda_1,\lambda_3)Y_y(\lambda_2,\lambda_4)\ ,
$$
where Eqs.\ (\ref{lagrancl1}) and (\ref{lagrancl2}) must be substituted for
$\lambda_1$, $\lambda_2$, $\lambda_3$ and $\lambda_4$.

Let us show that
\begin{eqnarray}\label{2m-1} \frac{\partial^{2m-1}}{\partial
\lambda_1^{2m-1}}\exp\left(\frac{\lambda_1^2}{4\lambda_3}\right) &=&
\left[\sum_{k=1}^mb_{m,k}\frac{\lambda_1^{2k-1}}{(2\lambda_3)^{m+k-1}}\right]
\exp\left(\frac{\lambda_1^2}{2\lambda_3}\right)\ , \\ \label{2m}
\frac{\partial^{2m}}{\partial
\lambda_1^{2m}}\exp\left(\frac{\lambda_1^2}{4\lambda_3}\right) &=&
\left[\sum_{k=1}^{m+1}a_{m,k}\frac{\lambda_1^{2k-2}}{(2\lambda_3)^{m+k-1}}\right]
\exp\left(\frac{\lambda_1^2}{4\lambda_3}\right)\ ,
\end{eqnarray} where $b_{m,k}$ and $a_{m,k}$ are positive integer
coefficients. It then follows that $Y_x$ is a polynomial of $Q$ and $\Delta
Q^2$ and $Y_y$ that of $P$ and $\Delta P^2$, because
$$
\frac{\lambda_1^{2k-1}}{(2\lambda_3)^{m+k-1}} =
\frac{\lambda_1^{2k-1}}{(2\lambda_3)^{2k-1}}\frac{1}{(2\lambda_3)^{m-k}} =
-Q^{2k-1}(\Delta Q^2)^{m-k}\ ,
$$
and similarly for $Y_y$.

The proof of Eqs.\ (\ref{2m-1}) and (\ref{2m}) by mathematical induction
consists of the following steps. First, we easily obtain:
\begin{eqnarray*} \frac{\partial}{\partial
\lambda_1}\exp\left(\frac{\lambda_1^2}{4\lambda_3}\right) &=&
\frac{\lambda_1}{2\lambda_3}\exp\left(\frac{\lambda_1^2}{4\lambda_3}\right)\ ,
\\ \frac{\partial^2}{\partial
\lambda_1^2}\exp\left(\frac{\lambda_1^2}{4\lambda_3}\right) &=&
\left[\frac{1}{2\lambda_3} +
\left(\frac{\lambda_1}{2\lambda_3}\right)^2\right]\exp\left(\frac{\lambda_1^2}{4\lambda_3}\right)\
,
\end{eqnarray*} which coincide with Eqs.\ (\ref{2m-1}) and (\ref{2m}) for $m =
1$.

Second, assuming the validity of Eq.\ (\ref{2m-1}), we calculate the even
derivative from the odd one:
$$
\frac{\partial^{2m}}{\partial
\lambda_1^{2m}}\exp\left(\frac{\lambda_1^2}{4\lambda_3}\right) =
\frac{\partial}{\partial
\lambda_1}\left[\left(\sum_{k=1}^mb_{m,k}\frac{\lambda_1^{2k-1}}{(2\lambda_3)^{m+k-1}}\right)
\exp\left(\frac{\lambda_1^2}{2\lambda_3}\right)\right]\ ,
$$
which, after a simple rearrangement, becomes (\ref{2m}) with
\begin{eqnarray*} a_{m,1} &=& b_{m,1}\ , \\ a_{m,k} &=& b_{m,k-1} +
(2k-1)b_{m,k}
\end{eqnarray*} for $k = 2,\ldots,m$, and
$$
a_{m,m+1} = b_{m,m}\ .
$$
Similarly,
$$
\frac{\partial^{2m+1}}{\partial
\lambda_1^{2m+1}}\exp\left(\frac{\lambda_1^2}{4\lambda_3}\right) =
\frac{\partial}{\partial
\lambda_1}\left[\left(\sum_{k=1}^{m+1}a_{m,k}\frac{\lambda_1^{2k-2}}{(2\lambda_3)^{m+k-1}}\right)
\exp\left(\frac{\lambda_1^2}{4\lambda_3}\right)\right]\ ,
$$
which becomes (\ref{2m-1}), if $m$ is replaced by $m + 1$, with
$$
b_{m+1,k} = a_{m,k} + 2ka_{m,k+1}
$$
for $k = 1,\ldots,m$, and
$$
b_{m+1,m+1} = a_{m,m+1}\ .
$$

Finally, substituting from Eq.\ (\ref{lagrancl1}) for $\lambda_1$ and
$\lambda_3$ into Eqs.\ (\ref{2m-1}) and (\ref{2m}), we obtain:
$$
\frac{\lambda_1^{2k-1}}{(2\lambda_3)^{m+k-1}} = -Q^{2k-1}\Delta Q^{2m-1k}\ ,
$$
and
$$
\frac{\lambda_1^{2k-2}}{(2\lambda_3)^{m+k-1}} = -Q^{2k-2}\Delta Q^{2m-1k +2}\
.
$$
In particular, the terms of the highest order in $\Delta Q$ are (after the
multiplication by $(-1)^{x+y}$)
$$
b_{m,1} Q\Delta Q^{2m-2}
$$
for $x = 2m-1$, and
$$
a_{m,1}\Delta Q^{2m}
$$
for $x = 2m$.

The whole procedure can be repeated in the same form for $P$ and $\Delta P$
thus showing Eqs.\ (\ref{2m-1}) and (\ref{2m}) with $\lambda_1$ and
$\lambda_3$ replaced by $\lambda_2$ and $\lambda_4$. Now, the claim of the
Lemma follows with $A_m = a_{m,1}$ and $B_m = b_{m,1}$ for all positive
integers $m$ and $n$, {\bf QED}.

\chapter{Classical limit} \setcounter{equation}{0} \setcounter{thm}{0}
\setcounter{assump}{0} \setcounter{sh}{0} \setcounter{df}{0}
\setcounter{lem}{0} Here, we first define and study quantum ME packets and
then compare their dynamical trajectories with those of their classical
counterparts for polynomial potential functions. The main result is a theorem,
on which our new notion of classical limit is based.

\section{Quantum ME packets} According to TH \ref{asobjectcq}, the conditions
defining the classical model determine the preparation of the corresponding
quantum model. The foregoing chapter introduced ME packets as models of
mechanical systems. The corresponding quantum models are defined by:
\begin{df}\label{dfold22} Let the quantum model $S_q$ of object ${\mathcal S}$
has spin 0, position ${\mathsf q}$ and momentum ${\mathsf p}$. State ${\mathsf
T}$ that maximizes von Neumann entropy under the conditions
\begin{equation}\label{12.1} tr[{\mathsf T}{\mathsf q}] = Q\ ,\quad
tr[{\mathsf T} {\mathsf q}^2] = Q^2 + \Delta Q^2\ ,
\end{equation}
\begin{equation}\label{12.2} tr[{\mathsf T}{\mathsf p}] = P\ ,\quad
tr[{\mathsf T} {\mathsf p}^2] = P^2 + \Delta P^2\ ,
\end{equation} where $Q$, $P$, $\Delta Q$ and $\Delta P$ are given numbers, is
called {\em quantum ME packet}.
\end{df} Then, TH \ref{asobjectcq} implies that the quantum ME packet for $Q$,
$P$, $\Delta Q$ and $\Delta P$ is the quantum model $S_q$ corresponding to the
classical ME packet with the same values of $Q$, $P$, $\Delta Q$ and $\Delta
P$.

Von Neumann entropy $\Sigma$ of a given state ${\mathsf T}$ (see, e.g.,
\cite{peres}, Section 9-1) is
\begin{equation}\label{vNentropy} \Sigma(\mathsf T) = -tr[{\mathsf
T}\ln({\mathsf T})].
\end{equation} As each ${\mathsf T}$ must have a discrete spectrum with
positive eigenvalues $t_k$ (see \cite{RS}, p.\ 209), we have
$$
\Sigma({\mathsf T}) = -\sum_k t_k \ln(t_k)\ .
$$

To calculate state ${\mathsf T}$, we pretend that all s.a.\ operators that
occur in the calculation are just $n \times n$ hermitian matrices, as it is
common in quantum mechanics courses. Real proofs are more difficult, but we
assume that they can be given. Then, we use the method of Lagrange multipliers
as in the classical case. The variational principle for the maximum entropy
yields the following equation:
\begin{equation}\label{12.5} d\Sigma -\lambda_0 d\ tr({\mathsf T})-\lambda_1
d\,tr({\mathsf T} {\mathsf q})-\lambda_2 d\,tr({\mathsf T} {\mathsf p})
-\lambda_3 d\,tr({\mathsf T} {\mathsf q}^2)-\lambda_4 d\,tr({\mathsf T}
{\mathsf p}^2) = 0\ .
\end{equation} Choosing an orthonormal basis $\{|n\rangle\}$ of the Hilbert
space ${\mathbf H}$ of ${\mathcal S}$, the differentials of the terms that are
linear in ${\mathsf T}$ can be brought to the form:
$$
d\,tr[{\mathsf T} {\mathsf x}] = \sum_{mn}x_{nm}dT_{mn}
$$
for any observable ${\mathsf x}$. Although not all elements of the matrix
$dT_{mn}$ are independent (it is a hermitian matrix), we can proceed as if
they were because the matrix $x_{nm}$ must also be hermitian. The only problem
is to calculate $d\Sigma$. We have the following
\begin{lem}\label{lemdS}
\begin{equation}\label{10.1} d\Sigma = -\sum_{mn}[\delta_{mn} + (\ln
T)_{mn}]dT_{mn}\ .
\end{equation}
\end{lem} {\bf Proof} Let ${\mathsf U}$ be a unitary matrix that diagonalizes
${\mathsf T}$,
$$
{\mathsf U}^\dagger {\mathsf T} {\mathsf U} = {\mathsf R}\ ,
$$
where ${\mathsf R}$ is a diagonal matrix with elements $R_n$. Then $\Sigma =
-\sum_n R_n\ln R_n$. Correction to $R_n$ if ${\mathsf T}$ changes to ${\mathsf
T} + d{\mathsf T}$ can be calculated by the first-order formula of the
stationary perturbation theory (see, e.g., \cite{ballent}, p.\ 276). This
theory is usually applied to Hamiltonians but it holds for any perturbed
hermitian operator. Moreover, the formula is exact for infinitesimal
perturbations. Thus,
$$
R_n \mapsto R_n + \sum_{kl}U^*_{kn}U_{ln}dT_{kl}\ .
$$
In this way, we obtain
\begin{multline*} d\Sigma = -\sum_n\left(R_n +
\sum_{kl}U^*_{kn}U_{ln}dT_{kl}\right)\\ \times \ln\left[R_n\left(1 +
\frac{1}{R_n}\sum_{rs}U^*_{rn}U_{sn}dT_{rs}\right)\right] + \sum_n R_n\ln
R_n\\ = -\sum_n\left[\ln R_n\sum_{kl}U^*_{kn}U_{ln}dT_{kl}
+\sum_{kl}U^*_{kn}U_{ln}dT_{kl}\right] \\ =-\sum_{kl}\left[\delta_{kl} +
(\ln{\mathsf T})_{kl}]\right]dT_{kl}\ ,
\end{multline*} {\bf QED}.

With the help of Lemma \ref{lemdS}, Eq.\ (\ref{12.5}) becomes
$$
tr\left((-1 - \ln{\mathsf T} -\lambda_0-\lambda_1 {\mathsf q} -\lambda_2
{\mathsf p}-\lambda_3 {\mathsf q}^2-\lambda_4{\mathsf p}^2)d{\mathsf T}\right)
= 0\ ,
$$
which must hold for any $d{\mathsf T}$. Hence, we have
\begin{equation}\label{12.6} {\mathsf T} = \exp(-\lambda_0-1-\lambda_1
{\mathsf q}-\lambda_2 {\mathsf p}-\lambda_3 {\mathsf q}^2-\lambda_4 {\mathsf
p}^2)\ .
\end{equation} This can be written as
$$
\exp(-\lambda_0-1)\exp(-\lambda_1 {\mathsf q}-\lambda_2 {\mathsf p}-\lambda_3
{\mathsf q}^2-\lambda_4 {\mathsf p}^2)
$$
$\exp(-\lambda_0-1)$ being just a number that can be considered as a
normalisation constant. Taking trace of Eq.\ (\ref{12.6}), we obtain
$$
e^{-\lambda_0-1} = \frac{1}{Z(\lambda_1,\lambda_2,\lambda_3,\lambda_4)}\ ,
$$
where $Z$ is the partition function,
\begin{equation}\label{13.1} Z(\lambda_1,\lambda_2,\lambda_3,\lambda_4) = tr[
\exp(-\lambda_1 {\mathsf q}-\lambda_2 {\mathsf p}-\lambda_3 {\mathsf
q}^2-\lambda_4 {\mathsf p}^2)]\ .
\end{equation} Thus, the state operator has the form
\begin{equation}\label{13.2} {\mathsf T} =
\frac{1}{Z(\lambda_1,\lambda_2,\lambda_3,\lambda_4)} \exp(-\lambda_1 {\mathsf
q}-\lambda_2 {\mathsf p}-\lambda_3 {\mathsf q}^2-\lambda_4 {\mathsf p}^2)\ .
\end{equation}

At this stage, the quantum theory begins to differ from the classical one. It
turns out that, for the case of non-commuting operators in the exponent of the
partition function, formula (\ref{43.1}) is not valid in general. We can only
show that it holds for the first derivatives.
\begin{lem}\label{lemderivexp} Let ${\mathsf A}$ and ${\mathsf B}$ be
Hermitian matrices. Then
\begin{equation}\label{ABcomm} \frac{d^n}{d\lambda^n}\exp({\mathsf A}+{\mathsf
B}\lambda) = {\mathsf B}^n\exp({\mathsf A}+{\mathsf B}\lambda)
\end{equation} if ${\mathsf A}$ and ${\mathsf B}$ commute and
\begin{equation}\label{14.1} \frac{d}{d\lambda}tr[\exp({\mathsf A}+{\mathsf
B}\lambda)] = tr[{\mathsf B}\exp({\mathsf A}+{\mathsf B}\lambda)]
\end{equation} in general.
\end{lem} {\bf Proof} Let us express the exponential function as a series:
\begin{multline}\label{dlambda} d[\exp({\mathsf A}+{\mathsf B}\lambda)] =
\sum_{n=0}^\infty\frac{1}{n!}[d({\mathsf A}+{\mathsf B}\lambda)^n] \\ =
\sum_{n=0}^\infty\frac{1}{n!}\left[\sum_{k=1}^n({\mathsf A}+{\mathsf
B}\lambda)^{k-1}{\mathsf B}({\mathsf A}+{\mathsf
B}\lambda)^{n-k}\right]d\lambda\ .
\end{multline} For the first part of the Lemma, ${\mathsf B}$ commutes with
${\mathsf A}+{\mathsf B}\lambda$ and can thus be brought to the left in each
of the products and we obtain (\ref{ABcomm}) by calculating higher derivatives
by the same mehtod.

For the second part, we take trace of both sides of Eq.\ (\ref{dlambda}) and
use the invariance of trace of a product with respect to any cyclic
permutation of the factors,
\begin{multline*} d\,tr[\exp({\mathsf A}+{\mathsf B}\lambda)] =
\sum_{n=0}^\infty\frac{1}{n!}tr[d({\mathsf A}+{\mathsf B}\lambda)^n] \\ =
\sum_{n=0}^\infty\frac{1}{n!}tr\left[\sum_{k=1}^n({\mathsf A}+{\mathsf
B}\lambda)^{k-1}{\mathsf B}({\mathsf A}+{\mathsf
B}\lambda)^{n-k}\right]d\lambda \\
=\sum_{n=0}^\infty\frac{1}{n!}\sum_{k=1}^ntr\left[{\mathsf B}({\mathsf
A}+{\mathsf B}\lambda)^{n-1}\right]d\lambda = tr[{\mathsf B}\exp({\mathsf
A}+{\mathsf B}\lambda)]d\lambda\ ,
\end{multline*} which is Eq.\ (\ref{14.1}), {\bf QED}.

The proof of Lemma \ref{lemderivexp} shows why formula (\ref{43.1}) is not
valid for higher derivatives than the first in the quantum case: the operator
$B$ does not commute with ${\mathsf A}+{\mathsf B}\lambda$ and cannot be
shifted from its position to the first position in product
$$
({\mathsf A}+{\mathsf B}\lambda)^k{\mathsf B}({\mathsf A}+{\mathsf
B}\lambda)^l\ .
$$
Only for the first derivative, it can be brought there by a suitable cyclic
permutation. However, each commutator $[{\mathsf B},({\mathsf A}+{\mathsf
B}\lambda)]$ is proportional to $\hbar$. Hence, formula (\ref{43.1}) with
higher derivatives is the leading term in the expansion of expectation values
in powers of $\hbar$.

Together with Eq.\ (\ref{13.1}), Lemma \ref{lemderivexp} implies that:
\begin{equation}\label{13.3} \frac{\partial \ln Z}{\partial\lambda_1} = -Q\
,\quad \frac{\partial \ln Z}{\partial\lambda_3} = -Q^2-\Delta Q^2
\end{equation} and
\begin{equation}\label{13.4} \frac{\partial \ln Z}{\partial\lambda_2} = -P\
,\quad \frac{\partial \ln Z}{\partial\lambda_4} = -P^2-\Delta P^2\ .
\end{equation} The values of the multipliers can then be calculated from Eqs.\
(\ref{13.3}) and (\ref{13.4}), if the form of the partition function is known.

Variational methods can find locally extremal values that are not necessarily
maxima. We can however prove that our state operator maximizes the
entropy. The proof is based on the generalized Gibbs' inequality,
$$
tr[{\mathsf T}\ln{\mathsf T} - {\mathsf T}\ln{\mathsf S}] \geq 0
$$
for all pairs $\{{\mathsf T},{\mathsf S}\}$ of state operators (for proof of
the inequality, see \cite{peres}, p.\ 264). The proof of maximality is then
analogous to the ``classical'' proof (see, e.g., \cite{Jaynes}, p.\ 357). The
first proof of maximality in the quantum case was given by von Neumann
\cite{JvN}.

The state operator (\ref{13.2}) can be inserted into formula (\ref{vNentropy})
to give the value of the maximum entropy,
\begin{equation}\label{15.1} \Sigma = \ln Z + \lambda_1\langle {\mathsf
q}\rangle + \lambda_2\langle {\mathsf p}\rangle + \lambda_3\langle {\mathsf
q}^2\rangle + \lambda_4\langle {\mathsf p}^2\rangle\ .
\end{equation} This, together with Eqs.\ (\ref{13.3}) and(\ref{13.4}) can be
considered as the Legendre transformation from the logarithm of partition
function to the entropy,
$$
\ln Z(\lambda_1,\lambda_2,\lambda_3,\lambda_4) \mapsto \Sigma(\langle {\mathsf
q}\rangle,\langle {\mathsf p}\rangle,\langle {\mathsf q}^2\rangle,\langle
{\mathsf p}^2\rangle )\ .
$$
One can observe that our modified Newtonian mechanics is quite similar to
ordinary thermodynamics. For example, in thermodynamics, the physical meaning
of the Lagrange multiplier $\lambda$ in the term $\lambda \langle E\rangle$ of
the variational principle is the inverse of some energy typical for the system
in the state of maximal entropy, such as the expected energy of a molecule. In
Newtonian mechanics, the meaning is that of the inverses of some typical
expected values of coordinates, momenta and of their squares in the maximum
entropy states.

\section{Diagonal representation} The exponent in Eq.\ (\ref{13.2}) can be
written in the form
\begin{equation}\label{19.1} \frac{\lambda_1^2}{4\lambda_3} +
\frac{\lambda_2^2}{4\lambda_4} -2\sqrt{\lambda_3\lambda_4}{\mathsf K}\ ,
\end{equation} where
\begin{equation}\label{19.2} {\mathsf K} =
\frac{1}{2}\sqrt{\frac{\lambda_3}{\lambda_4}}\left({\mathsf q} +
\frac{\lambda_1}{2\lambda_3}\right)^2 +
\frac{1}{2}\sqrt{\frac{\lambda_4}{\lambda_3}}\left({\mathsf p} +
\frac{\lambda_2}{2\lambda_4}\right)^2
\end{equation} is an operator acting on the Hilbert space of our
system. ${\mathsf K}$ has the form of the Hamiltonian\footnote{The operator
${\mathsf K}$ must not be confused with the Hamiltonian ${\mathsf H}$ of our
system, which can be arbitrary.} of a harmonic oscillator with coordinate
${\mathsf u}$ and momentum ${\mathsf w}$
\begin{equation}\label{16.1} {\mathsf u} = {\mathsf q} +
\frac{\lambda_1}{2\lambda_3}\ ,\quad {\mathsf w} = {\mathsf p} +
\frac{\lambda_2}{2\lambda_4}\ ,
\end{equation} that satisfy the commutation relation $[{\mathsf u},{\mathsf
w}] = i\hbar$. The oscillator has mass $M = \sqrt{\lambda_3/\lambda_4}$ and
frequency 1. The normalized eigenvectors $|k\rangle$ of the operator form a
basis in the Hilbert space of our system defining what we shall call {\em
diagonal representation}. The eigenvalues of ${\mathsf K}$ are $\hbar/2 +
\hbar k$. As is usual in dealing with a harmonic oscillator, we introduce the
``annihilation'' operator ${\mathsf A}$ such that
\begin{eqnarray}\label{commutA} {\mathsf A}{\mathsf A}^\dagger - {\mathsf
A}^\dagger{\mathsf A} &=& 1\ ,\\
\label{ua} {\mathsf u} &=& \sqrt{\frac{\hbar}{2M}}({\mathsf A}+{\mathsf
A}^\dagger)\ , \\
\label{qa} {\mathsf w} &=& -i\sqrt{\frac{\hbar M}{2}}({\mathsf A}-{\mathsf
A}^\dagger)\ , \\
\label{Hamilt} {\mathsf K} &=& \frac{\hbar}{2}({\mathsf A}^\dagger {\mathsf A}
+ {\mathsf A}{\mathsf A}^\dagger))\ , \\
\label{aact1} {\mathsf A}|k\rangle &=& \sqrt{k}|k-1\rangle\ , \\
\label{aact2} {\mathsf A}^\dagger|k\rangle &=& \sqrt{k+1}|k+1\rangle\ .
\end{eqnarray}

To calculate $Z$ in the diagonal representation is easy:
\begin{multline*} Z = tr\left[\exp\left(\frac{\lambda_1^2}{4\lambda_3} +
\frac{\lambda_2^2}{4\lambda_4} -2\sqrt{\lambda_3\lambda_4}{\mathsf
K}\right)\right] \\ = \sum_{k=0}^\infty\langle
k|\exp\left(\frac{\lambda_1^2}{4\lambda_3} + \frac{\lambda_2^2}{4\lambda_4}
-2\sqrt{\lambda_3\lambda_4}{\mathsf K}\right)|k\rangle \\ =
\exp\left(\frac{\lambda_1^2}{4\lambda_3} + \frac{\lambda_2^2}{4\lambda_4}
-\hbar\sqrt{\lambda_3\lambda_4}\right)\sum_{k=0}^\infty\exp(-2\hbar\sqrt{\lambda_3\lambda_4}k)\
.
\end{multline*} Summing the geometrical series at the right-hand side results
in the partition function for the quantum ME-packets of the form:
\begin{equation}\label{17.3} Z =
\frac{\exp\left(\frac{\lambda_1^2}{4\lambda_3} +
\frac{\lambda_2^2}{4\lambda_4}\right)}{2\sinh(\hbar\sqrt{\lambda_3\lambda_4})}\
.
\end{equation}

Now, we can express the Lagrange multipliers in terms of the expectation
values and variances. Eqs.\ (\ref{13.3}) and (\ref{13.4}) yield
\begin{equation}\label{lagranqu1} \lambda_1 = -\frac{Q}{\Delta
Q^2}\frac{\nu}{2}\ln\frac{\nu+1}{\nu-1}\ ,\quad \lambda_2 = -\frac{P}{\Delta
P^2}\frac{\nu}{2}\ln\frac{\nu+1}{\nu-1}\ ,
\end{equation} and
\begin{equation}\label{lagranqu2} \lambda_3 = \frac{1}{2\Delta
Q^2}\frac{\nu}{2}\ln\frac{\nu+1}{\nu-1}\ ,\quad \lambda_4 = \frac{1}{2\Delta
P^2}\frac{\nu}{2}\ln\frac{\nu+1}{\nu-1}\ ,
\end{equation} where $\nu$ is defined by Eq.\ (\ref{uncertainty}).

From Eqs.\ (\ref{15.1}), (\ref{lagranqu1}) and (\ref{lagranqu2}), we obtain
the entropy:
\begin{equation}\label{18.5} \Sigma = -\ln 2 + \frac{\nu+1}{2}\ln(\nu+1)
-\frac{\nu-1}{2}\ln(\nu-1)\ .
\end{equation} Thus, $\Sigma$ depends on $Q$, $P$, $\Delta Q$, $\Delta P$ only
via $\nu$. We have
$$
\frac{d\Sigma}{d\nu} = \frac{1}{2}\ln\frac{\nu+1}{\nu-1} > 0\ ,
$$
so that $\Sigma$ is an increasing function of $\nu$. Near $\nu = 1$,
$$
\Sigma \approx -\frac{\nu-1}{2}\ln(\nu-1)\ .
$$
Asymptotically ($\nu \rightarrow\infty$),
$$
\Sigma \approx \ln\nu+1-\ln 2\ .
$$

It is clear that the choice of Q and P cannot influence the entropy. The
independence of $\Sigma$ from $Q$ and $P$ does not contradict the Legendre
transformation properties. Indeed, usually, one would have
$$
\frac{\partial \Sigma}{\partial Q} = \lambda_1\ ,
$$
but here
$$
\frac{\partial \Sigma}{\partial Q} = \lambda_1 + 2\lambda_3 Q\ ,
$$
which is zero.

Eq.\ (\ref{19.2}) implies that
$$
-2\sqrt{\lambda_3\lambda_4}{\mathsf K} = -\lambda_3\left({\mathsf q} +
\frac{\lambda_1}{2\lambda_3}\right)^2 - \lambda_4\left({\mathsf p} +
\frac{\lambda_2}{2\lambda_4}\right)^2\ .
$$
Substituting for the Lagrange multipliers from Eqs.\ (\ref{lagranqu1}) and
(\ref{lagranqu2}), we obtain
$$
-2\sqrt{\lambda_3\lambda_4}{\mathsf K} = -\frac{\nu}{2}\ln \frac{\nu + 1}{\nu
- 1}\left[\frac{1}{2}\frac{({\mathsf q} - Q)^2}{{\Delta Q}^2} +
\frac{1}{2}\frac{({\mathsf p} - P)^2}{{\Delta P}^2}\right]\ .
$$
The resulting state operator, generalised to $n$ degrees of freedom, is then
described by the following
\begin{thm}\label{propold20} The state operator ${\mathsf T}[Q,P,\Delta
Q,\Delta P]$ of the ME packet of a system with $n$ degrees of freedom for
given expectation values and variances $Q_1,\cdots,Q_n$, $\Delta
Q_1,\cdots,\Delta Q_n$ of coordinates and $P_1,\cdots,P_n$, $\Delta
P_1,\cdots,$ $\Delta P_n$ of momenta, is
\begin{equation}\label{32.1} {\mathsf T}[Q,P,\Delta Q,\Delta P] =
\prod_{k=1}^n\left[\frac{2}{\sqrt{\nu_k^2-1}}\exp\left(-\frac{\nu_k}{2}\ln\frac{\nu_k+1}{\nu_k-1}{\mathsf
K}'_k\right)\right]\ ,
\end{equation} where
\begin{equation}\label{20.3} {\mathsf K}'_k = \frac{1}{2}\frac{({\mathsf
q}_k-Q_k)^2}{\Delta Q_k^2} + \frac{1}{2}\frac{({\mathsf p}_k-P_k)^2}{\Delta
P_k^2}
\end{equation} and
\begin{equation}\label{20.3b} \nu_k = \frac{2\Delta P_k\Delta Q_k}{\hbar}\ .
\end{equation}
\end{thm}

It may be interesting to observe that, strictly speaking, the state operator
(\ref{32.1}) is not a Gaussian distribution. Thus, it seems to be either a
counterexample to, or a generalization of, Jaynes' hypothesis (see the remark
after Theorem \ref{propold19}).

Let us study some further properties of quantum ME packets. In the diagonal
representation, we have for one degree of freedom, $n=1$:
\begin{equation}\label{20.2a} {\mathsf T}[Q,P,\Delta Q,\Delta P] =
\sum_{m=0}^\infty R_m|m\rangle\langle m|\ .
\end{equation} We easily obtain
\begin{equation}\label{20.2} R_m = \langle m|{\mathsf T}[Q,P,\Delta Q,\Delta
P]|m\rangle = 2\frac{(\nu-1)^m}{(\nu+1)^{m+1}} .
\end{equation} Hence,
$$
\lim_{\nu\rightarrow 1}R_m = \delta_{m0}\ ,
$$
and the state ${\mathsf T}[Q,P,\Delta Q,\Delta P]$ becomes $|0\rangle\langle
0|$. In general, states $|m\rangle$ depend on $\nu$. The state vector
$|0\rangle$ expressed as a function of $Q$, $P$, $\Delta Q$ and $\nu$ is
given, for any $\nu$, by
\begin{equation}\label{20.1} \psi(q) = \left(\frac{1}{\pi} \frac{\nu}{2\Delta
Q^2}\right)^{1/4} \exp\left[-\frac{\nu}{4\Delta Q^2}(q-Q)^2 +
\frac{iPq}{\hbar}\right]\ .
\end{equation} This is a Gaussian wave packet that corresponds to different
values of variances than ME packet (\ref{20.2a}): these values satisfy the
minimum uncertainty condition. For $\nu\rightarrow 1$, it remains regular and
the projection $|0\rangle\langle 0|$ becomes the state operator of the
original ME packet. Hence, Gaussian wave packets are special cases of quantum
ME packets.

The diagonal representation offers a method for calculating expectation values
of coordinates and momenta products in a quantum ME-packet state that can
replace formula (\ref{43.1}). Let us denote such a product X. We have
\begin{equation}\label{avX} \langle {\mathsf X}\rangle = \sum_{k=0}^\infty
R_k\langle k|{\mathsf X}|k\rangle\ .
\end{equation} To calculate $\langle k|{\mathsf X}|k\rangle$, we use Eqs.\
(\ref{ua}), (\ref{qa}) and (\ref{16.1}) to obtain
\begin{equation}\label{qandp} {\mathsf q} = Q +\frac{\Delta
Q}{\sqrt{\nu}}({\mathsf A}+{\mathsf A}^\dagger)\ ,\quad {\mathsf p} =
P-i\frac{\Delta P}{\sqrt{\nu}} ({\mathsf A}-{\mathsf A}^\dagger)\ .
\end{equation} After substituting these relations into ${\mathsf X}$,
${\mathsf X}$ can be expressed as a polynomial in ${\mathsf A}$ and ${\mathsf
A}^\dagger$ that will be denoted by $X({\mathsf A},{\mathsf A}^\dagger)$. Now,
we define a map ${\mathcal N}$ such that ${\mathcal N}(X({\mathsf A},{\mathsf
A}^\dagger))$ is a polynomial of the single variable ${\mathsf A}^\dagger
{\mathsf A}$ in two steps.
\begin{enumerate}
\item All monomials in $X({\mathsf A},{\mathsf A}^\dagger)$ that contain
different numbers of ${\mathsf A}$ and ${\mathsf A}^\dagger$ factors are
discarded. A polynomial $\bar{X}({\mathsf A},{\mathsf A}^\dagger)$ results.
\item Using the commutation relations (\ref{commutA}), each monomial in
$\bar{X}({\mathsf A},{\mathsf A}^\dagger)$ is reordered so that it becomes a
polynomial in a single variable ${\mathsf A}^\dagger {\mathsf A}$ and this is
the desired ${\mathcal N}(X({\mathsf A},{\mathsf A}^\dagger))$, which will be
denoted by $X_{\mathcal N}({\mathsf A}^\dagger {\mathsf A})$.
\end{enumerate} It follows that ${\mathcal N}$ is linear,
$$
{\mathcal N}\Bigl(X_1({\mathsf A},{\mathsf A}^\dagger) + X_2({\mathsf
A},{\mathsf A}^\dagger)\Bigr) = {\mathcal N}\Bigl(X_1({\mathsf A},{\mathsf
A}^\dagger)\Bigr) + {\mathcal N}\Bigl(X_2({\mathsf A},{\mathsf
A}^\dagger)\Bigr)
$$
and commutes with $\dagger$,
$$
{\mathcal N}\Bigl(X^\dagger({\mathsf A},{\mathsf A}^\dagger)\Bigr) =
\Bigl({\mathcal N}(X({\mathsf A},{\mathsf A}^\dagger)\Bigr)^\dagger\ .
$$
Examples:
$$
{\mathcal N}({\mathsf q}) = Q\ ,
$$
$$
{\mathcal N}({\mathsf p}^2) = P^2 + \frac{\Delta P^2}{\nu}(2{\mathsf
A}^\dagger {\mathsf A} + 1)\ .
$$
Returning to the original task of calculating expectation value of $X$, we
obtain
$$
\langle k|{\mathsf X}|k\rangle = X_{\mathcal N }(k)\ .
$$
In Eq.\ (\ref{avX}), there are, therefore, sums
$$
\sum_{k=0}^\infty k^nR_k\ .
$$
Substituting for $R_k$ from Eq.\ (\ref{20.2}), we arrive at
$$
\sum_{k=0}^\infty k^nR_k = \frac{2}{\nu+1}I_n\ ,
$$
where
$$
I_n(\nu) = \sum_{k=0}^\infty k^n\left(\frac{\nu-1}{\nu+1}\right)^k\ .
$$
We easily obtain
$$
I_n = \left(\frac{\nu^2-1}{2}\frac{d}{d\nu}\right)^n\frac{\nu+1}{2}\ .
$$
The desired expectation value value is then given by
\begin{equation}\label{average} \langle {\mathsf X} \rangle =
\frac{1}{\nu+1}X_{\mathcal N
}\left(\frac{\nu^2-1}{2}\frac{d}{d\nu}\right)(\nu+1)\ .
\end{equation} For example, we obtain:
\begin{eqnarray}\label{AA} \langle {\mathsf A}^\dagger {\mathsf A} \rangle &=&
\frac{\nu}{2} - \frac{1}{2}\ , \\ \label{AA2} \langle ({\mathsf A}^\dagger
{\mathsf A})^2 \rangle &=& \frac{\nu^2}{2} - \frac{\nu}{2}\ , \\ \label{AA3}
\langle ({\mathsf A}^\dagger {\mathsf A})^3 \rangle &=& \frac{3\nu^3}{4} -
\frac{3\nu^2}{4} - \frac{\nu}{4} + \frac{1}{4}\ .
\end{eqnarray} The calculation of the polynomial $X_{\mathcal N }$ for a given
${\mathsf X}$ and the evaluation of the right-hand side of Eq.\
(\ref{average}) are the two steps of the promised method.

\section{Polynomial potential function} Let the Hamiltonian of $S_q$ be that
of Eq.\ (\ref{comham}) and the unitary evolution group be ${\mathsf
U}(t)$. The dynamics in the Schr\"{o}dinger picture leads to the time
dependence of ${\mathsf T}$:
$$
{\mathsf T}(t) = {\mathsf U}(t){\mathsf T} {\mathsf U}(t)^\dagger\ .
$$
Substituting for ${\mathsf T}$ from Eq.\ (\ref{32.1}) and using a well-known
property of exponential functions, we obtain
\begin{equation}\label{32.2} {\mathsf T}(t) =
\frac{2}{\sqrt{\nu^2-1}}\exp\left(-\frac{\nu}{2}\ln\frac{\nu+1}{\nu-1}{\mathsf
U}(t){\mathsf K}'{\mathsf U}(t)^\dagger\right)\ .
\end{equation} As ${\mathsf K}'$ is not a Hamiltonian of $S_q$, ${\mathsf
U}(t){\mathsf K}'{\mathsf U}(t)^\dagger$ is difficult to calculate.

In the Heisenberg picture, ${\mathsf T}$ remains constant, while observables
are time dependent. We denote such time-dependent position and momentum
operators by $\bar{\mathsf q}$ and $\bar{\mathsf p}$ to distinguish them from
their initial values ${\mathsf q}$ and ${\mathsf p}$ at $t=0$. Operators
$\bar{\mathsf q}$ and $\bar{\mathsf p}$ satisfy the equations
\begin{equation}\label{33.1} i\hbar\frac{d\bar{\mathsf q}}{dt} = [\bar{\mathsf
q},{\mathsf H}]\ ,\quad i\hbar\frac{d\bar{\mathsf p}}{dt} = [\bar{\mathsf
p},{\mathsf H}]\ ,
\end{equation} which are solved by
$$
\bar{\mathsf q}(t;{\mathsf q},{\mathsf p}) = {\mathsf U}(t){\mathsf q}{\mathsf
U}(t)^\dagger\ ,\quad \bar{\mathsf p}(t;{\mathsf q},{\mathsf p}) = {\mathsf
U}(t){\mathsf p}{\mathsf U}(t)^\dagger
$$
for ${\mathsf q}=\bar{\mathsf q}(0;{\mathsf q},{\mathsf p})$ and ${\mathsf
p}=\bar{\mathsf p}(0;{\mathsf q},{\mathsf p})$. The resulting operators can be
written in the form of operator functions analogous to classical expressions
(\ref{36.65}) so that Eqs.\ (\ref{36.7}) and (\ref{36.8}) can again be used in
the operator form if $q$ and $p$ are replaced by ${\mathsf q}$ and ${\mathsf
p}$.

Let us assume that the potential function is given by Eq.\ (\ref{50.2}) in the
operator form:
$$
V(\bar{\mathsf q}) = \sum_{k=1}^N\frac{1}{k!}V_k\bar{\mathsf q}^k\ .
$$
Heisenberg equations of motion (\ref{motionq}) that result from Eq.\
(\ref{33.1}),
\begin{eqnarray}\label{mq} \frac{\partial \bar{\mathsf q}}{\partial t} &=&
\frac{1}{\mu}\bar{\mathsf p}\ , \\ \label{mp} \frac{\partial \bar{\mathsf
p}}{\partial t} &=& -V'(\bar{\mathsf q})\ ,
\end{eqnarray} can be used to calculate all time derivatives of the functions
$\bar{\mathsf q}$ and $\bar{\mathsf p}$. Here, the symbol $X'({\mathsf x})$
denotes the derivative of the polynomial $X({\mathsf x})$, for example,
$$
V'({\mathsf q}) = \sum_{k=0}^N \frac{1}{k!}V_{k+1}{\mathsf q}^k\ .
$$
For the expectation values and variances in a ME-packet state, we shall use
the same notation as in the classical case, that is,
$$
\bar{Q}(t) = \langle \bar{\mathsf q}\rangle\ ,\quad \bar{P}(t) = \langle
\bar{\mathsf p}\rangle\ ,\quad \Delta \bar{Q}(t) = \Delta \bar{\mathsf q}\
,\quad \Delta \bar{P}(t) = \Delta \bar{\mathsf p}\ ,
$$
so that $Q = \bar{Q}(0)$, etc. In the Heisenberg picture, expectation values
of time derivatives are time derivatives of expectation values. For instance,
we have
$$
\left(\frac{d\bar{Q}}{dt}\right)_0 = \left\langle \frac{d\bar{\mathsf
q}}{dt}\right\rangle_0\ ,
$$
the index $0$ indicating the value taken at $t=0$. Then, we can calculate all
time derivatives of expectation values $\bar{Q}$, $\bar{P}$ and variances
$\Delta \bar{Q}$ and $\Delta \bar{P}$ at $t = 0$ similarly as in Section
2.4. In the case of variances, we obtain from the definition of variance
($\Delta{\mathsf O} = \sqrt{\langle {\mathsf O}^2 \rangle - \langle {\mathsf
O} \rangle^2}$, see \cite{ballent}, p.\ 223):
$$
\frac{\partial}{\partial t}\Delta \bar{\mathsf q} = \frac{1}{2\Delta
\bar{\mathsf q}}\frac{\partial}{\partial t}(\langle \bar{\mathsf q}^2\rangle -
\langle \bar{\mathsf q}\rangle^2)
$$
and
$$
\frac{\partial}{\partial t}\Delta \bar{\mathsf p} = \frac{1}{2\Delta
\bar{\mathsf p}}\frac{\partial}{\partial t}(\langle \bar{\mathsf p}^2\rangle -
\langle \bar{\mathsf p}\rangle^2)\ .
$$
Eqs.\ (\ref{mq}) and (\ref{mp}) imply that
\begin{equation}\label{varq1} \frac{\partial \Delta \bar{Q}}{\partial t} =
\frac{1}{2\mu\Delta \bar{Q}}(\langle \bar{\mathsf q}\bar{\mathsf p} +
\bar{\mathsf p}\bar{\mathsf q}\rangle - 2\langle \bar{\mathsf
q}\rangle\langle\bar{\mathsf p})\rangle)
\end{equation} and
\begin{equation}\label{varp1} \frac{\partial \Delta \bar{P}}{\partial t} =
-\frac{1}{\Delta \bar{P}}\langle \bar{\mathsf p}V'(\bar{\mathsf q}) +
V'(\bar{\mathsf q})\bar{\mathsf p}\rangle + 2\langle \bar{\mathsf p}\rangle
\langle V'(\bar{\mathsf q})\rangle\ .
\end{equation}

To calculate any further, we need some properties of expectation values of
products of ${\mathsf q}$ and ${\mathsf p}$.
\begin{lem}\label{lemevenpow} Let $X_{mn}$ be a product of $m$ factors
${\mathsf q}$ and $n$ factors ${\mathsf p}$ in some ordering. Let $\langle
X_{mn} + X_{mn}^\dagger \rangle$ be the expectation value in the ME packet
defined by $Q$, $P$, $\Delta Q$ and $\Delta P$. Then
\begin{equation}\label{Xmnexp} \langle X_{mn} + X_{mn}^\dagger\rangle =
\bar{X}_{mn} + \bar{X}_{mn}^\dagger\ ,
\end{equation} where
\begin{equation}\label{barXmnexp} \bar{X}_{mn} =
\sum_{u=0}^{[m/2]}\sum_{v=0}^{[n/2]} (-1)^vQ^{m-2u}P^{n-2v}\Delta Q^{2u}\Delta
P^{2v} \nu^{-u-v} \langle Y_{(2u)(2v)}({\mathsf A} + {\mathsf
A}^\dagger,{\mathsf A} - {\mathsf A}^\dagger)\rangle
\end{equation} and $Y_{(2u)(2v)}({\mathsf A} + {\mathsf A}^\dagger,{\mathsf A}
- {\mathsf A}^\dagger)$ is a product of $2u$ factors ${\mathsf A} + {\mathsf
A}^\dagger$ and $2v$ factors ${\mathsf A} - {\mathsf A}^\dagger$ in some
ordering multiplied by some integer coefficient.
\end{lem} {\bf Proof} According to the method of diagonal representation of
Section 3.2, we have to use Eq.\ (\ref{qandp}) so that $\langle X_{mn}
\rangle$ becomes
\begin{equation}\label{barX1} \langle X_{mn} \rangle =
\sum_{x=0}^m\sum_{y=0}^n (-i)^yQ^{m-x}P^{n-y}\Delta Q^x\Delta P^y
\nu^{-(x+y)/2} \langle Y_{xy}({\mathsf A} + {\mathsf A}^\dagger,{\mathsf A} -
{\mathsf A}^\dagger)\rangle\ ,
\end{equation} where $Y_{xy}({\mathsf A} + {\mathsf A}^\dagger,{\mathsf A} -
{\mathsf A}^\dagger)$ is a product of $x$ factors ${\mathsf A} + {\mathsf
A}^\dagger$ and $y$ factors ${\mathsf A} - {\mathsf A}^\dagger$ in some
ordering.

The second step is to transform $Y_{xy}$ to a polynomial in ${\mathsf A}$ and
${\mathsf A}^\dagger$ and discard all terms in which the number of factors
${\mathsf A}$ differs from that of ${\mathsf A}^\dagger$. The result is
symbolised by $Y_{xy}({\mathsf A} + {\mathsf A}^\dagger,{\mathsf A} - {\mathsf
A}^\dagger) \mapsto Y^{(0)}_{xy}({\mathsf A},{\mathsf A}^\dagger)$. Clearly,
$Y_{xy}^{(0)}({\mathsf A},{\mathsf A}^\dagger) = 0$ if $x+y$ is odd. Hence,
only terms with even $x+y$ contribute to the expectation value.

The third step is to reorder $Y_{xy}^{(0)}({\mathsf A},{\mathsf A}^\dagger)$
using the commutation relation (\ref{commutA}) so that we obtain
$$
Y_{xy}^{(0)}({\mathsf A},{\mathsf A}^\dagger) = Y_{xy}^{(1)}({\mathsf
A}^\dagger{\mathsf A})\ ,
$$
where
$$
Y_{xy}^{(1)}({\mathsf A}^\dagger{\mathsf A}) = \sum_{k=0}^{(x+y)/2} z_k
({\mathsf A}^\dagger{\mathsf A})^k\ .
$$
It is easy to see that $z_k$ are integers. Operator $Y_{xy}^{(1)}({\mathsf
A}^\dagger{\mathsf A})$ is therefore self-adjoint. It follows that the terms
in Eq.\ (\ref{barX1}) with odd $y$'s cancel in $X_{mn} + X_{mn}^\dagger$,
\newline {\bf QED}.

As an example, we use Lemma \ref{lemevenpow} to simplify some terms in Eqs.\
(\ref{varq1}) and (\ref{varp1}):
\begin{equation}\label{QPlem} \langle \bar{\mathsf q}\bar{\mathsf p} +
\bar{\mathsf p}\bar{\mathsf q}\rangle_0 = 2QP
\end{equation} and
$$
\langle \bar{\mathsf p}V'(\bar{\mathsf q}) + V'(\bar{\mathsf q})(\bar{\mathsf
p})\rangle_0 = 2P \langle V'(\bar{\mathsf q})\rangle_0
$$
so that finally
\begin{equation}\label{dt1var} \left(\frac{d\Delta \bar{Q}}{dt}\right)_0 =
\left(\frac{d\Delta \bar{P}}{dt}\right)_0 = 0\ .
\end{equation} We then also have, as in the classical case:
$$
\left(\frac{d^K\Delta \bar{Q}}{dt^K}\right)_0 = \frac{1}{2\Delta
Q}\left(\frac{d^K(\langle \bar{q}^2\rangle - \langle
\bar{q}\rangle^2)}{dt^K}\right)_0\ ,
$$
and
$$
\left(\frac{d^K\Delta \bar{P}}{dt^K}\right)_0 = \frac{1}{2\Delta
P}\left(\frac{d^K(\langle \bar{p}^2\rangle - \langle
\bar{p}\rangle^2)}{dt^K}\right)_0\ .
$$

The case of potential function (\ref{36.1}) is solvable in quantum theory
exactly as in the classical one, and we can use it for comparison with the
classical dynamics as well as for a better understanding of the ME-packet
dynamics. Eqs.\ (\ref{33.1}) have then the solutions given by (\ref{37.1}) and
(\ref{37.2}) with functions $f_n(t)$ and $g_n(t)$ given by (\ref{37.4}) and
(\ref{37.5}) or (\ref{37.9a}) and (\ref{37.9b}). The calculation of the
expectation values and variances is analogous to the classical one and we
obtain Eqs.\ (\ref{38.3}) and Eq.\ (\ref{39.1}) again with the difference that
the term $2\langle qp\rangle$ on the right hand side of (\ref{38.4}) is now
replaced by $\langle {\mathsf q}{\mathsf p}+{\mathsf p}{\mathsf q}\rangle$,
which is given by Eq.\ (\ref{QPlem}). The result is again Eq.\
(\ref{39.1}). Similarly for ${\mathsf p}$, the results are given by Eqs.\
(\ref{39.2}) and (\ref{39.3}). Hence, quantum and classical dynamics coincide
for the polynomial potentials of the second and lower orders.

For a general polynomial potential $V$, we can calculate all time derivatives
of $\bar{Q}$, $\bar{P}$, $\Delta \bar{Q}$, $\Delta \bar{P}$ by an iterative
application of Eqs.\ (\ref{mq}) and (\ref{mp}). For example,
\begin{equation}\label{52.1} \left(\frac{d\bar{P}}{d t}\right)_0 = -\langle
V'\rangle\ ,
\end{equation}
\begin{equation}\label{52.2} \left(\frac{d^2\bar{P}}{dt^2}\right)_0 = -
\langle(V')^t\rangle\ ,
\end{equation} where $(V')^t$ is obtained from the polynomial $V'$ of
${\mathsf q}$ by discarding constant terms and replacing each ${\mathsf q}$ by
${\mathsf p}/\mu$ one by one. For example, if $N = 4$,
$$
(V')^t = \frac{1}{\mu}\left[V_2{\mathsf p} + \frac{V_3}{2}({\mathsf q}{\mathsf
p} + {\mathsf p}{\mathsf q}) + \frac{V_4}{6}({\mathsf q}^2{\mathsf p} +
{\mathsf q}{\mathsf p}{\mathsf q} + {\mathsf p}{\mathsf q}^2)\right]\ .
$$

The time derivatives of $Q(t)$ are then determined by the formula:
$$
\left(\frac{d^k\bar{Q}}{dt^k}\right)_0 =
\left\langle\frac{1}{\mu}\frac{d^{k-1}\bar{P}}{dt^{k-1}}\right\rangle_0\ .
$$

In an analogous way, we obtain for the second time derivatives of the
variances:
\begin{equation}\label{varq2} \left(\frac{d^2\Delta \bar{Q}}{dt^2}\right)_0 =
\frac{1}{\mu\Delta Q}(-\langle {\mathsf q} V'\rangle - \langle {\mathsf
q}\rangle \langle V'\rangle + \Delta P^2)
\end{equation} and
\begin{equation}\label{varp2} \left(\frac{d^2\Delta \bar{P}}{dt^2}\right)_0 =
\frac{1}{2\Delta P}[2\langle V^{\prime 2}\rangle - 2\langle V'\rangle^2 -
\langle (V')^t {\mathsf p} + {\mathsf p}(V')^t\rangle + 2\langle (V')^t\rangle
\langle {\mathsf p}\rangle]\ .
\end{equation}

An important property of Eq.\ (\ref{varq2}) is the proportionality of the
right-hand side to $\mu^{-1}$. Clearly, this property will be preserved for
all higher time derivatives because they are calculated by time derivatives of
Eq.\ (\ref{varq2}). Such derivatives are then applied to various polynomials
of ${\mathsf q}$ and ${\mathsf p}$ and lead at most to further negative powers
of $\mu$. The proportionality to $\mu^{-1}$ is confirmed by Eqs.\ (\ref{39.1})
and by motion of Gaussian wave packets. It also holds for classical evolution
of position variance. Hence, the spreading of macroscopic ME packets
(including Gaussian wave packets) can be very slow in general.

The next step is to calculate the expectation values of the products of
${\mathsf q}$ and ${\mathsf p}$ that occur in time derivatives of the averages
and variances. As we are mainly interested in differences between the quantum
and classical equations, we list only the lowest-order products that show
non-trivial quantum corrections:
\begin{equation}\label{corprod1} \langle {\mathsf q}^6\rangle = Q^6 +15 Q^4
\Delta Q^2 + 45 Q^2 \Delta Q^4 + 15 \Delta Q^6 + 9 \Delta Q^6 \nu^{-1} - 3
\Delta Q^6 \nu^{-3}\ ,
\end{equation}
\begin{multline}\label{corprod2} \langle {\mathsf q}^2{\mathsf p}^2 + {\mathsf
p}^2{\mathsf q}^2\rangle = 2Q^2P^2 + 2Q^2\Delta P^2 + 2 P^2\Delta Q^2 + 2
\Delta Q^2\Delta P^2 \\ - 4\Delta Q^2\Delta P^2 \nu^{-2}\ ,
\end{multline}
\begin{equation}\label{corprod3} \langle {\mathsf p}{\mathsf q}^2{\mathsf
p}\rangle = Q^2P^2 + Q^2\Delta P^2 + P^2\Delta Q^2 + \Delta Q^2\Delta P^2 +
2\Delta Q^2\Delta P^2 \nu^{-2}\ ,
\end{equation}
\begin{equation}\label{corprod4} \langle {\mathsf q}{\mathsf p}^2{\mathsf
q}\rangle = Q^2P^2 + Q^2\Delta P^2 + P^2\Delta Q^2 + \Delta Q^2\Delta P^2 +
2\Delta Q^2\Delta P^2 \nu^{-2}\ .
\end{equation} From these formulas, we can also infer quantum corrections of
one degree higher products because of Lemma \ref{lemevenpow}. For example,
$\langle{\mathsf q}^7\rangle = Q\langle{\mathsf q}^6\rangle$, etc.

In \cite{hajicek}, the first four time derivatives of $\bar{Q}$ and $\bar{P}$
at $t=0$ have been calculated for fourth-order potential functions and shown
to coincide with the corresponding classical expressions. A more interesting
question is, at which degree of the polynomial potential any quantum
corrections appear in a time derivative of some order higher than 4. One way
to answer the question is to look for the lowest time derivative of $\bar{P}$
that contains the ${\mathsf q}^6$ term. A short estimate shows that it is the
9-th derivative and a lengthy calculation yields the explicit form of the
term:
$$
\left(\frac{d^9\bar{P}}{dt^9}\right)_0 = \ldots
-\frac{125}{4}\frac{V_3^5}{\mu^4}{\mathsf q}^6\ .
$$
Hence, quantum dynamics differs from the classical one for polynomial
potentials of degree 3 and higher. This is the same number as that obtained in
Section 2.3, but the reason is now rather different from that considered
there, the corrections are different and appear at higher time derivatives.

\section{The Theorem} Eqs.\ (\ref{corprod1}) - (\ref{corprod4}) show that the
quantum corrections to the expectation values of ${\mathsf q}{\mathsf
p}$-products, at least for the products considered there, have the following
structure: the corrections are proportional to $\nu^{-1}$ and are multiplied
by factors of the form $\Delta Q^a\Delta P^b$, where $a$ and $b$ are some
integers. The classical expression that is corrected always contains at least
one term multiplied by the same factor $\Delta Q^a\Delta P^b$. Such a
structure suggests that the quantum corrections become negligible with respect
to the corrected classical expression in the limit $\Delta Q \rightarrow
\infty,\ \Delta P \rightarrow \infty$, at least in the cases listed, because
$\nu$ itself is proportional to $\Delta Q\Delta P$.

The aim of the present section is to prove that this structure is a general
property of polynomial potential functions of arbitrary high degree and of
arbitrary high time derivatives of the expectation values and variances.
\begin{thm}\label{thmclaslim} Let $X_{mn}$ be a product of $n$ factors of
${\mathsf q}$ and $m$ factors of ${\mathsf p}$ in an arbitrary ordering. Let
$\langle X_{mn} + X_{mn}^\dagger\rangle_q$ be the expectation value in the
quantum ME packet and $\langle q^mp^n\rangle_c$ that in the classical ME
packet, both packets being defined by the same values of $Q$, $P$, $\Delta Q$
and $\Delta P$. Then, first,
\begin{equation}\label{claslim1} \langle X_{mn} + X_{mn}^\dagger\rangle_q =
2\langle q^mp^n\rangle_c + 2 R[X_{mn}](Q,P,\Delta Q,\Delta P)\ ,
\end{equation} where $R[X_{mn}](Q,P,\Delta Q,\Delta P)$ is a function of $Q$,
$P$, $\Delta Q$ and $\Delta P$ depending on $\hbar$ via $\nu$. Second,
\begin{equation}\label{claslim2} \lim_{\Delta Q \rightarrow \infty,\ \Delta P
\rightarrow \infty} \frac{R[X_{mn}](Q,P,\Delta Q,\Delta P)}{\langle
q^mp^n\rangle_c} = 0\ .
\end{equation}
\end{thm} {\bf Proof} To show Eq.\ (\ref{claslim1}), we first compare the
quantum partition function (\ref{17.3}) with its classical counterpart
(\ref{22.1}). They differ by the denominators
$\sinh(\hbar\sqrt{\lambda_3\lambda_4})$ and
$\hbar\sqrt{\lambda_3\lambda_4}$. If
\begin{equation}\label{ll} \hbar\sqrt{\lambda_3\lambda_4} \ll 1\ ,
\end{equation} we can write
$$
\sinh(\hbar\sqrt{\lambda_3\lambda_4}) = \hbar\sqrt{\lambda_3\lambda_4}[1 +
O((\hbar\sqrt{\lambda_3\lambda_4})^2)]
$$
The leading term in the quantum partition function then is
$$
Z = \frac{\pi}{h}\frac{1}{\sqrt{\lambda_3\lambda_4}}
\exp\left(\frac{\lambda_1^2}{4\lambda_3} +
\frac{\lambda_2^2}{4\lambda_4}\right)\ .
$$
Comparing this with formula (\ref{22.1}) shows that the two expressions
coincide if we set
$$
v = h\ ,
$$
where $h = 2\pi\hbar$. Thus, quantum mechanics suggests a value for $v$. Next,
we have to express condition (\ref{ll}) in terms of the expectation values and
variances. Eqs.\ (\ref{lagranqu1}) and (\ref{lagranqu2}) imply
$$
\hbar\sqrt{\lambda_3\lambda_4} = \frac{1}{2}\ln\frac{\nu+1}{\nu-1}\ .
$$
Hence, condition (\ref{ll}) is equivalent to
\begin{equation}\label{gg} \nu \gg 1\ .
\end{equation}

It follows from the above consideration that the terms in the quantum
expectation values that are of the lowest order in $\hbar$ come from that part
of the quantum partition function that has the same form as the classical
partition function. It is true that the quantum expectation values cannot be
obtained from the quantum partition function by the same way as classical
expectation values are from the classical partition function. However, as it
follows from the proof of Lemma \ref{lemderivexp}, the terms of the lowest
order in $\hbar$ can be obtained so. Hence, the leading terms in the quantum
expectation values have the same form as the classical terms. Then, if we
replace $\hbar$ by $2\Delta Q\Delta P/\nu$, Eq.\ (\ref{claslim1}) will result.

To prove Eq.\ (\ref{claslim2}), let us write the contribution of $uv$-term to
$\bar{X}_{mn} + \bar{X}_{mn}^\dagger$ given by Eq.\ (\ref{barXmnexp}) as
follows
$$
(\bar{X}_{mn} + \bar{X}_{mn}^\dagger)_{uv} = 2(-1)^vQ^{m-2u}P^{n-2v}\Delta
Q^{2u}\Delta P^{2v}\nu^{-u-v}\langle Y_{(2u)(2v)}^{(1)}({\mathsf
A}^\dagger{\mathsf A})\rangle\ .
$$
To proceed, we need the following Lemma.
\begin{lem}\label{lemnuk} For any positive integer $k$,
$$
\langle({\mathsf A}^\dagger{\mathsf A})^k\rangle = \frac{k!}{2^k}\nu^k +
h_k(\nu)\ ,
$$
where $h_k$ is a polynomial of degree $k-1$.
\end{lem} {\bf Proof} We use the method of mathematical induction. First,
$$
\langle{\mathsf A}^\dagger{\mathsf A}\rangle = \frac{1}{2}\nu - \frac{1}{2}\ .
$$
Second, according to Eq.\ (\ref{average}),
\begin{multline*} \langle({\mathsf A}^\dagger{\mathsf A})^{k+1}\rangle =
\frac{1}{\nu + 1}\left(\frac{\nu^2 - 1}{2}\frac{d}{d\nu}\right)^{k+1}(\nu + 1)
\\ = \frac{\nu - 1}{2}\frac{d}{d\nu}\left[(1 + \nu)\frac{1}{1 +
\nu}\left(\frac{\nu^2 - 1}{2}\frac{d}{d\nu}\right)^k(\nu + 1)\right] =
\frac{\nu - 1}{2}\frac{d}{d\nu}\left[(1 + \nu)\langle ({\mathsf
A}^\dagger{\mathsf A})^k\rangle\right] \\ = \frac{\nu -
1}{2}\frac{d}{d\nu}\left[(1 + \nu)\left(\frac{k!}{2^k}\nu^k +
h_k(\nu)\right)\right] = \frac{(k + 1)!}{2^{k + 1}}\nu^{k + 1} + h_{k+1}\ ,
\end{multline*} {\bf QED}.

Using the Lemma, we obtain
$$
(\bar{X}_{mn} + \bar{X}_{mn}^\dagger)_{uv} = 2Q^{m-2u}P^{n-2v}\Delta
Q^{2u}\Delta P^{2v}\left[(-1)^v z_{u + v}\frac{(u + v)!}{2^{u + v}} +
\frac{1}{\nu^{u + v}}h^{(1)}_{u + v}(\nu)\right]\ ,
$$
where
$$
h^{(1)}_{u + v}(\nu) = \sum_{k=0}^{u + v - 1}z_k\left(\frac{k!}{2^k} +
h_k(\nu)\right) + z_{u + v}h_{u + v}(\nu)
$$
is a polynomial of $\nu$ of degree $u + v - 1$. The first term in the brackets
does not contain $\hbar$ while the second is proportional to $\hbar$. Hence,
the first term is the classical part and the second is the quantum
correction. If $z_{u+v} \neq 0$, the classical part is multiplied by the same
factor $\Delta Q^{2u}\Delta P^{2v}$ as the quantum correction to it is.

Suppose that $z_{u + v} = 0$. Then, the quantum corrections would be
multiplied by a higher power of $\Delta Q$ and $\Delta P$ than the corrected
classical part. But Lemma \ref{lemhopolyncl} implies that the corresponding
term cannot disappear in the classical part. This holds for any $u$ and $v$.

The term of the highest power of $\Delta Q\Delta P$ in Eq.\ (\ref{barXmnexp})
that gives a non-zero contribution to the quantum expectation value is $\Delta
Q^{2u}\Delta P^{2v}$ with
$$
u = \left[\frac{m}{2}\right]\ ,\quad v = \left[\frac{n}{2}\right]\ .
$$
The above argument shows that the highest power of $\Delta Q\Delta P$ in the
quantum correction to the expectation value cannot be higher. Then, Eq.\
(\ref{claslim2}) follows immediately, {\bf QED}.

Theorem \ref{thmclaslim}\vspace{.5cm} motivates the following:
\par\noindent {\bf High-Entropy Conjecture} {\it For all reasonable
potentials, the classical and quantum trajectories of ME packets satisfy:}
$$
\lim_{\Delta Q\rightarrow \infty,\Delta P\rightarrow \infty}\frac{Q_q(t) -
Q_c(t)}{Q_c(t)} = 0\ ,\quad \lim_{\Delta Q\rightarrow \infty,\Delta
P\rightarrow \infty}\frac{P_q(t) - P_c(t)}{P_c(t)} = 0\ ,
$$
{\it and}
$$
\lim_{\Delta Q\rightarrow \infty,\Delta P\rightarrow \infty}\frac{\Delta
Q_q(t) - \Delta Q_c(t)}{\Delta Q_c(t)} = 0\ ,\quad \lim_{\Delta Q\rightarrow
\infty,\Delta P\rightarrow \infty}\frac{\Delta P_q(t) - \Delta P_c(t)}{\Delta
P_c(t)} = 0\ ,
$$
{\it for all} $t$\vspace{.5cm} {\it for which the formulas make sense.}
\par\noindent That is: the fuzzier the classical and quantum ME packets are,
the closer their trajectories are to each other. But the entropy of an ME
packet is an increasing function of $\Delta Q\Delta P$. One can therefore say
that the classical limit is a high-entropy limit for mechanics, which is
similar to what is, in certain sense, also true for thermodynamics.

High-Entropy Conjecture has been proven only for polynomial potential
functions. More models ought to be studied, for example the Coulomb potential
and the corresponding theory of Kepler orbits.

\section{A model of classical rigid body} To show how the above theory of
classical properties works, we construct a one dimensional model of a free
rigid body. Large parts of this section follow \cite{PHJT}. The restrictions
to one dimension and absence of external forces enable us to calculate
everything explicitly---the model is completely solvable. The real object
${\mathcal S}$ is a very thin and very stiff solid rod free to move in the
direction of its length. Its classical model $S_c$ is a one-dimensional
continuum of mass $M$ and length $L$. Classical observables are internal
energy $E$, temperature, centre of mass position $X$, total momentum $P$ and
their variances $\Delta X$ and $\Delta P$. The one-dimensionality, zero
forces, the length and the total mass (structural parameters) are just part of
the definition of the model valid for any state of it.

The construction of its quantum model $S_q$ entails that, first, the
structural properties of the system must be given, second, its Hilbert space
and observables defined, third, the objective properties specified that
correspond to all classical properties and fourth, the suitable states are to
be determined by the preparation defined by the values $E$, $X$, $P$, $\Delta
X$ and $\Delta P$.

\begin{assump}\label{assrodham} $S_q$ is a linear chain of $N+1$ particles of
mass $\mu$ distributed along the $x$-axis with the Hamiltonian
\begin{equation}\label{Ham} {\mathsf H} = \frac{1}{2\mu}\sum_{n=1}^{N+1}
{\mathsf p}^{(n)2} + \frac{\kappa^2}{2}\sum_{n=2}^{N+1} ({\mathsf x}^{(n)} -
{\mathsf x}^{(n-1)} - \xi)^2\ ,
\end{equation} involving only nearest-neighbour elastic forces. Here operator
${\mathsf x}^{(n)}$ is the position, operator ${\mathsf p}^{(n)}$ the momentum
of the $n$-th particle, $\kappa$ the oscillator strength and $\xi$ the
equilibrium interparticle distance.
\end{assump} The parameters $N$, $\mu$, $\kappa$, $\xi$ and the label $n$ are
structural properties. The particles are not identical but distinguished by
their ordering within the chain. The minimal set of observables that generate
the whole algebra of observables of $S_q$ are ${\mathsf x}^{(n)}$ and
${\mathsf p}^{(n)}$, $n = 1,\ldots,N$. Their number is $2N + 2$. Thus, the
quantum model is much richer than the classical one, whose state is determined
just by 5 numbers. We shall construct some set of quantum observables and
obtain some relations between structural parameters of the two models, such as
between $\mu$ and $M$ or $\xi$ and $L$.

This kind of chain is in one respect different from most chains that are
studied in literature: the positions of the chain particles are dynamical
variables so that the chain can move as a whole and the invariance with
respect to (one-dimensional space) Galilean group is not disturbed. However,
the chain can still be solved by methods that are described in
\cite{Kittel,Rutherford}.

First, we find the variables ${\mathsf u}_n$ and ${\mathsf q}_n$ that
diagonalize the Hamiltonian and define thus the so-called normal modes. The
transformation is
\begin{equation}\label{xu} {\mathsf x}^{(n)} = \sum_{m=0}^{N}Y^m_n{\mathsf
u}_m + \left(n - \frac{N+2}{2}\right)\xi\ ,
\end{equation} and
\begin{equation}\label{pu} {\mathsf p}^{(n)} = \sum_{m=0}^{N}Y^m_n{\mathsf
q}_m\ ,
\end{equation} where the mode index $m$ runs through $0,1,\cdots,N$ and
$Y^m_n$ is an orthogonal matrix; for even $m$,
\begin{equation}\label{evenm} Y^m_n = A(m)\cos\left[\frac{\pi
m}{N}\left(n-\frac{N+2}{2}\right)\right],
\end{equation} while for odd $m$,
\begin{equation}\label{oddm} Y^m_n = A(m)\sin\left[\frac{\pi
m}{N}\left(n-\frac{N+2}{2}\right)\right]\ ,
\end{equation} and the normalization factors are given by
\begin{equation}\label{factor} A(0) = \frac{1}{\sqrt{N+1}}\ ,\quad A(m) =
\sqrt{\frac{2}{N+1}},\quad m>0\ .
\end{equation}

To show that ${\mathsf u}_n$ and ${\mathsf q}_n$ do represent normal modes, we
substitute Eqs.\ (\ref{xu}) and (\ref{pu}) into (\ref{Ham}) and obtain, after
some calculation,
$$
{\mathsf H} = \frac{1}{2\mu}\sum_{m=0}^{N}{\mathsf q}_m^2 +
\frac{\mu}{2}\sum_{m=0}^{N}\omega_m^2{\mathsf u}_m^2\ ,
$$
which is indeed diagonal. The mode frequencies are
\begin{equation}\label{spectr} \omega_m =
\frac{2\kappa}{\sqrt{\mu}}\sin\frac{m}{N}\frac{\pi}{2}\ .
\end{equation}

Consider the terms with $m=0$. We have $\omega_0=0$, and
$Y^0_n=1\sqrt{N+1}$. Hence,
$$
{\mathsf u}_0 = \sum_{n=1}^{N+1}\frac{1}{\sqrt{N+1}}{\mathsf x}^{(n)}\ , \quad
{\mathsf q}_0 = \sum_{n=1}^{N+1}\frac{1}{\sqrt{N+1}}{\mathsf p}^{(n)}\ ,
$$
so that
$$
{\mathsf u}_0 = {\mathsf X}\sqrt{N+1}\ ,\quad {\mathsf q}_0 = \frac{{\mathsf
P}}{\sqrt{N+1}}\ ,
$$
where ${\mathsf X}$ is the centre-of-mass coordinate of the chain and
${\mathsf P}$ is its total momentum. The ``zero'' terms in the Hamiltonian
then reduce to
\begin{equation}\label{bulkham} \frac{1}{2(N+1)\mu}{\mathsf P}^2\ .
\end{equation} Thus, the ``zero mode'' describes a straight, uniform motion of
the chain as a whole. The fact that the centre of mass degrees of freedom
decouple from other, internal, ones is a consequence of the Galilean
invariance.

The other modes are harmonic oscillators called ``phonons'' with frequencies
$\omega_m$, $m = 1,2,\dots,N$. Important observables are the energy of the
phonons,
\begin{equation}\label{intenop} {\mathsf H}_{\text{int}} =
\frac{1}{2\mu}\sum_{m=1}^{N}{\mathsf q}_m^2 +
\frac{\mu}{2}\sum_{m=1}^{N}\omega_m^2{\mathsf u}_m^2\ ,
\end{equation} and the length of the body,
\begin{equation}\label{length} {\mathsf L} = {\mathsf x}^{(N)} - {\mathsf
x}^{(1)}\ .
\end{equation}

We assume that
$$
E = \langle {\mathsf H}_{\text{int}} \rangle\ ,\quad M = (N+1)\mu\ ,
$$
$$
L = \langle {\mathsf L} \rangle\ , \quad X = \langle {\mathsf X} \rangle\ ,
\quad P = \langle {\mathsf P} \rangle\ , \quad \Delta X = \langle \Delta
{\mathsf X} \rangle\ , \quad \Delta P = \langle \Delta {\mathsf P} \rangle\ .
$$

The next point is the choice of suitable states. We write the Hilbert space of
$S_q$ as
$$
{\mathbf H} = {\mathbf H}_{\text{CM}} \otimes {\mathbf H}_{\text{int}}\ ,
$$
where ${\mathbf H}_{\text{CM}}$ is constructed from the wave functions
$\Psi(X)$ and ${\mathbf H}_{\text{int}}$ has the phonon eigenstates as a
basis.
\begin{assump}\label{assrodstate} The suitable states have the form
$$
{\mathsf T}_{\text{CM}} \otimes {\mathsf T}_{\text{int}}\ .
$$
Internal state ${\mathsf T}_{\text{int}}$ maximises the entropy under the
condition of fixed expectation value of the internal energy,
\begin{equation}\label{intenergav} \text{Tr}({\mathsf T}_{\text{int}}{\mathsf
H}_{\text{int}}) = E\ ,
\end{equation} where Tr is the partial trace over ${\mathbf
H}_{\text{int}}$. The external state ${\mathsf T}_{\text{CM}}$ is the ME
packet for given expectation values $X$, $P$, $\Delta X$ and $\Delta P$.
\end{assump}

For ${\mathsf T}_{\text{int}}$, the condition of maximum entropy has to do
with the preparation. Physically, the thermodynamic equilibrium can settle
down spontaneously starting from an arbitrary state only if some weak but
non-zero interaction exists both between the phonons and between the rod and
the environment. We assume that this can be arranged so that the interaction
can be neglected in the calculations of the present section. It turns then out
that all other classical internal properties are functions of the classical
internal energy.

The mathematics associated with the maximum entropy principle is variational
calculus as in Sections 2.3 and 3.1. The condition of fixed expectation value
of energy is expressed with the help of Lagrange multiplier denoted by
$\lambda$:
$$
d\Sigma - \lambda_0 d\,tr({\mathsf T}_{\text{int}}) - \lambda d\,tr({\mathsf
T}_{\text{int}}{\mathsf H}_{\text{int}}) = 0\ .
$$
Substituting from Eq.\ (\ref{10.1}) for the differential of entropy, we obtain
$$
{\mathsf T}_{\text{int}} = \exp(-1 - \lambda_0 - \lambda {\mathsf
H}_{\text{int}})\ .
$$
Hence, ${\mathsf T}_{\text{int}}$ is a Gibbs state for some temperature $T$,
where
$$
\lambda(E) = \frac{1}{kT}
$$
and $k$ is Boltzmann constant (actually, this is a general theorem, see, e.g.,
\cite{Jaynes}).

The values of $\lambda_0$ and $\lambda$ are obtained from the normalisation
condition, $Tr({\mathsf T}_{\text{int}}) = 1$, and Eq.\ (\ref{intenergav}) for
the expectation value of energy. The normalisation condition yields
$$
\exp(1 + \lambda_0) = Z(\lambda)\ ,
$$
where
\begin{equation}\label{partifuncphon} Z(\lambda) = Tr\Bigl(\exp(-\lambda
{\mathsf H}_{\text{int}})\Bigr)
\end{equation} is the partition function of the phonons. We have then
$$
\frac{1}{Z(\lambda)}\frac{dZ(\lambda)}{d\lambda} = E\ ,
$$
which gives the relation between the expectation value energy and the
temperature. From the first part of Lemma \ref{lemderivexp}, it follows that
$$
\frac{1}{Z(\lambda)}\frac{d^2Z(\lambda)}{d\lambda^2} = Tr({\mathsf
T}_{\text{int}}{\mathsf H}^2_{\text{int}})\ .
$$

Calculations are simplified if the variables ${\mathsf u}_m$ and ${\mathsf
q}_m$ are transformed to the annihilation and creation operators ${\mathsf
a}_m$ and ${\mathsf a}^\dagger_m$ of the phonons:
$$
{\mathsf u}_m = \sqrt{\frac{\hbar}{2\mu\omega_m}}({\mathsf a}^\dagger_m +
{\mathsf a}_m)\ ,\quad {\mathsf q}_m =
i\sqrt{\frac{\hbar\mu\omega_m}{2}}({\mathsf a}^\dagger_m - {\mathsf a}_m)
$$
so that $[{\mathsf a}_m,{\mathsf a}^\dagger_m] = 1$. Then,
$$
{\mathsf H}_{\text{int}} = \sum_{m=1}^N\hbar\omega_m\left({\mathsf
a}^\dagger_m{\mathsf a}_m + \frac{1}{2}\right)\ .
$$
The eigenvalues of the phonon number operator ${\mathsf a}^\dagger_m{\mathsf
a}_m$, the phonon occupation numbers, will be denoted by $n_m$. Then, the
spectrum of ${\mathsf H}_{\text{int}}$ is built from the mode frequencies by
the formula
\begin{equation}\label{phonons} E_{\text{int}} = \sum_{m=1}^{N}n_m
\hbar\omega_m\ .
\end{equation}

Let us set
$$
{\mathbf H}_{\text{int}} = \prod_{m=1}^N\otimes {\mathbf H}_m\ ,
$$
where ${\mathbf H}_m$ is the Hilbert space of the phonons of type $m$. For
calculation of traces of operators ${\mathsf A}$ of the form
$$
{\mathsf A} = \sum_{m=1}^N{\mathsf A}_m\ ,
$$
over ${\mathbf H}_{\text{int}}$, where ${\mathsf A}_m$ is an operator on
${\mathbf H}_m$, we use the formula
$$
Tr({\mathsf A}) = \prod_{m=1}^N tr^{(m)}({\mathsf A}_m)\ ,
$$
where $tr^{(m)}$ is the trace over the Hilbert space of the photon of kind
$m$.

An easy calculation gives
$$
Z(\lambda) = \prod_{m=1}^N Z_m(\lambda)\ ,
$$
where
$$
Z_m(\lambda) = \frac{\exp(-\frac{1}{2}\lambda\hbar\omega_m)}{1 -
\exp(-\lambda\hbar\omega_m)}\ .
$$
Then,
\begin{equation}\label{factort} {\mathsf T}_{\text{int}} = \prod_{m=1}^N
{\mathsf T}^{(m)}\ ,
\end{equation} where
$$
{\mathsf T}^{(m)} = Z_m^{-1}\exp\left[-\lambda\hbar\omega_m \left({\mathsf
a}^\dagger_m{\mathsf a}_m + \frac{1}{2}\right)\right]\ .
$$
As it is well-known, the internal energy has itself a very small relative
variance, $\Delta E/E$, in the Gibbs state if $N$ is large.

The diagonal matrix elements of ${\mathsf u}_m$ between the energy eigenstates
$|n_m\rangle$ that we shall need then are
\begin{equation}\label{averu} \langle n_m|{\mathsf u}_m|n_m\rangle = 0,\quad
\langle n_m| {\mathsf u}^2_m|n_m\rangle = \frac{\hbar}{2\mu\omega_m}(2n_m +
1).
\end{equation}

The length can be expressed in terms of modes ${\mathsf u}_m$ using
Eq.~(\ref{xu}),
$$
{\mathsf L} = N\xi + \sum_{m=0}^{N}(Y^m_N-Y^m_1){\mathsf u}_m\ .
$$
The differences on the right-hand side are non-zero only for odd values of
$m$, and equal then to $-2Y^m_1$. We easily find, using Eqs.~(\ref{oddm}) and
(\ref{factor}):
\begin{equation}\label{L} {\mathsf L} = N\xi - \sqrt{\frac{8}{N}}\
\sum_{m=1}^{[(N+1)/2]}(-
1)^m\cos\left(\frac{2m-1}{N+1}\frac{\pi}{2}\right)\,{\mathsf u}_{2m-1}\ ,
\end{equation} where $[(N+1)/2]$ is the entire part of $(N+1)/2$.

The expectation value of the length operator in the Gibbs state can then be
obtained using Equqtions (\ref{averu}) and (\ref{factort}),
\begin{equation}\label{averL} \langle {\mathsf L}\rangle = N\xi.
\end{equation} It is a function of objective properties $N$, $\xi$ and $E$.

Eq.\ (\ref{L}) is an important result. It shows that contributions to the
length are more or less evenly distributed over all odd modes. Such a
distribution leads to a very small variance of ${\mathsf L}$ in Gibbs
states. A lengthy calculation \cite{PHJT} using Eq.\ (\ref{averu}) gives for
large $N$
\begin{equation} \frac{\Delta {\mathsf L}}{\langle {\mathsf L}\rangle} \approx
\frac{2\sqrt{3}}{\pi\kappa\xi\sqrt{\lambda}}\frac{1}{\sqrt{N+1}}.
\end{equation}

Thus, the small relative variance for large $N$ need not be assumed from the
start and it guarantees the approximative match between the quantum and
classical models of the real object. The only assumptions are values of some
structural parameters and that an expectation value of energy is fixed. We
have obtained even more information, viz.\ the internal-energy dependence of
the length (in this model, the length is a structure parameter and the
dependence is trivial). This is an objective relation that can be in principle
tested by measurements.

Similar results can be obtained for further thermodynamic properties such as
specific heat, elasticity coefficient\footnote{If we extend the classical
model so that it contains the elasticity coefficient, we could calculate the
coefficient for an extended quantum model, in which the rod would be placed
into a non-homogeneous ``gravitational'' field described by, say, a quadratic
potential. This would again give a solvable model.} etc. All these quantities
are well known to have small variances in Gibbs states. The reason is that the
contributions to these quantities are evenly distributed over the normal modes
and the modes are mechanically and statistically independent.

The mechanical properties of the system are the centre of mass and the total
momentum. The contributions to them are evenly distributed over all atoms, not
modes: the bulk motion is mechanically and statistically independent of all
other modes and so its variances will not be small in Gibbs states defined by
a fixed expectation value of the total energy. Still, generalized statistical
methods of Chapter 2 can be applied to it.

First, we assume that the real rod we are modelling cannot possess a sharp
trajectory and that satisfactory models of it can be ME packets in both
Newtonian and quantum mechanics. Then, according to Assumption
\ref{assrodstate} and Theorem \ref{propold19}, the external state of the
classical model can be chosen as
\begin{equation}\label{MErodc} \rho = \frac{v}{2\pi}\frac{1}{\langle \Delta
X\rangle\langle \Delta P\rangle}\exp\left[-\frac{(X-\langle
X\rangle)^2}{2\langle \Delta X\rangle^2} -\frac{(P-\langle
P\rangle)^2}{2\langle \Delta P\rangle^2}\right]\ .
\end{equation} Similarly, Theorem \ref{propold20} implies that the external
state of the quantum model can be chosen as
\begin{equation}\label{MErodq} {\mathsf T}_{\text{CM}}=
\frac{2}{\sqrt{\nu^2-1}}\exp\left(-\frac{\nu}{2}\ln\frac{\nu+1}{\nu-1}{\mathsf
K}'\right)\ ,
\end{equation} where
$$
{\mathsf K}' = \frac{1}{2}\frac{({\mathsf q}-Q)^2}{\Delta Q^2} +
\frac{1}{2}\frac{({\mathsf p}-P)^2}{\Delta P^2}\ .
$$

The Hamiltonian for the bulk motion of both models is given by Eq.\
(\ref{bulkham}). Thus, as explained in Section 3.3, the quantum trajectory
coincides with the classical one exactly. (Recall that trajectory has been
defined as the time dependence of expectation values and variances.)

Hopefully, this simple rod example has sufficiently illustrated how our idea
of model construction works in the case of classical properties and we can
finish the comparison of classical and quantum models here.

\chapter{Measurement and exchange symmetry} \setcounter{equation}{0}
\setcounter{thm}{0} \setcounter{assump}{0} \setcounter{sh}{0}
\setcounter{df}{0} \setcounter{lem}{0} Let systems $S^{(1)}$ and $S^{(2)}$ of
different types in states ${\mathsf T}^{(1)}$ and ${\mathsf T}^{(2)}$ be
considered as one composite system $S$. Then the state ${\mathsf T}$ of $S$ is
the tensor product,
$$
{\mathsf T} = {\mathsf T}^{(1)} \otimes {\mathsf T}^{(2)}\ ,
$$
see \cite{peres}, p.\ 115. However, if the systems to be composed are of the
same type then the situation is different:
\begin{quote} Any [registration] performed on the [composite] quantum system
treats all [indistinguishable] subsystems in the same way, and it is
indifferent to a permutation of the labels that we attribute to the individual
subsystems for computational purposes.
\end{quote} (Peres \cite{peres}, p.\ 126.) Thus, the particle labels or
tensor-product positions have no physical importance: they are superfluous
variables similar to gauge in gauge field theories. This introduces another
group into the foundation of quantum mechanics: the system-permutation group.

In the present chapter, we are going to show how the RCU interpretation of
quantum mechanics can be adapted to exchange symmetry. In particular, we
review the results of \cite{incomplete}, especially on the incompleteness of
registration apparatuses, on the interpretation of preparation processes and
of quantum observables. Our understanding of systems, states and observables
will thus reach its definitive shape.

\section{Indistinguishable systems} Here, we collect briefly some well-known
mathematics (see, e.g., Section 5-4 of \cite{peres} or Chapter 17 of
\cite{ballent}) in the form that will be useful for further development and
use it to state physical assumptions concerning indistinguishable particles in
the framework of RCU interpretation.

We begin by an account of the action of the group of system permutations on
tensor products of Hilbert spaces. Let ${\mathbf S}_N$ be the permutation
group of $N$ objects, that is, each element $g$ of ${\mathbf S}_N$ is a
bijective map $g : \{1,\cdots,N\} \mapsto \{1,\cdots,N\}$, the inverse element
to $g$ is the inverse map $g^{-1}$ and the group product of $g_1$ and $g_2$ is
defined by the composition of the maps, $(g_1 g_2)(k) = g_1(g_2(k))$, $k \in
\{1,\cdots,N\}$.

Given a Hilbert space $\mathbf H$, let us denote by ${\mathbf H}^N$ the tensor
product of $N$ copies of $\mathbf H$,
$$
{\mathbf H}^N = {\mathbf H} \otimes {\mathbf H} \otimes \cdots \otimes
{\mathbf H}\ .
$$
On ${\mathbf H}^N$, the permutation group ${\mathbf S}_N$ acts as follows. Let
$|\psi_k\rangle \in {\mathbf H}$, $k=1,\cdots,N$, be $N$ vectors. Then
$$
|\psi_1\rangle \otimes \cdots \otimes |\psi_N\rangle \in {\mathbf H}^N
$$
and
\begin{equation}\label{perm} {\mathsf g} (|\psi_1\rangle \otimes \cdots
\otimes |\psi_N\rangle) = |\psi_{g(1)}\rangle \otimes \cdots \otimes
|\psi_{g(N)}\rangle\ .
\end{equation} $\mathsf g$ is linear, preserves the inner product of ${\mathbf
H}^N$ and is, therefore, bounded and continuous. Hence, it can be extended by
linearity and continuity to the whole of ${\mathbf H}^N$. The resulting
operator on ${\mathbf H}^N$ is denoted by the same symbol $\mathsf g$ and is a
unitary operator by construction. The action (\ref{perm}) thus defines a
unitary representation of the group ${\mathbf S}_N$ on ${\mathbf H}^N$.

Consider a particle with Hilbert space ${\mathbf H}_s$ defined at the
beginning of Section 1.4. The wave functions representing vectors of ${\mathbf
H}_s$ have the form $\psi(\vec{x},m)$ in the $Q$-representation, in which the
position and the third component of spin are diagonal. As it is well known,
other representations exist, e.g.\ the $P$-representation, in which the
momentum and the first component, say, of spin are diagonal and we have
functions $\psi(\vec{p},m)$, or a representation associated with some
Hilbert-space basis $\{|n\rangle\}$ and we have functions of discrete numbers
$n$ so that $\psi(n)$ are coefficients in the expansion of the vector into the
basis. We can work in any representation by replacing the arguments of the
wave function by a single shorthand $\lambda = (\vec{x},m)$, or $\lambda =
(\vec{p},m')$, or $\lambda = n$, etc. The scalar product of two vectors
$|\psi\rangle$ and $|\phi\rangle$ can then be written as the Stieltjes
integral over $\lambda$ (see, e.g., \cite{RS}, p.\ 19):
$$
\langle \psi|\phi \rangle = \int d\lambda\, \psi^*(\lambda) \phi(\lambda)
$$
so that
$$
\int d\lambda\, \psi^*(\lambda) \phi(\lambda) = \sum_{m=-s}^s\int_{{\mathbb
R}^3}d^3x\,\psi^*(\vec{x},m) \phi(\vec{x},m) = \sum_n \psi^*(n) \phi(n)\ ,
$$
etc. This method not only shortens formulas but is also manifestly
representation independent.

Next, consider two such particles. The wave functions that are associated with
vectors of ${\mathbf H}_s \otimes {\mathbf H}_s$ have the form
$$
\psi(\lambda^{(1)},\lambda^{(2)})\ ,
$$
and the wave function associated with vector $|\psi_1\rangle \otimes
|\psi_2\rangle$ is
$$
\psi_1(\lambda^{(1)}) \psi_2(\lambda^{(2)})\ .
$$
For a rigorous justification of writing tensor products in this way, see
\cite{RS}, p.\ 51. The only non-trivial element of group ${\mathbf S}_2$, $g$,
exchanges the two numbers 1 and 2 and its action on $|\psi_1\rangle \otimes
|\psi_2\rangle$ is represented by
$$
g\bigl(\psi_1(\lambda^{(1)}) \psi_2(\lambda^{(2)})\bigr) =
\psi_2(\lambda^{(1)}) \psi_1(\lambda^{(2)}) = \psi_1(\lambda^{(2)})
\psi_2(\lambda^{(1)})\ .
$$
The extension of this operation to the whole of ${\mathbf H}_s \otimes
{\mathbf H}_s$ leads to the following operation on general wave functions:
$$
g\bigl(\psi(\lambda^{(1)},\lambda^{(2)})\bigr) =
\psi(\lambda^{(2)},\lambda^{(1)})\ .
$$
Hence, it exchanges the system labels.

Group ${\mathbf S}_2$ has just two irreducible representations: the symmetric
one on the symmetric wave functions, and the alternating one on the
antisymmetric wave functions (see, e.g., \cite{wigner}, Chapter 14). Clearly,
(A) each wave function can be written as a sum of a symmetric and an
antisymmetric one,
$$
\psi(\lambda^{(1)},\lambda^{(2)}) =
\frac{1}{2}\bigl(\psi(\lambda^{(1)},\lambda^{(2)}) +
\psi(\lambda^{(2)},\lambda^{(1)})\bigr) +
\frac{1}{2}\bigl(\psi(\lambda^{(1)},\lambda^{(2)}) -
\psi(\lambda^{(2)},\lambda^{(1)})\bigr)
$$
and (B), each symmetric wave function, $\psi_s(\lambda^{(1)},\lambda^{(2)})$
is orthogonal to each antisymmetric one,
$\psi_a(\lambda^{(1)},\lambda^{(2)})$,
$$
\int d\lambda^{(1)}d\lambda^{(2)}\,\psi^*_s(\lambda^{(1)},\lambda^{(2)})
\psi_a(\lambda^{(1)},\lambda^{(2)}) = 0\ .
$$
It then follows that the whole Hilbert space is an orthogonal sum of two
subspaces, each containing only states transforming under one irreducible
representation of group ${\mathbf S}_2$.

In general, all vectors of ${\mathbf H}^N$ that transform according to a fixed
irreducible unitary representation ${\mathcal R}$ of ${\mathbf S}_N$ form a
closed linear subspace of ${\mathbf H}^N$ that will be denoted by ${\mathbf
H}^N_{\mathcal R}$. The representations being unitary, the subspaces ${\mathbf
H}^N_{\mathcal R}$ are orthogonal to each other (see, e.g.,
\cite{wigner}). Let us denote by ${\mathsf \Pi}^{N}_{\mathcal R}$ the
orthogonal projection operator,
$$
{\mathsf \Pi}^{N}_{\mathcal R} : {\mathbf H}^N \mapsto {\mathbf H}^N_{\mathcal
R}\ .
$$
The index $N$ at ${\mathsf \Pi}^{N}_{\mathcal R}$ ought not to be confused
with a power (in fact, the projections are idempotent).

The location order of a given state in a tensor product or the index at the
variables $\lambda^{(k)}$ can be considered as an information about the
identity of the corresponding system. As already mentioned, such information
has no physical meaning and any permutation thereof is just a kind of gauge
transformation\footnote{A different and independent part (ignored here) of the
theory of identical particles is that states of two identical systems can also
be swapped in a physical process of continuous evolution, and can so entail a
non-trivial phase factor at the total state (anyons, see, e.g.\ Ref.\
\cite{wilczek}).}. The formalism will be gauge invariant if only
one-dimensional unitary representations of ${\mathbf S}_N$ are allowed because
only these transform vectors by a phase factor multiplication and thus do not
change the corresponding states. Another motivation for this restriction is
the cluster separability (see \cite{peres}, p.\ 128). ${\mathbf S}_N$ has
exactly two one-dimensional unitary representations: the symmetric (trivial)
one, $g \mapsto {\mathsf 1}$, and the alternating one $g \mapsto
\eta(g){\mathsf 1}$ for each $g \in {\mathbf S}_N$, where $\eta(g) = 1$ for
even and $\eta(g) = -1$ for odd permutation $g$ \cite{wigner}. If ${\mathcal
R}$ is the symmetric (alternating) representation we use symbol ${\mathbf
H}^N_+ $ (${\mathbf H}^N_-$) for ${\mathbf H}^N_{\mathcal R}$. Let us
introduce index $\tau$ with values $-1$ and $+1$ (in indices, we just write
$+$ or $-$) and let ${\mathsf \Pi}_\tau^{N}$ be the orthogonal projection on
${\mathbf H}^N_\tau$. The projection symmetrize or antisymmetrize in
dependence of $\tau$ and can therefore be called
``$\tau$-symmetrisation''. Note that the usual operation of symmetrisation or
antisymmetrisation (we shall speak of $\tau$-symmetrisation in general) on a
vector $\Psi \in{\mathbf H}^N$, such as
$$
|\psi\rangle \otimes |\phi\rangle \mapsto (1/2)(|\psi\rangle \otimes
|\phi\rangle \pm |\phi\rangle \otimes |\psi\rangle)
$$
for ${\mathbf H}^2$, is nothing but ${\mathsf \Pi}^{N}_+\Psi$ or ${\mathsf
\Pi}^{N}_-\Psi$, respectively.

An important property of the subspaces ${\mathbf H}^N_\tau$ is their
invariance with respect to tensor products of unitary transformations. Let
${\mathsf U}$ be a unitary transformation on ${\mathbf H}$, then ${\mathsf U}
\otimes {\mathsf U} \otimes \ldots \otimes {\mathsf U}$ is a unitary
transformation on ${\mathbf H} \otimes {\mathbf H} \otimes \ldots \otimes
{\mathbf H}$ and each subspace ${\mathbf H}^N_\tau$ is invariant with respect
to it. Hence, ${\mathsf U} \otimes {\mathsf U} \otimes \ldots \otimes {\mathsf
U}$ acts as a unitary transformation on ${\mathbf H}^N_\tau$ for each
$\tau$. A unitary representation of a group ${\mathbf G}$ on ${\mathbf
H}^N_\tau$ that is constructed in this way from a representation of ${\mathbf
G}$ on ${\mathbf H}$ is the tensor product of representations (see e.g.\
\cite{rao}, Section 5.17; there, tensor products are called ``Kronecker
products'').
\begin{thm}\label{propsymG} Let ${\mathsf G}$ be the generator of a
one-parameter Lie group $g(t)$ of unitary operators on ${\mathbf H}$. Then,
the generator $\tilde{\mathsf G}$ of $g(t)$ on ${\mathbf H}^N_\tau$ for the
tensor product of the representations is given by
\begin{equation}\label{symG} \tilde{\mathsf G} = {\mathsf G} \otimes {\mathsf
1} \otimes \ldots \otimes {\mathsf 1} + {\mathsf 1} \otimes {\mathsf G}
\otimes {\mathsf 1} \otimes \ldots \otimes {\mathsf 1} + \ldots + {\mathsf 1}
\otimes \ldots \otimes {\mathsf 1} \otimes {\mathsf G}\ ,
\end{equation} where ${\mathsf 1}$ is the unit operator on ${\mathbf H}$.
\end{thm} The Theorem follows easily from the definition of group
generators. Observe that the form of the generator is independent of whether
the space is symmetric or antisymmetric.

Now, we are ready to formulate the basic assumption of standard quantum
mechanics concerning identical subsystems. From relativistic quantum field
theory (see, e.g., \cite{Weinberg}), we take over the following result.
\begin{assump}\label{rlold12} Let ${\mathcal S}^N$ be a quantum system
composite of $N$ subsystems ${\mathcal S}$ of the same type, each with Hilbert
space ${\mathbf H}_s$. Then, the Hilbert space of ${\mathcal S}^N$ is
$({\mathbf H}_s)_\tau^N$ with $\tau = (-1)^{2s}$. The definition of system
${\mathcal S}^N$ is finished if a representation of group $\bar{\mathbf
G}^+_\mu$ (see Section 1.4) on $({\mathbf H}_s)_\tau^N$ is chosen.
\end{assump} As it is well known, systems with integer spin are called {\em
bosons} and those with half-integer spin are called {\em fermions}.
\begin{assump}\label{assymG} Let ${\mathsf G}$ be the generator of subgroup
$g(t)$ of $\bar{\mathbf G}^+_\mu$ on ${\mathbf H}$ that does not contain time
translations. Then, the generator $\tilde{\mathsf G}$ of $g(t)$ on ${\mathbf
H}^N_\tau$ is given by Eq.\ (\ref{symG}). For time translation subgroup,
suitable potential function can be added to the right-hand side of Eq.\
(\ref{symG}). The potential function must commute with all other generators of
the group and be an operator on ${\mathbf H}^N_\tau$.
\end{assump} The constructed representation of $\bar{\mathbf G}^+_\mu$ on
${\mathbf H}^N_\tau$ for $N > 1$ is not irreducible and non-equivalent
representations of it on ${\mathbf H}^N_\tau$ can differ by potential
functions, that is addition terms in the time-translation operator that
commute with all other generators of the group (see e.g.\ \cite{KP}). Hence,
to define a system of $N$ identical particles, one has also to specify its
Hamiltonian.

For example, a possible Hamiltonian of three identical fermions with mass
$\mu$ and charge $e$ is
\begin{equation}\label{threeH} -\frac{\hbar^2}{2\mu}\left(\bigtriangleup^{(1)}
+ \bigtriangleup^{(2)} + \bigtriangleup^{(3)}\right) + V(\vec{x}^{(1)},
\vec{x}^{(2)}) + V(\vec{x}^{(3)}, \vec{x}^{(1)}) + V(\vec{x}^{(2)},
\vec{x}^{(3)})\ ,
\end{equation} where $\bigtriangleup$ is the Laplacean and
\begin{equation}\label{threeV} V(\vec{x}^{(k)}, \vec{x}^{(l)}) =
\frac{e^2}{4\pi\epsilon_0 |\vec{x}^{(k)} - \vec{x}^{(l)}|}
\end{equation} is a potential function that is invariant with respect to
translations, rotations and boosts. We can see that both the differential
operator and the potential function are invariant with respect of all
permutation of three numbers 1, 2 and 3. This is necessary and sufficient for
the Hamiltonian to be an operator on ${\mathbf H}^3_-$

The first term in the Hamiltonian is defined by a one-particle operator
$-\hbar^2/(2\mu)\bigtriangleup$ and the rest is defined by a two-particle
operator $V(\vec{x}^{(1)}, \vec{x}^{(2)})$.

For states and observables, we have:
\begin{assump}\label{propold12} Possible states of system ${\mathcal S}^N$
composite of $N$ systems of the same type with spin $s$ are state operators on
${\mathbf H}^N_\tau$ (elements of ${\mathbf T}({\mathbf H}^N_\tau)$) and the
observables of ${\mathcal S}^N$ are self-adjoint operators on ${\mathbf
H}^N_\tau$, where $\tau = (-1)^{2s}$. The probabilities of registration
outcomes are given by the corresponding Born rule.
\end{assump}

The well-known rules represented by Assumptions \ref{rlold12}, \ref{assymG}
and \ref{propold12} lead to deeper understanding of quantum systems. From TH's
\ref{shobject} and \ref{rhmacro}, we can conclude that, for any particle of a
given type, there is a huge number of particles of the same type somewhere in
the world. In the present chapter, we have learnt: If we label a system so
that it is distinguished from another system of the same type, then such a
label is just a mathematical tool and does not carry any physical
information. Hence, a quantum system is just an auxiliary mathematical
notion. A physical quantum object is defined by its type and its prepared
state. This state can be ``occupied'' by any system of the type. We are
going to make this statement clearer in Section 4.3.

\section{Born rule} Here, we are going to study the connection between an
observable and a meter that registers the observable. A new understanding of
this connection can be achieved if the exchange symmetry is taken into
account.

An observable is an $n$-tuple $\{{\mathsf O}_1,\ldots,{\mathsf O}_n\}$ of
commuting self-adjoint operators ${\mathsf O}_k : {\mathbf H}_s \mapsto
{\mathbf H}_s$, $k = 1,\ldots,n$. Let $\sigma \subset {\mathbb R}^n$ be the
spectrum of $\{{\mathsf O}_1,\ldots,{\mathsf O}_n\}$, ${\mathcal B}({\mathbb
R}^n)$ the set of Borel subsets of ${\mathbb R}^n$ and let ${\mathsf \Pi}(X)$,
$X \in {\mathcal B}({\mathbb R}^n)$, describe the spectral measure of
$\{{\mathsf O}_1,\ldots,{\mathsf O}_n\}$ (see e.g.\ \cite{RS}, p.\ 228). In
particular, ${\mathsf \Pi}(X)$ is an orthogonal projection on ${\mathbf H}_s$
for each $X \in {\mathcal B}({\mathbb R}^n)$,
$$
{\mathsf \Pi}(X)^\dagger = {\mathsf \Pi}(X)\ ,\quad {\mathsf \Pi}(X)^2 =
{\mathsf \Pi}(X)\ ,
$$
and satisfies the normalization condition,
\begin{equation}\label{normaliz} {\mathsf \Pi}({\mathbb R}^n) = {\mathsf 1}\ .
\end{equation}

A relation between observables and registrations is expressed by the {\em Born
rule}. According to the usual notion of it, the probability that a value of
the observable within $X$ will be obtained by registration on state ${\mathsf
T} \in {\mathbf T}({\mathbf H}_s)$ is $tr({\mathsf T}{\mathsf \Pi}(X))$. In
practice, the Born rule means that the relative frequencies of the values
obtained by many registration by the same meter ${\mathcal M}$ on the same
state ${\mathsf T}$ must tend to the probabilities given by the Born rule if
the number or the registrations increases. We are going to formulate this in
the following cautious way.
\begin{df}\label{dfAregO} Given state ${\mathsf T}$ and $X \in {\mathcal
B}({\mathbb R}^n)$, let us denote by $\omega[{\mathcal M},{\mathsf T}](X)$ the
frequencies of finding values of observable $\{{\mathsf O}_1,\ldots,{\mathsf
O}_n\}$ within $X$ obtained from many registrations by meter ${\mathcal M}$ on
state ${\mathsf T}$. Let $\omega[{\mathcal M},{\mathsf T}](X)$ tends to
$tr({\mathsf T} {\mathsf \Pi}(X))$ as the number of registration events
increases,
\begin{equation}\label{born2} \omega[{\mathcal M},{\mathsf T}](X) \mapsto
tr({\mathsf T} {\mathsf \Pi}(X))\ ,
\end{equation} for some states ${\mathsf T}$ and all $X \in {\mathcal
B}({\mathbb R}^n)$. Then we say that meter ${\mathcal M}$ measures $\{{\mathsf
O}_1,\ldots,{\mathsf O}_n\}$.
\end{df} We must also clarify, what are the relative frequencies:
\begin{df}\label{dfrelfr} The relative frequencies are defined by the number
of registration events in a given series of registrations:
$$
\omega[{\mathcal M},{\mathsf T}](X) = \frac{N[{\mathcal M},{\mathsf
T}](X)}{N[{\mathcal M},{\mathsf T}]({\mathbb R}^n)}\ .
$$ where $N[{\mathcal M},{\mathsf T}](X)$ is the number of registration events giving the registered value within $X$ while $N[{\mathcal M},{\mathsf T}]({\mathbb R}^n)$ is the number of all registration events.
\end{df} Observe that $N[{\mathcal M},{\mathsf T}]({\mathbb R}^n)$ need not be
the number of all individual particles in state ${\mathsf T}$ shot at
${\mathcal M}$.

Definition \ref{dfAregO} differs from the common version by a weaker
requirement on the states: the frequency agrees with the Born rule on ``some
states'' but not necessarily on ``all states''. Indeed, ``all states'' seems
to be the understanding by various books, such as \cite{ludwig1,peres} at
least implicitly, and \cite{BLM} quite explicitly (``probability
reproducibility condition'', p.\ 29). To explain Definition \ref{dfAregO}, let
us consider some examples.
\begin{enumerate}
\item The position $\vec{\mathsf x}$ of particle $S$ with Hilbert space
${\mathbf H}_s$ is registered by some detector with active volume $D$ (see,
e.g., \cite{leo}). If the detector gives a response (clicks) then we conclude
that a particle has been detected inside $D$. If the wave function of the
detected particle is $\psi(\vec{x})$ then the probability that the particle
will be found inside the detector is
$$
{\mathrm P}(D) = \int_{{\mathbb R}^3}d^3x \chi_D(\vec{x})|\psi(\vec{x})|^2\ ,
$$
where $\chi_D(\vec{x})$ is the characteristic function of $D$. Hence,
${\mathrm P}(D) = 0$ if $\text{supp}\,\psi(\vec{x}) \cap D = \emptyset$. The
integral on the right-hand side represents the trace (\ref{born2}) with
${\mathsf T} = |\psi\rangle\langle\psi|$ and $\Pi(X) = \xi_X(\vec{x})$.

A real meter ${\mathcal M}$ registering position is composed of several such
sub-detectors, ${\mathcal M}_1,\ldots,{\mathcal M}_n$ with disjoint active
volumes $D_1,\ldots,D_n$ and the frequency of finding the particle inside
$D_k$ then satisfies
$$
\omega[{\mathcal M}_k,\psi] \mapsto \int_{{\mathbb R}^3}d^3x
\chi_{D_k}(\vec{x})|\psi(\vec{x})|^2\ .
$$
Such detector gives some information only about the subset of ${\mathcal
B}({\mathbb R}^n)$ generated from $\{D_1,\ldots,D_n\}$ by set unions. The
meter does not register the whole spectrum but only the part
$\sigma'_{\vec{x}} \subset \sigma_{\vec{x}}$ defined by
$$
\sigma'_{\vec{x}} = \bigcup_{k=1}^n D_k\ .
$$
Hence, for all states $\psi$ such that
\begin{equation}\label{suppsi} \text{supp}\,\psi(\vec{x}) \subset
\sigma'_{\vec{x}}
\end{equation} the detector satisfies the Born rule for all $X \in {\mathcal
B}({\mathbb R}^3)$ because zero probability for $S$ being outside of
$\sigma'_{\vec{x}}$ results from both the Born rule and the registrations by
${\mathcal M}$. However, for the states that do not satisfy Eq.\
(\ref{suppsi}), the meter still gives zero probability for $S$ being outside
of $\sigma'_{\vec{x}}$ contradicting the Born rule.

The meter does not react at all to any wave function satisfying
$$
\text{supp}\,\psi(\vec{x}) \subset {\mathbb R}^3 \setminus \bigcup_{k=1}^n
D_k\ .
$$
Not to react to state $\psi$ means that
$$
\omega[{\mathcal M}_k,\psi] = 0
$$
for all $k$.
\item Next, consider an electron and its energy. A meter that can register
energy is, e.g., a proportional counter. Such a meter reacts to a free
particle if its energy is larger than some non-zero threshold. Moreover, if
the electron is bound, as in a hydrogen atom, then it is impossible to
register its energy directly, that is by a meter that reacts to the state of
electron as a subsystem of the atom. This is different from registrations of
the energy of the atom itself. All energy levels of its bounded states can be
registered indirectly (by scattering of photons, say), or directly (at least
in principle, see \cite{pauli}). On top of that, the spectrum of free electron
is continuous, and no meter can distinguish any two eigenvalues, if they are
too close to each other. Hence, it does not cover the whole spectrum but only
a finite number of energy intervals of a sufficient width.

To summarize, such a meter gives information only about some part of
${\mathcal B}({\mathbb R})$ and it will not react to a number of states. This
example shows that the problem need not be caused just by the geometric
arrangement of the experiment as in point 1.
\item The Stern-Gerlach apparatus as described in Section 1.3 can register the
spin observable only if the particle arriving at it can pass through the
opening between the magnets. Although it gives information about the whole of
${\mathcal B}({\mathbb R})$, it will not react to a large set of states. This
example shows that the problem can arise even for a meter that registers the
whole spectrum.
\end{enumerate} Let us compare this with the well-known cases (see, e.g.,
\cite{BLM}) of meters that do not register the whole spectrum. For instance, a
real meter can only discriminate between sufficiently different values of an
observable ${\mathsf O}$ with a continuous spectrum that is, it registers only
some coarse-grained version of the spectrum. Thus, one introduces a finite
partition the space ${\mathbb R}^n$,
$$
{\mathbb R}^n = \bigcup_{l=1}^n X_l\ ,
$$
and considers only projections ${\mathsf \Pi}(X_l)$, $l = 1,2,\ldots,n$, which
define a new observable with spectrum $\{1,2,\ldots,n\}$ (for details, see
\cite{BLM}, p.\ 35). Notice that the idea is to modify the observable so that
the correspondence between observable and meter via the Born rule is
improved. The meter then does react to all states of the system and satisfies
the Born rule corresponding to the corrected observable. Still, the method
does not work for the cases above, in which the Borel sets that are controlled
by the apparatus do not cover the spectrum. Such an apparatus does not react
to states corresponding to the part of the spectrum that is not covered, so
that the difficulty with the states also occurs.

The possibility that a meter may control only a (sometimes rather small)
proper subset of the whole Hilbert space, as the apparatuses of the above
examples do, does not seem to be even mentioned in the literature. This might
be due to the belief that, as in most cases of non-ideal real circumstances,
the shortcoming of real apparatuses is a natural way of practical things which
just must be taken into account if necessary in each particular instance and
that some real apparatuses might be arbitrarily close to the ideal or, at
least, that continuous improvement of techniques will make apparatuses ever
better. The main aim of the present chapter is to show that quantum mechanics
of indistinguishable particles sets an objective limit to this: an apparatus
that were ideal in this sense could not register its observable at all.

\section{Incompleteness of registration apparatuses} Let us now simplify
things by considering observables described by a single operator ($n =
1$). The foregoing section motivates the following definition.
\begin{df}\label{dfcomplapp} Let $S$ be a quantum system with Hilbert space
${\mathbf H}$ and let observable ${\mathsf O}$ be a s.a.\ operator on
${\mathbf H}$ with spectrum $\sigma$ and spectral measure ${\mathsf \Pi}(X)$,
$X \in {\mathcal B}({\mathbb R})$. Let meter ${\mathcal M}$ register
observable ${\mathsf O}$. We say that ${\mathcal M}$ is {\em complete} if Eq.\
(\ref{born2}) holds true for all ${\mathsf T} \in {\mathbf T}({\mathbf H})$
and $X \in {\mathcal B}({\mathbb R})$.

If there is any state ${\mathsf T}$ for which the frequency $\omega[{\mathcal
M},{\mathsf T}]({\mathbb R})$ of registering any value by ${\mathcal M}$ is
zero, meter ${\mathcal M}$ is called {\em incomplete}. Let the subset of
states for which this is the case be denoted by ${\mathbf T}({\mathbf
H})_{{\mathcal M}0}$ and the subset of states for which Eq.\ (\ref{born2})
holds by ${\mathbf T}({\mathbf H})_{\mathcal M}$. The convex set ${\mathbf
T}({\mathbf H})_{\mathcal M}$ is called the {\em domain} of ${\mathcal M}$.
\end{df} Thus, the three examples in the foregoing section describe incomplete
apparatuses. We are now going to prove that a complete meter cannot work.

Let registrations by meter ${\mathcal M}$ be performed on a system $S$ with
Hilbert space ${\mathbf H}$. Suppose that meter ${\mathcal M}$ registers
observable ${\mathsf O}$ with spectral measure ${\mathsf \Pi}(X)$, $X \in
{\mathcal B}({\mathbb R})$, and is complete. Then, because of the
normalisation condition (\ref{normaliz}), we must have $tr({\mathsf T}
{\mathsf \Pi}({\mathbb R})) = 1$ for {\em any state} ${\mathsf T}$. This means
that {\em any} registration on ${\mathsf T}$ by ${\mathcal M}$ must give {\em
some} result.

Then, according to the theory of indistinguishable systems, ${\mathcal M}$
must also register some values on any state ${\mathsf T}'$ of any system $S'$
of the same type as $S$. Clearly, this is a difficulty: the measurement of
observable ${\mathsf O}$ of $S$ by ${\mathcal M}$ is disturbed by the
existence of a system of the same type as $S$ anywhere else in the world, even
if it is localised arbitrarily far away from $S$ because it cannot be
distinguished from $S$ by ${\mathcal M}$. According to Trial Hypotheses
\ref{shobject} and \ref{rhmacro}, for most microsystems $S$, the world
contains a huge number of systems of the same type so that a horrible noise
must disturb any registration by a complete meter.

To show the problem in more detail, let us consider two distant laboratories,
A and B. Let ${\mathsf O}$ be a non-degenerate discrete observable of $S$ with
eigenstates $|k\rangle$ and eigenvalues $o_k$. Let state $|k\rangle$ be
prepared in A and $|l\rangle$ in B so that $k \neq l$ and let ${\mathsf O}$ be
registered in laboratory A by complete meter ${\mathcal M}$. Using Fock space
formalism (see e.g.\ \cite{peres}, Section 5-6), we have
\begin{equation}\label{FockA} {\mathsf O} = \sum_n o_n{\mathsf
a}^\dagger_n{\mathsf a}_n\ ,
\end{equation} where ${\mathsf a}_k$ is an annihilation operator of state
$|k\rangle$. Such an observable perfectly expresses the fact that the meter
cannot distinguish particles of the same type. The state prepared by the two
laboratories is
\begin{equation}\label{Fockstate} {\mathsf a}^\dagger_k{\mathsf
a}^\dagger_l|0\rangle\ .
\end{equation} For the average $\langle {\mathsf O} \rangle$ of (\ref{FockA})
in state (\ref{Fockstate}), the standard theory of measurement gives
$$
\langle {\mathsf O} \rangle = \langle 0|{\mathsf a}_l{\mathsf a}_k\left(\sum_n
a_n{\mathsf a}^\dagger_n{\mathsf a}_n\right){\mathsf a}^\dagger_k{\mathsf
a}^\dagger_l|0\rangle\ .
$$
Using the relation
$$
{\mathsf a}_r{\mathsf a}^\dagger_s = \eta {\mathsf a}^\dagger_s{\mathsf a}_r +
\delta_{rs}\ ,
$$
where $\eta = 1$ for bosons and $\eta = -1$ for fermions, we can bring all
annihilation operators to the right and all creation ones to the left
obtaining
$$
\langle {\mathsf O} \rangle = a_k + a_l\ .
$$
The result is independent of the distance between the laboratories. Thus, the
measurement in A by any complete meter depends on what is done in B.

Let us next suppose that ${\mathcal M}$ is incomplete in such a way that the
state of any system of the same type as $S$ that may occur in the environment
of $S$ lies in ${\mathbf T}({\mathbf H})_{{\mathcal M}0}$. Apparently, such an
assumption can be checked experimentally by looking at the level of noise of
the meter. Then, if we prepare a copy of system $S$ in a state that lies
within ${\mathbf T}({\mathbf H})_{\mathcal M}$ the registration of $S$ by
${\mathcal M}$ cannot be disturbed by the systems in the environment. In fact,
this must be the way of how all quantum measurement are carried out. We can
say that objective properties of our environment require certain kind of
incompleteness of registration apparatus ${\mathcal M}$ in order that
${\mathcal M}$ can work in this environment.

Accordingly, the course of any successful measurement must be as
follows. First, a registration apparatus ${\mathcal M}$ for a system $S$ with
Hilbert space ${\mathbf H}$ is constructed and checked. In particular, the
level of its noise must be sufficiently low. From the construction of the
meter, we can infer some set ${\mathbf T}_{\mathcal M}$ of states on which the
meter is able make registrations. ${\mathbf T}_{\mathcal M}$ might be smaller
than the whole domain,
$$
{\mathbf T}_{\mathcal M} \subset {\mathbf T}({\mathbf H})_{\mathcal M}
$$
(the domain is often difficult to specify). Second, systems $S$ is prepared in
one of such states. The registration by ${\mathcal M}$ will then not be
disturbed and the probabilities of the results can be calculated by formula
(\ref{born2}).

The states of ${\mathbf T}_{\mathcal M}$ must therefore be in some sense
sufficiently different from the states of all systems of the same type as $S$
that occur in the environment of $S$. Let us try to express this idea
mathematically. This can be done in the simplest way, if we choose a
particular representation so that the wave function of an extremal state
$|\psi\rangle$ will be $\psi(\lambda)$ and the kernel of a state of arbitrary
external object will be
$T(\lambda^{(1)},\ldots,\lambda^{(N)};\lambda^{(1)\prime},\ldots,\lambda^{(N)\prime})$.

Then we propose the following definition:
\begin{df}\label{dfss} Let system $S$ with Hilbert space ${\mathbf H}$ be
prepared in state $|\psi\rangle \langle \psi| \in {\mathbf T}({\mathbf
H})$. Let the environment consists of macroscopic objects with quantum models
and well-defined quantum states. Let $O$ be the quantum model associated with
such an object, $O_S$ the subsystem of $O$ containing all subsystems of $O$
that are indistinguishable from $S$ and let ${\mathsf T}$ be the state of
$O_S$. If
\begin{equation}\label{dfsseq} \int d\lambda^{(1)\prime}
T(\lambda^{(1)},\ldots,\lambda^{(N)};
\lambda^{(1)\prime},\ldots,\lambda^{(N)\prime})\psi(\lambda^{(1)\prime}) = 0
\end{equation} holds for any object in the environment, then $|\psi\rangle
\langle \psi|$ is said to have {\em separation status}.
\end{df} The definition can easily be extended to states of $S$ that are not
extremal. Some motivation of the definition is as follows. Suppose that there
is a system in the environment in a state $\phi$ and that $\langle \phi| \psi
\rangle \neq 0$. Then,
$$
\phi = c_1\psi + c_2 \psi^\bot\ ,
$$
with non-zero $c_1$ and $\langle \psi^\bot| \psi \rangle = 0$, and the meter
would react to the $\psi$-part of $\phi$.

The above ideas also have some relevance to the meaning of preparation
processes in quantum measurements. Stating such meaning extends the Minimum
Interpretation, for which preparations and registrations are primitive notions
(see, e.g., \cite{peres}, p.\ 12).
\begin{sh}\label{shprep} Any preparation of a single microsystem must yield a
state having a separation status.
\end{sh}

A separation status of a microsystem is a property that is uniquely determined
by a preparation. Hence, it belongs to objective properties of quantum systems
(Section 1.2). But it is a property that is a necessary condition for any
other objective property because each preparation must create a separation
status. Moreover, only a separation status makes a quantum system
distinguishable from each other system in the environment except for those
that are used in the experiment being performed. Thus, a quantum object can
come into being, namely in a preparation process, and can expire, viz. if it
loses its separation status.

In this connection, the question of a separation status of a macroscopic body
can arise. It seems that any macroscopic body in our environment is
distinguishable, as a quantum system, from any other quantum system just
because of its composition. It is very implausible that there can be,
somewhere, a macroscopic system that is identical, in the quantum mechanical
sense, to a given macroscopic body.

We can understand the role of incompleteness of meters better if we compare
quantum apparatuses with classical ones. To this aim, we construct a simple
model of an eye. Indeed, an eye is a classical registration device, either by
itself or as a final part of other classical apparatuses.

Our model contains of an optically sensitive surface (retina) that can
register visible light (i.e., with a wave-length between 0.4 and 0.75
$\mu$m). It can distinguish between some small intervals of the visible wave
lengths and between small spots where the retina is hit by photons.

The retina covers one side of a chamber that has walls keeping light away
except for a small opening at the side opposite to the retina wall\footnote{An
eye with a small opening instead of a lens occurs in some animals such as
nautilus.}. The radius of the circular opening, even if very small, is much
larger than the wave length of visible light so that this light waves suffer
only a negligible bending as they pass the opening. Hence, the assumed
insensitivity of the retina to smaller wave lengths is an incompleteness that
helps to make the picture sharp.

Another aspect of incompleteness is that only the light that can pass the
opening will be registered. Again, this is important: if the retina were
exposed to all light that can reach it from the neighbourhood, only a smeared,
more or less homogeneous signal would result. Because of the restrictions, a
well-structured colour picture of the world in front of the eye will appear on
the retina.

Clearly classical apparatuses must also be incomplete in order to yield a
non-trivial information. Of course, they give information about value
distributions obtained simultaneously for a great number of particles rather
than an information on a single particle.

\section{Observables} There are two ways of how one could react to the
necessary incompleteness of registration apparatuses. First, one can try to
modify the definition of the observable that is registered by such a meter so
that the results of the registrations and the probabilities calculated from
the Born rule coincide, similarly as it has been done above for coarse-grained
version of the spectrum. Second, one can leave the observables as they are and
accept the fact that each meter can register its observable only partially as
in Definition \ref{dfcomplapp}. In our previous work
(\cite{hajicekC1,hajicek2,survey}), we have tried the first way. It turned
out, however (see \cite{incomplete}), that the modification that was necessary
for an observable to describe how a real meter worked was messy. Not only the
notion of observable became rather complicated but also just some idealized
kinds of meters could be captured in this way.

The mentioned idealized kind of incomplete meter can be described as
follows. Such a meter ${\mathcal M}$ determines a closed linear subspace
${\mathbf H}_{\text{ss}}$ of ${\mathbf H}$ so that, instead of Eq.\
(\ref{born2}), we have
$$
\omega[{\mathcal M},{\mathsf T}](X) \mapsto tr\Biggl(\Bigl({\mathsf
\Pi}_{\text{ss}}{\mathsf T}{\mathsf \Pi}_{\text{ss}}\Bigr){\mathsf
\Pi}(X)\Biggr)
$$
for all ${\mathsf T} \in {\mathbf T}({\mathbf H})$ and $X \in {\mathcal
B}({\mathbb R})$, where ${\mathsf \Pi}_{\text{ss}}$ is the orthogonal
projection onto ${\mathbf H}_{\text{ss}}$. Then,
\begin{equation}\label{specdomain} {\mathbf T}({\mathbf H})_{\mathcal M} =
{\mathbf T}({\mathbf H}_{\text{ss}})
\end{equation} and
$$
{\mathbf T}({\mathbf H})_{{\mathcal M}0} = {\mathbf T}([{\mathsf 1} - {\mathsf
\Pi}_{\text{ss}}]{\mathbf H})
$$
because any element ${\mathsf T}$ of ${\mathbf T}({\mathbf H}_{\text{ss}})$
satisfies
\begin{equation}\label{THA} {\mathsf T} = {\mathsf \Pi}_{\text{ss}}{\mathsf
T}{\mathsf \Pi}_{\text{ss}}\ .
\end{equation}

The construction of the corresponding modified observable can go as
follows. First, we have
$$
tr\Biggl(\Bigl({\mathsf \Pi}_{\text{ss}}{\mathsf T}{\mathsf
\Pi}_{\text{ss}}\Bigr){\mathsf \Pi}(X)\Biggr) = tr\Biggl({\mathsf
T}\Bigl({\mathsf \Pi}_{\text{ss}}{\mathsf \Pi}(X){\mathsf
\Pi}_{\text{ss}}\Bigr)\Biggr)\ .
$$
Next, consider operator ${\mathsf \Pi}_{\text{ss}}{\mathsf \Pi}(X){\mathsf
\Pi}_{\text{ss}}$. It is bounded by ${\mathsf 1}$ and self adjoint because
${\mathsf \Pi}_{\text{ss}}$ and ${\mathsf \Pi}(X)$ are. It is obviously
positive. Thus, it is an {\em effect} (a positive operator bounded by 1, see
\cite{davies}). A collection of effects ${\mathsf E}(X)$, $X \in {\mathcal
B}({\mathbb R})$, with certain properties (including the normalisation
condition ${\mathsf E}({\mathbb R}) = {\mathsf 1}$) is called a {\em
positive-operator valued measure} (POVM) (see \cite{davies} or \cite{BLM}) and
generalizes the notion of spectral measure. The collection of effects
${\mathsf \Pi}_{\text{ss}}{\mathsf \Pi}(X){\mathsf \Pi}_{\text{ss}}$ for all
$X \in {\mathcal B}({\mathbb R})$ is not a POVM, however, because we have,
instead of the above normalisation condition,
$$
{\mathsf \Pi}_{\text{ss}}{\mathsf \Pi}({\mathbb R}){\mathsf \Pi}_{\text{ss}} =
{\mathsf \Pi}_{\text{ss}}
$$
Such a quantity could be called ``truncated POVM''. Thus, the notion of
observable had to be changed from a self-adjoint operator to a truncated POVM.

However, the above model of incomplete meter is too simple. For instance, some
of the examples listed in Section 4.2 cannot be described by it. Indeed,
consider the Stern-Gerlach meter that is arranged in such a way that it can
react to particles moving within a thin tube around the third axis of
coordinates $x_1,x_2,x_3$. The particle that can be registered must thus
arrive at the magnets only within some small subset of the (1,2)-plane, the
third component of its momentum must satisfy
$$
p_3 \in (a_3,b_3)\ ,
$$
which must be large enough to lie above the detector threshold, and
$$
p_1 \in (-c_1,c_1)\ ,\quad p_2 \in (-c_2,c_2)\ ,
$$
where $c_k < \epsilon$ for $k = 1,2$ and for sufficiently small
$\epsilon$. However, these conditions can be satisfied, by any wave packets,
only approximately. Then, the Born rule will also be satisfied only
approximately. Now, a linear superposition of such packets need not be again
such a packet. The above conditions mean that the wave function (in $Q$- or
$P$-representation) of the registered particle must satisfy inequalities of
the form
$$
|\psi(\lambda)|^2 < \epsilon'
$$
for some fixed values of $\lambda$ determined by the arrangement, where
$\lambda$ stands either for $\vec{x}$ or for $\vec{p}$, and $\epsilon'$ is a
small positive number. Suppose that another wave function, $\phi$, also
satisfies the condition. Then it only follows, for all $c$ and $c'$ satisfying
$|c|^2 + |c'|^2 = 1$, that
$$
|c \psi(\lambda) + c' \phi(\lambda)|^2 < 2\epsilon'\ .
$$
Hence, the packets need not form a closed linear subspace of ${\mathbf H}$.

These problems do not afflict the second way because the approach using
incomplete meters works even if the domain of a meter does not satisfy Eq.\
(\ref{specdomain}). In fact, the knowledge of the whole domain ${\mathbf
T}({\mathbf H})_{\mathcal M}$ of a meter is not necessary for the construction
of a model of a registration by ${\mathcal M}$ because it is sufficient to
know only those elements of ${\mathbf T}({\mathbf H})_{\mathcal M}$ that are
prepared for the experiment.

These are the reasons why we adopt the second way in the present review. Then,
the Minimum Interpretation of observables, as represented e.g.\ by Assumption
\ref{propold12}, has the following refinement.
\begin{sh}\label{shobsmeter} For any observable ${\mathsf O}$ of a system
${\mathcal S}$, there is a meter that can register ${\mathsf O}$. There is no
meter that can register ${\mathsf O}$ on any state.
\end{sh} There might be cases in which there is a meter ${\mathcal M}_{\mathsf
T}$ registering ${\mathsf O}$ on ${\mathsf T}$ for every given state ${\mathsf
T}$. Still, as each of these meters must be incomplete, a non-trivial set of
meters would then be needed for a complete registration of ${\mathsf O}$.

Everything that has been said in this and the foregoing sections can easily be
extended if the notion of observable is generalized from a self-adjoint
operator to a POVM as in \cite{ludwig1,davies,BLM}. Indeed, POVM's can be
considered as a generalization of observables: an observable is uniquely
determined by its spectral decomposition and a spectral decomposition is a
special case of POVM. Thus, some physicists started to define the notion of
observable as POVM. However, there are important differences between the
applications of the two notions: the observables are more handy for purposes
that cannot be helped by POVM's (for instance, basing quantum mechanics on
normed algebras of bounded observables---$C^*$-algebras \cite{haag}), or the
relation of observables to the spacetime transformations) while POVM's are
more closely related to particular registration methods. In this book, we keep
the old notion of observables (as self-adjoint operators) and use POVM's where
it is advantageous in connection with registrations.

\section{Tensor-product method} We have seen in the foregoing sections that
the disturbance of measurement by environmental particles can be avoided, if
the meter is incomplete and the measured system is prepared in a state with a
separation status.

The present section is going to study the resulting mathematics of quantum
measurement theory in more detail. In particular, we shall consider two {\em
ways of description} of composite-system states. The first way works with the
tensor product of the environmental and the registered system states and the
second one with the $\tau$-symmerized state of the whole composite system as
required by rules of the theory of indistinguishable systems. On the one hand,
the second way of description is in any case the correct one so that we have
to show that the two descriptions lead to the same measurable results. On the
other hand, the first way is the only practically feasible one because it does
not require the knowledge of the environment state.

To develop the two descriptions, let us consider system $S$ and its
environment $E$ with the system $E_S$ of all its subsystems that are
indistinguishable from $S$. Let ${\mathbf H}$ be the Hilbert space of $S$ and
let $E_S$ consist of $N$ subsystems so that the Hilbert space of $E_S$ is
${\mathbf H}^N_\tau$. For the sake of simplicity, we assume that the states of
$S$ and $E_S$ are extremal. The proof for general states is analogous. Let
$\psi(\lambda)$ be the wave function of $S$. The wave function of $E_S$ has
the form
$$
\Psi(\lambda^{(1)},\ldots,\lambda^{(N)}) \in {\mathbf H}^N_\tau\ .
$$
Then the wave functions of the two descriptions are
\begin{equation}\label{firstway} \Psi(\lambda^{(1)}, \ldots, \lambda^{(N)})
\psi(\lambda^{(N+1)})
\end{equation} and
\begin{equation}\label{secndway} N_{\text{exch}}{\mathsf
\Pi}^{N+1}_\tau\Bigl(\Psi(\lambda^{(1)}, \ldots, \lambda^{(N)})
\psi(\lambda^{(N+1)})\Bigr)\ ,
\end{equation} where ${\mathsf \Pi}^{N+1}_\tau : {\mathbf H}^N_\tau \otimes
{\mathbf H}_\tau \mapsto {\mathbf H}_\tau^{N+1}$ is an $\tau$-symmetrisation
defined in Section 4.1. Orthogonal projections do not preserve
normalization. Hence, the projection must be followed by a normalization
factor, which we will denote by $N_{\text{exch}}$ standing before the
projection symbol. Of course, $N_{\text{exch}}$ depends on the projection and
the wave function being projected, but we just write $N_{\text{exch}}$ instead
of $N({\mathsf \Pi}^N_\tau, \Psi\psi)$ to keep equations short.

As $\Psi$ is already $\tau$-symmetric and normalised, the expression
(\ref{secndway}) can be rewritten as follows:
\begin{multline}\label{secway} N_{\text{exch}}{\mathsf
\Pi}^{N+1}_\tau\Bigl(\Psi(\lambda^{(1)}, \ldots, \lambda^{(N)})
\psi(\lambda^{(N+1)})\Bigr) \\ = N'\sum_{K=1}^{N+1}(\tau)^{N+1-K}
\Psi(\lambda^{(K+1)},\ldots,\lambda^{(N+1)},\lambda^{(1)},\ldots,\lambda^{(K-1)})\psi(\lambda^{(K)})\
,
\end{multline} where $N' \neq N_{\text{exch}}$ is a suitable normalisation
factor. This relation will simplify some subsequent calculations.

Eq.\ (\ref{secndway}) shows that we can recover the second description from
the first one, but if the two descriptions are to be equivalent, the first
description must be derivable from the second one, too. For this aim, the
separation status is necessary. Let state $\psi(\lambda)$ be prepared with
separation status and let ${\mathsf \Pi}_\psi = |\psi\rangle \langle
\psi|$. Eq.\ (\ref{dfsseq}) then implies
$$
N' = \frac{1}{\sqrt{N+1}}\ .
$$

Now, we make use the of fact that the operators on ${\mathbf H}$ can act on
different wave functions (elements of ${\mathbf H}$) in a product and that
this action can be specified by the argument of the function. For example, if
we have product $\phi(\lambda^{(1)})\phi'(\lambda^{(2)})$ and operator
${\mathsf O} : {\mathbf H} \mapsto {\mathbf H}$, operator ${\mathsf O}^{(1)} :
{\mathbf H} \otimes {\mathbf H} \mapsto {\mathbf H} \otimes {\mathbf H}$ is
defined by
$$
{\mathsf O}^{(1)}\Bigr[\phi(\lambda^{(1)})\phi'(\lambda^{(2)})\Bigl] =
({\mathsf O}\phi)(\lambda^{(1)})\phi'(\lambda^{(2)})
$$
while ${\mathsf O}^{(2)}$ by
$$
{\mathsf O}^{(2)}\Bigr[\phi(\lambda^{(1)})\phi'(\lambda^{(2)})\Bigl] =
\phi(\lambda^{(1)})({\mathsf O}\phi')(\lambda^{(2)})\ .
$$
From the definition of separation status we then obtain that
$$
{\mathsf \Pi}^{(k)}_\psi\Psi((\lambda^{(1)}),\ldots,(\lambda^{(N)})) = 0
$$
for any $k = 1,\ldots,N$.

With this notation, we can achieve our aim: Eq.\ (\ref{secway}) implies that
\begin{multline}\label{piA} {\mathsf
\Pi}^{(N+1)}_\psi\Biggl(N_{\text{exch}}{\mathsf
\Pi}^{N+1}_\tau\Bigl(\Psi(\lambda^{(1)}, \ldots, \lambda^{(N)})
\psi(\lambda^{(N+1)})\Bigr)\Biggr) \\ = \nu_\psi \Psi(\lambda^{(1)}, \ldots,
\lambda^{(N)}) \psi(\lambda^{(N+1)})\ ,
\end{multline} where $\nu_\psi$ is a suitable normalisation factor. Observe
that this operation is naturally described by tensor products of
$\tau$-symmetriz\-ed wave functions rather than by the Fock-space
formalism. The exchange symmetry is not violated because we can use ${\mathsf
\Pi}^{(K)}_\psi$ for any fixed $K = 1,\dots, N$ instead of ${\mathsf
\Pi}^{(N+1)}_\psi$ and the result will again be the above wave function with
rearranged arguments.

The next point is to give an account of registration by an incomplete
meter. We construct two observables that represent the meter, each for one of
the two description ways, and show that the two ways lead to the same
results. We work with a simple model to show the essential points; the general
situation can be dealt with in an analogous way.

Let meter ${\mathcal M}$ register observable ${\mathsf O} : {\mathbf H}
\mapsto {\mathbf H}$ that is additive, discrete and non-degenerate. Let its
eigenvalues be $o_k$ and eigenfunctions be $\psi_k(\lambda)$, $k \in {\mathbb
N}$. Let ${\mathcal M}$ be incomplete in the way that it reacts only to
$\psi_k$ if $k = 1,\ldots,K$ for some $K \in {\mathbb N}$. Hence, the subspace
${\mathbf H}_{\text{ss}}$ is spanned by vectors $\psi_k(\lambda)$, $k =
1,\ldots,K$, and the projection onto it is
$$
{\mathsf \Pi}_{\text{ss}} = \sum_{k=1}^K{\mathsf \Pi}_k\ ,
$$
where
$$
{\mathsf \Pi}_k = |\psi_k\rangle \langle \psi_k|\ .
$$
The action of the meter can now be described as follows. Let us prepare state
$\psi$ with a separation status. Hence, $\psi \in {\mathbf H}_{\text{ss}}$ and
its decomposition into the eigenstates of ${\mathsf O}$ is
$$
\psi = \sum_{k=1}^K c_k\psi_k
$$
with $\sum_{k=1}^K |c_k|^2 = 1$. Then the probability ${\mathrm P}_k$ of
registering $o_k$ on $\psi$ is
$$
{\mathrm P}_k = \langle \psi|{\mathsf \Pi}_k |\psi\rangle|^2\ ,
$$
or
$$
{\mathrm P}_k = |c_k|^2
$$
for $k \leq K$ and ${\mathrm P}_k = 0$ for $k > K$.

Let us start with the first way, Eq.\ (\ref{firstway}). We define the
corresponding observable by restricting the action of ${\mathsf O}$ or
${\mathsf \Pi}_k$ to the second factor:
\begin{multline}\label{obsfirstway} ({\mathsf 1} \otimes {\mathsf
\Pi}_k)\Bigl[\Psi(\lambda^{(1)}, \ldots, \lambda^{(N)}) \otimes
\psi(\lambda^{(N+1)})\Bigr] \equiv {\mathsf
\Pi}_k^{(N+1)}\Bigl[\Psi(\lambda^{(1)}, \ldots, \lambda^{(N)})
\psi(\lambda^{(N+1)})\Bigr] \\ = c_k\Psi(\lambda^{(1)}, \ldots, \lambda^{(N)})
\psi_k(\lambda^{(N+1)})
\end{multline} for $k \leq K$. Eq.\ (\ref{obsfirstway}) specifies the Born
rule of the observable. Now, coming to the second way of description, Eq.\
(\ref{secndway}), we use the fact that the observable is additive. For
example, it acts on product $\phi(\lambda^{(1)})\phi'(\lambda^{(2)})$ as
follows
$$
({\mathsf O}^{(1)} + {\mathsf
O}^{(2)})\Bigl(\phi(\lambda^{(1)})\phi'(\lambda^{(2)})\Bigr)\ .
$$
Then, to define the observable registered by ${\mathcal M}$, we need the
action of its projection ${\mathsf \Pi}'_k$ for eigenvalue $o_k$, $k \in
{\mathbb N}$. Let us try:
\begin{equation}\label{piprimek} {\mathsf \Pi}'_k = \sum_{l=1}^{N+1}({\mathsf
\Pi}_k{\mathsf \Pi}_{\text{ss}})^{(l)}\ .
\end{equation} Observe that operators ${\mathsf \Pi}_k$ and ${\mathsf
\Pi}_{\text{ss}}$ commute. Then, using Eq.\ (\ref{secway}), we obtain
\begin{multline*} \sum_{l=1}^{N+1}({\mathsf \Pi}_k{\mathsf
\Pi}_{\text{ss}})^{(l)}\Bigl[N_{\text{exch}}{\mathsf
\Pi}^{N+1}_\tau\Bigl(\Psi(\lambda^{(1)}, \ldots, \lambda^{(N)})
\psi(\lambda^{(N+1)})\Bigr)\Bigr]= \sum_{l=1}^{N+1}({\mathsf \Pi}_k{\mathsf
\Pi}_{\text{ss}})^{(l)} \\
\left[\frac{1}{\sqrt{N+1}}\sum_{K=1}^{N+1}(\tau)^{N+1-K}
\Psi(\lambda^{(K+1)},\ldots,\lambda^{(N+1)},\lambda^{(1)},\ldots,\lambda^{(K-1)})\psi(\lambda^{(K)})\right]
\\ = \frac{1}{\sqrt{N+1}}\sum_{K=1}^{N+1}(\tau)^{N+1-K}
\Psi(\lambda^{(K+1)},\ldots,\lambda^{(N+1)},\lambda^{(1)},\ldots,\lambda^{(K-1)})c_k\psi_k(\lambda^{(K)})
\\ = c_kN_{\text{exch}}{\mathsf \Pi}^{N+1}_\tau\Bigl(\Psi(\lambda^{(1)},
\ldots, \lambda^{(N)}) \psi_k(\lambda^{(N+1)})\Bigr)
\end{multline*} because operator $({\mathsf \Pi}_k{\mathsf
\Pi}_{\text{ss}})^{(l)}$ annihilates the state to the right if the argument
$\lambda^{(l)}$ is in function $\Psi$ and gives $c_k\psi_k(\lambda^{(l)})$ if
the argument is in $\psi$. Thus, the Born rules for the observables of the two
ways of description coincide.

In general, operator (\ref{piprimek}) is not a projection because the product
$$
({\mathsf \Pi}_k{\mathsf \Pi}_{\text{ss}})^{(m)}({\mathsf \Pi}_k{\mathsf
\Pi}_{\text{ss}})^{(n)}
$$
does not in general vanish for $m \neq n$ and then $({\mathsf \Pi}'_k)^2 \neq
{\mathsf \Pi}'_k$. However, on the subspace of ${\mathsf
\Pi}^{N+1}_\tau({\mathbf H}^N_\tau \otimes {\mathbf H}_{ss})$ with which we
are working, the product is non-zero only if $m = n$, so that it is a
projection under these conditions.

The last question is whether the dynamical evolutions for the two ways of
description are compatible. First, we define the corresponding
Hamiltonians. Let ${\mathsf H} : {\mathbf H}^N_\tau \otimes {\mathbf H}
\mapsto {\mathbf H}^N_\tau \otimes {\mathbf H}$ be a Hamiltonian for the first
way of description and let us assume that
$$
{\mathsf H}{\mathsf \Pi}^{N+1}_\tau = {\mathsf \Pi}^{N+1}_\tau{\mathsf H}\ .
$$
Such a Hamiltonian leaves the subspace ${\mathsf \Pi}^{N+1}_\tau ({\mathbf
H}^N_\tau \otimes {\mathbf H})$ invariant and can also be viewed as a
Hamiltonian for the second way of description. Then, the two
Schr\"{o}dinger equations that we are going to compare are:
\begin{equation}\label{schrodfirstway} {\mathbf H}[\Psi(\lambda^{(1)}, \ldots,
\lambda^{(N)}) \psi(\lambda^{(N+1)})] = i\hbar\frac{\partial}{\partial
t}[\Psi(\lambda^{(1)}, \ldots, \lambda^{(N)}) \psi(\lambda^{(N+1)}]
\end{equation} for the first way of description and
\begin{equation}\label{schrodsecndway} {\mathbf H}{\mathsf
\Pi}^{N+1}_\tau[\Psi(\lambda^{(1)}, \ldots, \lambda^{(N)})
\psi(\lambda^{(N+1)}] = i\hbar\frac{\partial}{\partial t}{\mathsf
\Pi}^{N+1}_\tau[\Psi(\lambda^{(1)}, \ldots, \lambda^{(N)})
\psi(\lambda^{(N+1)}]
\end{equation} for the second way (the normalization factors cancel).

The compatibility can only be proved if the evolution preserves the separation
status. Mathematically, this means that the Hamiltonian must commute with the
projections defining the status:
\begin{equation}\label{hamss} {\mathsf H}{\mathsf \Pi}^{(k)}_{\text{ss}} =
{\mathsf \Pi}^{(k)}_{\text{ss}}{\mathsf H}
\end{equation} for all $k = 1,\ldots,N+1$. Then, the projections are conserved
and their eigenspaces are stationary. In the case under study, this implies
that the time derivative commutes with the projections, too:
\begin{equation}\label{timess} \frac{\partial}{\partial t}{\mathsf
\Pi}^{(k)}_{\text{ss}} = {\mathsf
\Pi}^{(k)}_{\text{ss}}\frac{\partial}{\partial t}
\end{equation} for all $k = 1,\ldots,N+1$.

Now, the proof of the compatibility is very simple: applying projection
${\mathsf \Pi}^{(N+1)}_{\text{ss}}$ to both sides of Eq.\
(\ref{schrodsecndway}) and using Eqs.\ (\ref{piA}) (where ${\mathsf
\Pi}^{(N+1)}_\psi$ can be replaced by ${\mathsf \Pi}^{(N+1)}_{\text{ss}}$),
(\ref{hamss}) and (\ref{timess}), we obtain Eq.\ (\ref{schrodfirstway}).

For processes, in which e.g.\ a measured system loses its separation status,
the two evolutions are not compatible and the second way equation must be
used. Such processes occur during registration. Many examples of such
registrations have been given in \cite{hajicek4,hajicek5}, but the separation
was defined in a different way. We shall adapt the examples to the present
definition of separation status in Chapter 5.

\chapter{Taking detectors seriously} \setcounter{equation}{0}
\setcounter{thm}{0} \setcounter{assump}{0} \setcounter{sh}{0}
\setcounter{df}{0} This chapter has two aims. First, it reformulates the
standard quantum theory of measurement so that it becomes compatible with the
ideas of Chapters 3 and 4. Second, it states three trial hypotheses about the
course of state reduction that make the theory more specific and definite than
the standard one is. In particular, it postulates objective conditions under
which the state reduction occurs. The chapter gives an account of the theory
that has gradually evolved from some ideas of references
\cite{hajicek2,hajicekC1,hajicek4} and were described in \cite{reduction2}.

The Trial Hypotheses of Section 1.2 associate quantum states with individual
quantum systems. As it is well known, this idea has an undesired lateral
effect: the {application of Schr\"{o}dinger equation to the process of
measurement leads to results that contradict some evidence, see
\cite{ballent}, Section 9.2 and \cite{peres}, p.\ 374. This motivated the
introduction of a new phenomenon, the so-called {\em state reduction}: a
non-linear correction to Schr\"{o}dinger equation.

Let us first briefly recapitulate some current ideas concerning the state
reduction.

\section{Standard theory of measurement} Here, we give a short review of the
theory of measurement as it is employed in the analysis of many measurements
today and as it is described in, e.g., \cite{WM,bragin,svensson}---this is
what we call ``standard theory of measurement''.

The standard theory considers two systems: {\em object
system}\footnote{``object system'' is a notion of the standard theory of
measurement which has nothing to do with our notions of ``object'' and
``system''} $S$ with Hilbert space ${\mathbf H}$ on which the measurement is
done, and {\em meter} or {\em apparatus} $M_q$ that performs the
registration. Quantum states refer to individual quantum systems. A
measurement process is split into three steps.
\begin{enumerate}
\item Initially, quantum system $S$ is prepared in state ${\mathsf
T}^S_{\text{i}}$ and the meter $M_q$ is prepared in state ${\mathsf
T}^M_{\text{i}}$. The preparations of the object system and the meter are
assumed to be independent from each other so that the composite $S + M_q$ is
then in state ${\mathsf T}^S_{\text{i}} \otimes {\mathsf
T}^M_{\text{i}}$. (The standard theory ignores the possibility that $M_q$
contains subsystems of the same type as $S$.)
\item A quantum interaction between $S$ and $M_q$ suitably entangles
them. Such an interaction can be theoretically represented by unitary map
$\bar{\mathsf U}$, called {\em measurement coupling}, but more general forms
of evolution are possible (see, e.g., \cite{BLM}). There is then
mathematically well-defined evolution of system $S +M_q$ during a finite time
interval. The end result of the evolution is
$$
\bar{\mathsf U} ({\mathsf T}^S_{\text{i}} \otimes {\mathsf T}^M_{\text{i}})
\bar{\mathsf U}^\dagger\ .
$$
\item Finally, {\em reading} the meter gives some value $r \in {\mathbb R}$ of
the measured quantity. Observational facts motivate one of the most important
assumptions of the standard theory: Each individual reading gives a definite
value. This has been called {\em objectification requirement} \cite{BLM}. The
most important assumption of the standard theory is that, after the reading of
value $r$ in an individual measurement, the object system $S$ is in a
well-defined state,
$$
{\mathsf T}^S_{{\text{out}},r}\ ,
$$
called {\em conditional} or {\em selective} state.  In quantum mechanics, the
fact that an arbitrary large number of identical copies of $S$ is available
enables to carry out a whole ensemble of equivalent individual
measurements. If the equivalent measurements are repeated many times
independently from each other, each reading $r \in {\mathbb R}$ occurs with a
definite frequency, ${\mathrm P}_r$. (A more general assumption can be
adopted: the final state is conditional on reading the registered value within
Borel set $X$, etc., see \cite{BLM}.)
\end{enumerate}

A special case of a conditional state is given by Dirac's postulate:
\begin{quote} A measurement always causes a system to jump in an eigenstate of
the observed quantity.
\end{quote} Such a measurement is called {\em projective} and it is the
particular case when ${\mathsf T}^S_{{\text{out}},r} = |r\rangle\langle r|$
where $|r\rangle$ is an eigenvector of a s.a.\ operator for an eigenvalue $r$.

The average of all conditional states after registrations, a proper mixture,
$$
\sum_r+_s {\mathrm P}_r {\mathsf T}^S_{{\text{out}},r}\ ,
$$
is called {\em unconditional} or {\em non-selective} state (the sign ``$+_s$''
means that the sum represents a proper mixture of states, see Section 1.3). It
is usually described by the words: ``make measurements but ignore the
results''. One also assumes that
\begin{equation}\label{uncond} \sum_r {\mathrm P}_r {\mathsf
T}^S_{{\text{out}},r} = tr^M\Big(\bar{\mathsf U} ({\mathsf T}^S_{\text{i}}
\otimes {\mathsf T}^M_{\text{i}}) \bar{\mathsf U}^\dagger\Big)\ ,
\end{equation} where $tr^M$ denotes a partial trace defined by any orthonormal
frame in the Hilbert space of the meter.

In the standard theory, the reading is a mysterious procedure. If the meter is
considered as a quantum system then to observe it, another meter is needed, to
observe this, still another is and the resulting series of measurements is
called {\em von-Neumann chain}. At some (unknown) stage including the
processes in the mind (brain?) of observer, there is the so-called {\em
Heisenberg cut} that gives the definite value $r$. Moreover, the conditional
state cannot, in general, result by a unitary evolution. The transition
\begin{equation}\label{stredstand} \bar{\mathsf U} ({\mathsf T}^S_{\text{i}}
\otimes {\mathsf T}^M_{\text{i}}) \bar{\mathsf U}^\dagger \mapsto \sum_r +_s
({\mathrm P}_r {\mathsf T}^S_{{\text{out}},r} \otimes {\mathsf
T}^M_{{\text{out}},r})\ ,
\end{equation} where ${\mathsf T}^M_{{\text{out}},r}$ is the signal state of
the meter indicating that outcome of the registration is $r$, is not a unitary
transformation and its existence contradicts the dynamical principles that
have been postulated for quantum mechanics. This is the state reduction.

The standard theory is deliberately left incomplete. First, the time and
location of the Heisenberg cut is not known. Second, if there are two
different kinds of dynamics in quantum mechanics, there ought to be also
objective conditions under which each of them is applicable. The standard
theory of measurement identifies no such {\em objective} conditions. It is
just assumed to be valid under the subjective condition that a physicist
performs a measurement. The difference between the physical process of the
measurement and other physical processes remains obscure.

The standard theory ignores such questions and focuses on what effectively
happens with the object system. A general measurement is then
phenomenologically described by two mathematical quantities. The first is a
{\em state transformer} ${\mathcal O}_r : {\mathbf T}({\mathbf H}^S) \mapsto
{\mathbf L}_r({\mathbf H}^S)$. ${\mathcal O}_r$ enables us to calculate
${\mathsf T}^S_{{\text{out}},r}$ from ${\mathsf T}^S_{\text{i}}$ by
$$
{\mathsf T}^S_{{\text{out}},r} = \frac{{\mathcal O}_r({\mathsf
T}^S_{\text{i}})}{tr\Big({\mathcal O}_r({\mathsf T}^S_{\text{i}})\Big)}\ .
$$
${\mathcal O}_r$ is a so-called {\em completely positive map}. Such maps have
the form \cite{kraus}
\begin{equation}\label{kraus} {\mathcal O}_r({\mathsf T}) = \sum_k {\mathsf
O}_{rk}{\mathsf T}{\mathsf O}_{rk}^\dagger
\end{equation} for any state operator ${\mathsf T}$, where ${\mathsf O}_{rk}$
are some operators satisfying
$$
\sum_{rk} {\mathsf O}_{rk}^\dagger{\mathsf O}_{rk} = {\mathsf 1}\ .
$$
Eq.\ (\ref{kraus}) is called {\em Kraus representation}. A given state
transformer ${\mathcal O}_r$ does not determine, via Eq.\ (\ref{kraus}), the
operators ${\mathsf O}_{rk}$ uniquely.

The second quantity is an operator ${\mathsf E}_r$ called {\em effect} giving
the probability to read value $r$ by
$$
{\mathrm P}_r = tr\Big({\mathcal O}_r({\mathsf T}^S_{\text{i}})\Big) =
tr({\mathsf E}_r {\mathsf T}^S_{\text{i}})\ .
$$
The set $\{{\mathsf E}_r\}$ of effects ${\mathsf E}_r$ for all $r \in {\mathbf
R}$ is a POVM (see \cite{davies}). As shown there, every POVM satisfies two
conditions: positivity,
$$
{\mathsf E}_r \geq {\mathsf 0}\ ,
$$
for all $r \in {\mathbf R}$, and normalisation,
$$
\sum_{r\in{\mathbf R}} {\mathsf E}_r = {\mathsf 1}\ .
$$

One can show that ${\mathcal O}_r$ determines the effect ${\mathsf E}_r$ by
$$
{\mathsf E}_r = \sum_{k} {\mathsf O}_{rk}^\dagger{\mathsf O}_{rk}\ .
$$

In the standard theory, the state transformer of a given registration contains
all information that is necessary for further analysis and for classification
of measurements. Such a classification is given in \cite{WM}, p.\ 35. Thus,
the formalism of the state transformers and POVMs can be considered as the
core of the standard theory.

However, the analysis of Chapter 4 has shown that there is no well-defined
decomposition of the system $S + M_q$ into subsystems $S$ and $M_q$ after the
registered system and the apparatus become entangled by the interaction if the
detector is not {\em clean}, that is if it contains particles of the same type
as the object system. Then, there is neither a well-defined conditional state
nor any sense in which Eq.\ (\ref{uncond}) can be understood. Something like a
conditional state could perhaps be found before the interaction between the
measured system and the detector: this part of the measuring process could be
called ``premeasurement'' (see \cite{BLM}). It seems therefore that the
standard theory describes the premeasurement onto which some results of the
whole measurement process are ``grafted'' post hoc via Eq.\
(\ref{stredstand}), or to the whole measurement if the screens and detectors
are clean. But this is a condition that cannot be fulfilled. Even if one
starts with a clean apparatus, it becomes very quickly polluted during the
measurement.

The standard theory is, so to speak, sufficient for all practical purposes
(abbreviation FAPP introduced by John Bell) but, as already explained, it is
not complete and not strictly correct. The phenomena of reading a meter and of
state reduction need a dynamics similar to that of Eq.\ (\ref{stredstand})
that is different from the usual unitary evolution. Even if one accepted that
Schr\"{o}dinger equation needs correction under some objective
conditions, the theory remains incomplete until such objective conditions are
formulated and specific corrections are proposed.

Some attempts to solve these difficulties start from the assumption that the
transition (\ref{stredstand}) is not observable because the registration of
observables that would reveal the difference is either very difficult or that
such observables do not exist. One can then deny that the transition
(\ref{stredstand}) really takes place and so assume that the objectification
is only apparent (no-collapse scenario). There are three most important
no-collapse approaches:
\begin{enumerate}
\item Quantum decoherence theory \cite{Zeh,Zurek,schloss}. The idea is that
system $S +M_q$ composed of a quantum system and an apparatus cannot be
isolated from environment $E_q$. Then the unitary evolution of $S + M_q + E_q$
leads to a non-unitary evolution of $S + M_q$ that can erase all correlations
and interferences from $S + M_q$ hindering the objectification
\cite{Zeh,Zurek,schloss} (see discussion in Refs.\
\cite{d'Espagnat,Ghirardi,bub,BLM}).
\item Superselection sectors approach \cite{Hepp,Primas,wanb}. Here, classical
properties are described by superselection observables of $M_q$ which commute
with each other and with all other observables of $M_q$. Then, the state of
$M_q$ after the measurement is equivalent to a suitable proper mixture.
\item Modal interpretation \cite{bub}. One assumes that there is a subset of
orthogonal-projection observables that, first, can have determinate values in
the state of $S + M_q$ before the registration in the sense that the
assumption does not violate contextuality (see e.g.\ \cite{peres}, Chapter 7)
and second, that one can reproduce all important results of ordinary quantum
mechanics with the help of these limited set of observables. Thus, one must
require that the other observables are not registered.
\end{enumerate}

Other attempts (collapse scenario) do assume that the reduction is a real
process and postulate a new dynamics that leads directly to something
analogous to (\ref{stredstand}) accepting the consequence that some
measurement could disprove this postulate. An example of the collapse scenario
is known as Dynamical Reduction Program \cite{GRW,pearle}. It postulates new
universal, unique dynamics that is non-linear and stochastic. Both the unitary
evolution and the state reduction result as some approximations. The physical
idea is that of spontaneous localisation. That is, linear superpositions of
different positions spontaneously decay, either by jumps \cite{GRW} or by
continuous transitions \cite{pearle}. The form of this decay is chosen
judiciously to take a very long time for microsystems, so that the standard
quantum mechanics is a good approximation, and a very short time for
macrosystems, so that a state reduction results. In this way, a simple
explanation of the definite positions of macroscopic systems and of the
pointers of registration apparatuses is achieved.

One of the important ideas of the Dynamical Reduction Program is to make the
state reduction well-defined by choosing a particular frame for it: the
$Q$-representa\-tion. This leads to breaking of the symmetry with respect to
all unitary transformations that was not only a beautiful but also a practical
feature of standard quantum mechanics.

Another example of collapse scenario is our approach (see Refs.\
\cite{hajicek2,survey,hajicekW1,hajicek5}). Its aim is to postulate the
existence of state reductions that does not break the unitary symmetry (even
if it itself is a non-unitary state transformation) and to formulates
hypotheses about the conditions, origin and form of state reduction.

The above review of attempts given is rather short and incomplete. However,
the only purpose of it is to specify the position that is taken by the attempt
of the present book among them and to focus on our ideas, which are rather
different from all others.

\section{Reformulation of the standard theory} This section explains our
approach with the help of two models. It starts by trying to make the theory
compatible with the results of Chapters 3 and 4. These results suggest that it
is difficult or impossible to identify the state of the measured system after
the measurement. Then, the notions of the conditional state and of the state
transformer become problematic. We must, therefore, avoid the need for the
definite state of the registered system after the registration and replace the
definition of the state reduction given by Eq.\ (\ref{stredstand}) by an
equation where the conditional state does not appear.

\subsection{Stern-Gerlach story retold}

Hence, we have to modify the textbook description (e.g., \cite{peres}, pp.\ 14
and 375 or \cite{ballent}, p.\ 230) of the Stern-Gerlach experiment. There are
two changes. First, we take more seriously the role of real detectors in the
experiment. The detector is assumed to be an object with both classical and
quantum model that gives information on the registered quantum object via its
classical properties. Hence, it has to satisfy the assumptions of Section 2.1
on classical properties. Second, the description is made compatible with the
consequences of the exchange symmetry for the measurement process that were
explained in Chapter 4 so that it can make use of changes of separation
status.

The original experiment measures the spin of silver atoms. A silver atom
consists of 47 protons and 61 neutrons in the nucleus and of 47 electrons
around it. This leads to some complications that can be dealt with technically
but that would obscure the ideas we are going to illustrate. Just constructing
a simple model, we replace the silver atom by a neutral spin 1/2 particle.

Let the particle be denoted by $S$ and its Hilbert space by ${\mathbf H}$. Let
$\vec{\mathsf x}$ be its position, $\vec{\mathsf p}$ its momentum and
${\mathsf S}_z$ the $z$-component of its spin with eigenvectors $|j\rangle$
and eigenvalues $j \hbar/2$, where $j = \pm 1$ (see e.g.\ \cite{ballent},
Section 7.4).

Let ${\mathcal M}$ be a Stern-Gerlach apparatus (see Section 1.3) with an
inhomogeneous magnetic field oriented so that it separates different
$z$-components of spin of $S$ arriving there. To calculate the evolution of
$S$ in the magnetic field, we use the modified Schr\"{o}dinger equation
that describes the interaction between the particle and external field, as it
is done, e.g., in \cite{peres}, p.\ 375.

Let the detector of the apparatus be a photo-emulsion film ${\mathcal D}$ with
energy threshold $E_0$. Its emulsion grains are not macroscopic in the sense
that each would contain about $10^{23}$ molecules---they contain only about
$10^{10}$ in average. Still, the chemical and thermodynamic process in them
can be described with a sufficient precision by classical chemistry and
phenomenological thermodynamics. They have classical states and classical
properties. The emulsion grains that are hit by $S$ run through a process of
change and of modification and the modification can be made directly
visible. ${\mathcal D}$ is a macroscopic object formed by such grains. Let its
classical model be $D_c$ and its quantum one be $D_q$ with Hilbert space
${\mathbf H}^{\mathcal D}$. According to our theory of classical properties in
Chapter 3, the quantum states of the grains, and so of the whole $D_q$, must
be some high-entropy states. The usual description of meters by wave functions
is thus not completely adequate.

First, let $S$ be prepared at time $t_1$ in a definite spin-component state,
\begin{equation}\label{sin} |\text{in},j\rangle = |\vec{p},\Delta
\vec{p}\rangle \otimes |j \rangle\ ,
\end{equation} where $|\vec{p},\Delta \vec{p}\rangle$ is a Gaussian wave
packet with the expectation value $\vec{p}$ and variance $\Delta \vec{p}$ of
momentum. To make the mathematics easier, we shall also work with the
formalism of wave functions and kernels explained in Section 4.1. The wave
function of state (\ref{sin}) in an arbitrary representation will be denoted
by $\psi_j(\lambda)$. Let system $D_q$ be prepared in metastable state
${\mathsf T}^{\mathcal D}$ at $t_1$. We assume that $D_q$ consists of $N$
particles of which $N_1$ are indistinguishable from $S$. Hence, the kernel of
${\mathsf T}^{\mathcal D}$ is
$$
T^{\mathcal
D}(\lambda^{(1)},\ldots,\lambda^{(N_1)},\lambda^{(N_1+1)},\ldots,\lambda^{(N)};
\lambda^{(1)\prime},\ldots,\lambda^{(N_1)\prime},\lambda^{(N_1+1)\prime},\ldots,\lambda^{(N)\prime})\
,
$$
where the function $T^{\mathcal D}$ is antisymmetric both in variables
$\lambda^{(1)},\ldots,\lambda^{(N_1)}$ and $\lambda^{(1)\prime},\ldots$,
$\lambda^{(N_1)\prime}$. The initial state of the composite $S + D_q$ then is
\begin{equation}\label{initSD1} \bar{\mathsf T}_j =
N^2_{\text{exch}}\bar{\mathsf
\Pi}^{N_1+1}_-\Bigl(\psi_j(\lambda^{(0)})\psi^*_j(\lambda^{(0)\prime})T^{\mathcal
D}(\lambda^{(1)},\ldots,\lambda^{(N)};
\lambda^{(1)\prime},\ldots,\lambda^{(N)\prime})\Bigr)\bar{\mathsf
\Pi}^{N_1+1}_-\ ,
\end{equation} where $\bar{\mathsf \Pi}^{N_1+1}_-$ denotes the
antisymmetrisation in the variables $\lambda^{(0)},\ldots,\lambda^{(N_1)}$ (or
$\lambda^{(0)\prime},\ldots,\lambda^{(N_1)\prime}$). It is an orthogonal
projection acting on Hilbert space ${\mathbf H} \otimes {\mathbf H}^{\mathcal
D}$ (see Section 4.1).

We also assume that the direction of $\vec{p}$ is suitably restricted and its
magnitude respects the energy threshold $E_0$. Such states lie in the domain
of the apparatus ${\mathcal M}$, see Section 4.4. According to our theory of
meters in Chapter 4, states in the domain of ${\mathcal M}$ have a separation
status before their registration by ${\mathcal M}$. Hence, state (\ref{sin})
has a separation status at $t_1$ and so the system $S$ represents initially an
individual quantum object with an objective state. From Definition \ref{dfss}
of separation status, it follows that
\begin{equation}\label{DSss1} \int d\lambda^{(k)}\,
\psi^*_j(\lambda^{(k)})T^{\mathcal D}(\lambda^{(1)},\ldots,\lambda^{(N_1)};
\lambda^{(1)\prime},\ldots,\lambda^{(N_1)\prime}) = 0
\end{equation} for any $k = 1,\ldots,N_1$, and
\begin{equation}\label{DSss2} \int d\lambda^{(l)\prime}\,
\psi_j(\lambda^{(l)\prime})T^{\mathcal
D}(\lambda^{(1)},\ldots,\lambda^{(N_1)};
\lambda^{(1)\prime},\ldots,\lambda^{(N1)\prime}) = 0
\end{equation} for any $l = 1,\ldots,N_1$.

Eqs.\ (\ref{DSss1}) and (\ref{DSss2}) enable us to rewrite state
(\ref{initSD1}) in a more explicit form. To this aim, we need the following
Lemma:
\begin{lem}\label{lemNPia} Let $F_n(\lambda^{(1)},\ldots,\lambda^{(N)})$, $n =
1,\ldots,K$, be $K$ functions of $N$ variables that satisfy:
\begin{enumerate}
\item Function $F_n$ is antisymmetric in the variables
$\lambda^{(1)},\ldots,\lambda^{(N_1)}$ for all $n$ and for some $N_1 < N$.
\item For some functions $\psi_j(\lambda)$, $j=1,\ldots,L$, such that $\int
d\lambda\, \psi^*_j(\lambda)\psi_j(\lambda) = 1$,
\begin{equation}\label{lem1} \int d\lambda^{(k)}\,
\psi_j(\lambda^{(k)})F_n(\lambda^{(1)},\ldots,\lambda^{(N)}) = 0
\end{equation} for all $j$, $n$ and $k = 1,\ldots,N_1$.
\item $\{F_n\}$ is an orthonormal set,
\begin{equation}\label{lem2} \int d^N\lambda\,
F^*_{n'}(\lambda^{(1)},\ldots,\lambda^{(N)})
F_n(\lambda^{(1)},\ldots,\lambda^{(N)}) = \delta_{nn'}
\end{equation} for all $n,n'$.
\end{enumerate} Let function $\bar{F}_{jn}$ of $N+1$ variables
$\lambda^{(0)},\lambda^{(1)},\ldots,\lambda^{(N)}$ be defined by
\begin{equation}\label{lem3}
\bar{F}_{jn}(\lambda^{(0)}\lambda_1,\ldots,\lambda^{(N)}) =
\frac{1}{\sqrt{N_1+1}} \sum_{k=0}^{N_1} (-1)^{kN_1} \psi_j(\lambda^{(k)})
F_n[\lambda^{(0)}\mapsto \lambda^{(k)}]\ ,
\end{equation} where
$$
F_n[\lambda^{(0)}\mapsto \lambda^{(k)}] =
F_n(\lambda^{(k+1)},\ldots,\lambda^{(N_1)},\lambda^{(0)},\ldots,\lambda^{(k-1)},\lambda^{(N_1+1)},\ldots,\lambda^{(N)})\
.
$$
Then functions $\bar{F}_{jn}$ are antisymmetric in variables
$\lambda^{(0)},\lambda^{(1)},\ldots,\lambda^{(N_1)}$ and satisfy:
\begin{equation}\label{lem4} \int d^{N+1}\lambda\,
\bar{F}^*_{jn}(\lambda^{(0)},\lambda^{(1)},\ldots,\lambda^{(N)})
\bar{F}_{jn'}(\lambda^{(0)},\lambda^{(1)},\ldots,\lambda^{(N)}) = \delta_{nn'}
\end{equation} for all $j$, $n$ and $n'$.
\end{lem} The set $\lambda^{(a)},\ldots,\lambda^{(b)}$ for any integers $a$
and $b$ is empty if $a > b$ and contains all entries $\lambda^{(c)}$ for $a
\leq c \leq b$ in the increasing index order if $a \leq b$.\par\noindent {\bf
Proof} Function $\bar{F}_{jn}$ is antisymmetric because $F_n$ is and the sum
in (\ref{lem3}) contains already exchanges of $\lambda^{(0)}$ and
$\lambda^{(k)}$ for all $k > 0$ with the proper signs (see Section 4.1). To
show Eq.\ (\ref{lem4}), we substitute Eq.\ (\ref{lem3}) into the right-hand
side of Eq.\ (\ref{lem4}):
\begin{multline*} \int d^{N+1}\lambda\, \bar{F}^*_{jn'}\bar{F}_{jn} =
\frac{1}{N_1+1} \int d^{N+1}\lambda \sum_{k=0}^{N_1}
(-1)^{kN_1}\sum_{l=0}^{N_1} (-1)^{lN_1} \\ \times \psi_j(\lambda^{(k)})
F_n[\lambda^{(0)}\mapsto \lambda^{(k)}]
\psi^*_j(\lambda_l)F^*_{n'}[\lambda^{(0)}\mapsto \lambda^{(l)}]\ .
\end{multline*} The terms
$$
\int d^{N+1}\lambda\, \psi_j(\lambda^{(k)}) F_n[\lambda^{(0)}\mapsto
\lambda^{(k)}] \psi^*_j(\lambda^{(1l)})F^*_{n'}[\lambda^{(0)}\mapsto
\lambda^{(l)}]
$$
vanish for any $k \neq l$ because of Eq.\ (\ref{lem1}). The remaining terms
$$
\int d^{N+1}\lambda\, \psi_j(\lambda^{(k)}) F_n[\lambda^{(0)}\mapsto
\lambda^{(k)}] \psi^*_j(\lambda^{(k)})F^*_{n'}[\lambda^{(0)}\mapsto
\lambda^{(k)}]
$$
are equal to $\delta_{nn'}$ for all $k$ because of the normalisation of
$\psi_j$ and Eq.\ ({\ref{lem2}), {\bf QED}.

State (\ref{initSD1}) has then the following kernel:
\begin{multline}\label{kernbarTj}
\bar{T}_j(\lambda^{(0)},\ldots,\lambda^{(N)};
\lambda^{(0)\prime},\ldots,\lambda^{(N)\prime}) =
\frac{1}{N_1+1}\sum_{k=0}^{N_1} (-1)^{kN_1}\sum_{l=0}^{N_1} (-1)^{lN_1}
\psi_j(\lambda^{(k)})\psi^*_j(\lambda^{(l)\prime}) \\ T^{\mathcal
D}(\lambda^{(k+1)},\ldots,\lambda^{(N_1)},\lambda^{(0)},\ldots,\lambda^{(k-1)},\lambda^{(N_1+1)},\ldots,\lambda^{(N)};
\\
\lambda^{(l+1)\prime},\ldots,\lambda^{(N_1)\prime},\lambda^{(0)\prime},\ldots,\lambda^{(l-1)\prime},\lambda^{(N_1+1)\prime},\ldots,\lambda^{(N)\prime})\
.
\end{multline} Kernel $\bar{T}_j$ can be shown to be antisymmetric in
variables $\lambda^{(0)},\ldots,\lambda^{(N_1)}$ and
$\lambda^{(0)\prime},\ldots,$ $\lambda^{(N_1)\prime}$ and to have trace equal
1 by the same methods as those used to prove Lemma \ref{lemNPia}. Eqs.\
(\ref{DSss1}) and (\ref{DSss2}) expressing the separation status of
$|\psi\rangle$ play an important role in the derivation of formula
(\ref{kernbarTj}).

The initial state of $S + D_q$ does not contain any modified emulsion
grains. Extremal states with this property form a subspace of the Hilbert
space $\bar{\mathsf \Pi}^{N_1+1}_-({\mathbf H} \otimes {\mathbf H}^{\mathcal
D})$ of $S + D_q$. Let us denote the projection to this subspace by
$\bar{\mathsf \Pi}[\emptyset]$. Thus, we have
\begin{equation}\label{nograins} tr(\bar{\mathsf T}_j\bar{\mathsf
\Pi}[\emptyset]) = 1\ .
\end{equation}

The process of registration includes the interaction of $S$ with the magnetic
field and with system $D_q$ as well as the resulting modification of the
emulsion grains. We assume that meter ${\mathcal M}$ is {\em ideal}: each copy
of $S$ that arrives at the emulsion $D_q$ modifies at least one emulsion
grain.

The registration is assumed to be a quantum evolution described by a unitary
group $\bar{\mathsf U}(t)$, the so-called {\em measurement coupling} see
\cite{BLM}. We assume that $\bar{\mathsf U}(t)$ commutes with $\bar{\mathsf
\Pi}^{N_1+1}_-$, see Section 4.1. Let $t_2$ be the time at which the
modification of the hit grains is finished and let $\bar{\mathsf U} =
\bar{\mathsf U}(t_2 - t_1)$. We are going to derive some important properties
of $\bar{\mathsf U}{\mathsf T}_j\bar{\mathsf U}^\dagger$, and for this we need
a technical trick that transforms calculations with kernels into that with
wave functions.

Let
\begin{equation}\label{specdecTD} {\mathsf T}^{\mathcal D} = \sum_n a_n
|n\rangle \langle n|
\end{equation} be the spectral decomposition of ${\mathsf T}^{\mathcal
D}$. Then, $0 \leq a_n \leq 1$ for each $n \in {\mathbb N}$ and $\sum_n a_n =
1$. In $\lambda$-representation, state $|n\rangle$ has the wave function
$\varphi_n(\lambda^{(1)},\ldots,\lambda^{(N)})$.

Eqs.\ (\ref{kernbarTj}) and (\ref{specdecTD}) imply that
\begin{equation}\label{sdbarTj} \bar{T}_j(\lambda^{(0)},\ldots,\lambda^{(N)};
\lambda^{(0)\prime},\ldots,\lambda^{(N)\prime}) = \sum_na_n
\bar{\Psi}_{jn}(\lambda^{(0)},\ldots,\lambda^{(N)})\bar{\Psi}^*_{jn}(\lambda^{(0)\prime},\ldots,\lambda^{(N)\prime})\
,
\end{equation} where
\begin{equation}\label{compinvec}
\bar{\Psi}_{jn}(\lambda^{(0)},\ldots,\lambda^{(N)}) = \frac{1}{\sqrt{N_1+1}}
\sum_{k=0}^{N_1} (-1)^{kN_1}
\psi_j(\lambda^{(k)})\varphi_n[\lambda^{(0)}\mapsto \lambda^{(k)}]\ .
\end{equation}
\begin{lem}\label{lemsdbarTj} Eq.\ (\ref{sdbarTj}) is the spectral
decomposition of state $\bar{\mathsf T}_j$.
\end{lem} {\bf Proof} Conditions (\ref{DSss1}) and (\ref{DSss2}) on
$|\psi_j\rangle$ and $\bar{\mathsf T}^{\mathcal D}$ imply
$$
\sum_n a_n \int d\lambda^{(k)} \int d\lambda^{(k)\prime}\,
\psi^*_j(\lambda^{(k)})\psi_j(\lambda^{(k)\prime})\,
\varphi_n(\lambda^{(1)},\ldots,\lambda^{(N)})
\varphi_n^*(\lambda^{(1)\prime},\ldots,\lambda^{(N)\prime}) = 0$$ for all $k=
1,\dots,N_1$. However, the integral defines a positive kernel
$$
K_n(\lambda^{(1)},\ldots,\lambda^{(k-1)}\lambda^{(k+1)},\ldots,\lambda^{(N)};\lambda^{(1)\prime},\ldots,\lambda^{(k-1)\prime}\lambda^{(k+1)\prime},\ldots,\lambda^{(N)\prime})
$$
for each $n$ and a sum with positive coefficients of such kernels can be zero
only if each such kernel itself vanishes. Hence, we have
\begin{equation}\label{wfSDss} \int d\lambda_k\,
\psi^*(\lambda^{(k)})\varphi_n(\lambda^{(1)},\ldots,\lambda^{(N)}) = 0
\end{equation} for each $n$ and all $k = 1,\ldots,N$.

From Lemma \ref{lemNPia}, it then follows now that
$$
\langle \bar{\Psi}_{jn}|\bar{\Psi}_{jn'}\rangle = \delta_{nn'}\ .
$$
This implies Lemma \ref{lemsdbarTj}, {\bf QED}.

A simple consequence of Lemma \ref{lemsdbarTj} is the following. Combining
Eqs.\ (\ref{nograins}) and (\ref{sdbarTj}), we obtain
$$
tr\Bigl(\sum_na_n |\bar{\Psi}_{jn}\rangle \langle\bar{\Psi}_{jn}|\bar{\mathsf
\Pi}[\emptyset] \Bigl) = \sum_na_n tr\Bigl((\bar{\mathsf
\Pi}[\emptyset]|\bar{\Psi}_{jn}\rangle) (\langle\bar{\Psi}_{jn}|\bar{\mathsf
\Pi}[\emptyset]) \Bigl) = 1\ .
$$
But operator $\bar{\mathsf \Pi}[\emptyset]|\bar{\Psi}_{jn}\rangle
\langle\bar{\Psi}_{jn}|\bar{\mathsf \Pi}[\emptyset]$ is positive so that its
trace must be non-negative. As the sum of $a_n$'s is already 1, we must have
$$
tr(\bar{\mathsf \Pi}[\emptyset]|\bar{\Psi}_{jn}\rangle
\langle\bar{\Psi}_{jn}|\bar{\mathsf \Pi}[\emptyset]) = 1
$$
or
$$
\langle\bar{\mathsf
\Pi}[\emptyset]|\bar{\Psi}_{jn}|\bar{\Psi}_{jn}|\bar{\mathsf
\Pi}[\emptyset]\rangle = 1
$$
for each $n$. However,
$$
|\bar{\Psi}_{jn}\rangle = \bar{\mathsf \Pi}[\emptyset]|\bar{\Psi}_{jn}\rangle
+ ({\mathsf 1} - \bar{\mathsf \Pi}[\emptyset])|\bar{\Psi}_{jn}\rangle
$$
and
$$
\langle \bar{\mathsf \Pi}[\emptyset]|\bar{\Psi}_{jn}|({\mathsf 1} -
\bar{\mathsf \Pi}[\emptyset])|\bar{\Psi}_{jn}\rangle = 0
$$
so that
$$
1 = \langle \bar{\Psi}_{jn}|\bar{\Psi}_{jn}\rangle = \langle \bar{\mathsf
\Pi}[\emptyset]|\bar{\Psi}_{jn}|\bar{\mathsf
\Pi}[\emptyset]|\bar{\Psi}_{jn}\rangle + \langle ({\mathsf 1} - \bar{\mathsf
\Pi}[\emptyset])|\bar{\Psi}_{jn}|({\mathsf 1} - \bar{\mathsf
\Pi}[\emptyset])|\bar{\Psi}_{jn}\rangle\ .
$$
Hence,
\begin{equation}\label{PiPsi} \bar{\mathsf
\Pi}[\emptyset]|\bar{\Psi}_{jn}\rangle = |\bar{\Psi}_{jn}\rangle\ .
\end{equation}

Let us now return to the time evolution of $\bar{\mathsf T}_j$ within
$\bar{\mathsf \Pi}^{N_1+1}_-({\mathbf H} \otimes {\mathbf H}^{\mathcal D})$
from $t_1$ to $t_2$. System $S + D_q$ is composed of two disjoint subsystems,
$S'$ and $D_q'$, $S'$ containing $S$ and all $N_1$ particles of $D_q$ that are
indistinguishable from $S$. Then, $\bar{\mathsf \Pi}^{N_1+1}_-({\mathbf H}
\otimes {\mathbf H}^{\mathcal D}) = ({\mathbf H})^{N_1+1}_- \otimes {\mathbf
H}^{\mathcal D\prime}$. The evolution defines states $\bar{\mathsf
T}_{j}(t_2)$ of $S + D_q$ by:
\begin{equation}\label{evolTt1} \bar{U}\bar{\mathsf T}_j\bar{U}^\dagger =
\bar{\mathsf T}_{j}(t_2)\ .
\end{equation} Evolution $\bar{\mathsf U}$ includes a thermodynamic relaxation
of $S + D_q$ and a loss of separation status of $S$. Thus, in general, quantum
system $S$ does not represent an individual quantum object after the
registration. The individual states that could be ascribed to $S$ as its
objective properties are not well defined (see Section 4.1) at $t = t_2$. We
can say that they do not exist. However, the whole composite $S + D_q$ is a
quantum object, prepared in the measurement experiment, hence one can consider
its individual states as its objective properties.

Accordingly, states $\bar{\mathsf T}_{j}(t_2)$ also describe the modified
emulsion grains, which can be called {\em detector signals}. The signals are
concentrated within two strips of the film, each strip corresponding to one
value of $j$. The two space regions, $R_+$ and $R_-$, of the two strips are
sufficiently separated and help to determine, in the present case, what is
generally called a pointer observable: the occurrence of a modified emulsion
grain within $R_+$ or $R_-$. Let the projections onto the subspaces of
$({\mathbf H})^{N_1+1}_- \otimes {\mathbf H}^{\mathcal D\prime}$ containing
the corresponding extremal states be $\bar{\mathsf \Pi}[R_j]$.

We avoid specifying $\bar{\mathsf U}(t)$ e.g.\ by writing the Hamiltonian of
system $S + {\mathcal D}_q$. Instead, we express the condition that the meter
registers ${\mathsf S}_z$ through properties of end states ${\mathsf
T}_{j}(t_2)$ as follows:
\begin{equation}\label{registcond1} tr\Bigl(\bar{\mathsf \Pi}[R_j]\bar{\mathsf
T}_{k}(t_2)\Bigr) = \delta_{jk}\ .
\end{equation}

If we substitute Eqs.\ (\ref{evolTt1}) and (\ref{sdbarTj}) into
(\ref{registcond1}), we obtain
$$
\sum_na_n tr(\bar{U}|\bar{\Psi}_{kn}\rangle \langle
\bar{\Psi}_{kn}|\bar{U}^\dagger\bar{\mathsf \Pi}[R_j]) = \delta_{jk}\ .
$$
By the same argument as that leading to formula (\ref{PiPsi}), we then have
\begin{equation}\label{PiDjPsik} \bar{\mathsf
\Pi}[R_j]|\bar{\Psi}_{kn}(t_2)\rangle =
\delta_{jk}|\bar{\Psi}_{kn}(t_2)\rangle\ ,
\end{equation} where
$$
|\bar{\Psi}_{kn}(t_2)\rangle = \bar{U}|\bar{\Psi}_{kn}\rangle\ .
$$
Hence, the state $\bar{U}|\bar{\Psi}_{kn}\rangle$ contains modified emulsion
grains in the region $R_k$ and no such grains in the region $R_l$ for each $n$
and $l \neq k$.

Suppose next that the initial state of $S$ at $t_1$ is
\begin{equation}\label{initS2} |\text{in}\rangle = \sum_j c_j
|\text{in},j\rangle
\end{equation} with
$$
\sum_j|c_j|^2 = 1\ .
$$

The linearity of $\bar{\mathsf U}$ implies the following form of the
corresponding end state $\bar{\mathsf T}(t_2) \in {\mathbf
T}\Bigl(\bar{\mathsf \Pi}^{N_1+1}_-({\mathbf H} \otimes {\mathbf H}^{\mathcal
D})\Bigr)$:
\begin{multline}\label{endstat2} \bar{\mathsf T}(t_2) =
N^2_{\text{exch}}\bar{\mathsf U}\bar{\mathsf \Pi}^{N_1+1}_-\left[
\left(\sum_jc_j|\text{in},j \rangle\right) \left(\sum_{j'}c^*_{j'}\langle
\text{in},j'|\right)\otimes {\mathsf T}^{\mathcal D}\right]\bar{\mathsf
\Pi}^{N_1+1}_-\bar{\mathsf U}^\dagger \\ = \sum_{jj'}c_jc^*_{j'} \bar{\mathsf
T}_{jj'}(t_2)\ ,
\end{multline} Operators $\bar{\mathsf T}_{jj'}(t_2)$ act on the Hilbert space
$\bar{\mathsf \Pi}^{N_1+1}_-({\mathbf H} \otimes {\mathbf H}^{\mathcal D})$ of
$S + D_q$ and are defined by
\begin{equation}\label{barTjj'} \bar{\mathsf T}_{jj'}(t_2) =
N_{\text{exch}}\bar{\mathsf U}\bar{\mathsf \Pi}^{N_1+1}_-(|\text{in},j \rangle
\langle \text{in},j'|\otimes {\mathsf T}^{\mathcal D})\bar{\mathsf
\Pi}^{N_1+1}_-\bar{\mathsf U}^\dagger\ .
\end{equation} They are state operators only for $j' = j$. Eqs.\
(\ref{evolTt1}) and (\ref{initSD1}) imply that
$$
{\mathsf T}_{jj}(t_2) = {\mathsf T}_{j}(t_2)\ .
$$

If we substitute the spectral decomposition (\ref{specdecTD}) of ${\mathsf
T}^{\mathcal D}$ into Eq.\ (\ref{barTjj'}), we obtain for the kernel of
operator ${\mathsf T}_{jj'}(t_2)$
\begin{multline*} T_{jj'}(t_2) = \sum_n a_n \bar{\mathsf U} \\ \times
\Bigl(\sum_{k=0}^{N_1}(-1)^{N_1k}\psi_j(\lambda^{(k)})\varphi_n(\lambda^{(k+1)},\ldots,\lambda^{(N_1)},\lambda^{(0)},\ldots,\lambda^{(k-1)},\lambda^{(N_1+1)},\ldots,\lambda^{(N)})\Bigr)
\\ \times \Bigl(\sum_{l=0}^{N_1}
(-1)^{N_1l}\psi^*_{j'}(\lambda^{(l)}\prime)\varphi^*_n(\lambda^{(l+1)\prime},\ldots,\lambda^{(N_1)\prime},\lambda^{(0)\prime},\ldots,\lambda^{(l-1)\prime},\lambda^{(N_1+1)\prime},\ldots,\lambda^{(N)\prime})\Bigr)\bar{\mathsf
U}^\dagger \\ = \sum_n a_n \Bigl(\bar{\mathsf
U}\bar{\Psi}_{jn}(\lambda^{(0)},\ldots,\lambda^{(N)})\Bigr)\Bigl(\bar{\Psi}^*_{j'n}(\lambda^{(0)\prime},\ldots,\lambda^{(N)\prime})\bar{\mathsf
U}^\dagger\Bigr)\ ,
\end{multline*} or
\begin{equation}\label{decbarTjj'} {\mathsf T}_{jj'}(t_2) = \sum_n a_n
|\bar{\Psi}_{jn}(t_2)\rangle \langle \bar{\Psi}_{j'n}(t_2)|\ .
\end{equation} Eq.\ (\ref{decbarTjj'}) is, of course, not the spectral
decomposition of ${\mathsf T}_{jj'}(t_2)$ because this operator is not
self-adjoint, but it can be used to show that Eq.\ (\ref{PiDjPsik}) implies:
\begin{equation}\label{compoutvec} tr\Bigl(\bar{\mathsf
\Pi}[R_k]|\bar{\Psi}_{jn}(t_2)\rangle \langle \bar{\Psi}_{j'n}(t_2)|\Bigr) =
\delta_{kj}\delta_{kj'}\ .
\end{equation} Then, because of the orthonormality of state vectors
$|\bar{\Psi}_{jn}(t_2)\rangle$, it follows that
\begin{equation}\label{registcond3} tr\Bigl(\bar{\mathsf \Pi}[R_j]\bar{\mathsf
T}_{kl}(t_2)\Bigr) = \delta_{jk}\delta_{jl}
\end{equation} and
\begin{equation}\label{PRC} tr\Bigl(\bar{\mathsf \Pi}[R_j] \bar{\mathsf
T}(t_2)\Bigr) = |c_j|^2\ .
\end{equation} The significance of Eq.\ (\ref{PRC}) is that the modified
grains will be found in the strip $j$ with the probability given by the Born
rule for registering the spin $j$ in the state (\ref{initS2}).

Eq.\ (\ref{endstat2}) can be written as
\begin{equation}\label{decom101} \bar{\mathsf T}(t_2) = \bar{\mathsf
T}_{\text{end}1} + \bar{\mathsf T}_{\text{end}0}\ ,
\end{equation} where
\begin{equation}\label{decom102} \bar{\mathsf T}_{\text{end}1} = \sum_j
|c_j|^2 \bar{\mathsf T}_j(t_2)\ ,\quad \bar{\mathsf T}_{\text{end}0} =
\sum_{j\neq j'} c_j c^*_{j'} \bar{\mathsf T}_{jj'}(t_2)\ .
\end{equation} It follows that
\begin{equation}\label{decom103} tr(\bar{\mathsf T}_{\text{end}1}) = 1\ ,\quad
tr(\bar{\mathsf T}_{\text{end}0}) = 0\ .
\end{equation} Eq.\ (\ref{decom102}) says that $\bar{\mathsf T}_{\text{end}1}$
is a convex combination of quantum states that differ from each other by
expectation values of operator $\bar{\mathsf \Pi}[R_j]$.

Finally, we have to analyse more closely what is observed in Stern-Gerlach
experiment. The basic fact is that there are modified emulsion grains at some
definite positions at the film after each registration. This is represented by
definite states of the classical model $D_c$ of the film. A basic assumption
about classical models is that their states are objective, that is, they exist
before being observed and the observation only reveals them (see Chapter 2). A
state $T_c$ of $D_c$ can be described by specifying the positions of the
modified grains. Then we can express the fact that the modified grains lie in
strip $R_j$ by the (epistemic) classical state represented by composed
expression $T_c \subset R_j$. Quantum mechanics can only give us the
probabilities ${\mathrm P}(T_c)$ that state $T_c$ is observed:
$$
{\mathrm P}(T_c) = tr\Bigl(\bar{\mathsf T}(t_2)\bar{\mathsf \Pi}[R_j]\Bigr)\ .
$$

According to Minimum Interpretation, state $\bar{\mathsf T}(t_2)$ just
describes the statistics of the ensemble of particular measurements on system
$S + D_q$ and does not refer to anything existing before the registration and
concerning each individual system. Minimum Interpretation does not even say
that a prepared individual system is an element of a definite ensemble of
measurements: such ensemble is only defined if the registered observable is
fixed.

According to RCU Interpretations, state $\bar{\mathsf T}(t_2)$ is a property
referring directly to each individual composite system $S + {\mathcal D}_q$
immediately before the registration. Moreover, two quantum states
$\bar{\mathsf T}_j(t_2)$, $j = 1,2$, are in a bijective relation with two
classical states $T_c \subset R_j$, $j = 1,2$. The observation that the
classical state of $D_c$ is $T_c \subset R_j$ implies, therefore, that the
quantum state of $S + D_q$ must be $\bar{\mathsf T}_j(t_2)$ already before the
(classical) observation. Hence, the state of the individual composite system
$S + D_q$ immediately before the registration must be a proper mixture of
states $\bar{\mathsf T}_j(t_2)$ each of which has a definite value of $j$:
\begin{equation}\label{endst2} \sum_j\ +_s\ |c_j|^2 \bar{\mathsf T}_{j}(t_2)
\end{equation} instead of (\ref{endstat2}) that results by unitary, linear
evolution law of quantum mechanics. Observe that the transition from state
(\ref{endstat2}) to (\ref{endst2}) is non-linear but preserving the norm of
the state. The additional ``evolution'' from state (\ref{endstat2}) to state
(\ref{endst2}) that must then be caused in some way by the registration, is
the state reduction.

This section was rather technical because it was to describe registrations in
a way that avoids the concept of conditional state. This lead to a new
formalism represented by Eqs.\ (\ref{decom101}) and (\ref{endst2}). The state
reduction is now defined as the transition between the states described by
these two equations.

\subsection{Screen} Screens are used in most preparation procedures. For
example, in optical experiments \cite{RSH}, polarisers, such as Glan-Thompson
ones, are employed. A polariser contains a crystal that decomposes the coming
light into two orthogonal-polarisation parts. One part disappears inside an
absorber and the other is left through. Similarly, the Stern-Gerlach
experiment (see Section 2.3) can be modified so that the beam corresponding to
spin down is blocked out by an absorber and the other beam is left through. In
the interference experiment \cite{tono}, there are several screens, which are
just walls with openings. Generally, a screen is a macroscopic body that
decomposes the incoming, already prepared, beam into one part that disappears
inside the body and the other that goes through.

Here, a simple model of screen is constructed and its physics is studied. Let
the particle $S$ interacting with the screen have mass $\mu$ and spin 0 and
the screen have the following geometry:
\begin{assump}\label{asgeomscr} The screen is at $x_3 = 0$ and the half-spaces
$x_3 < 0$ and $x_3 > 0$ are empty. There is a opening $D$ in the screen, that
is $D$ is an open subset of the plane $x_3 = 0$, not necessary connected
(e.g., two slits). Finally, let the screen be stationary, that is the geometry
is time independent.
\end{assump} For the interaction between the particle and the screen, we
assume:
\begin{assump}\label{asdynscr} Inside the half-spaces $x_3 < 0$, $x_3 > 0$,
the wave function $\psi(\vec{x},t)$ of $S$ satisfies the free
Schr\"{o}dinger equation,
\begin{equation}\label{schrodfree} i\hbar\frac{\partial
\psi(\vec{x},t)}{\partial t} = -\frac{\hbar^2}{2\mu}\left(\frac{\partial^2
\psi(\vec{x},t)}{\partial x_1^2} + \frac{\partial^2 \psi(\vec{x},t)}{\partial
x_2^2} + \frac{\partial^2 \psi(\vec{x},t)}{\partial x_3^2}\right)\ .
\end{equation}

Let us denote the part of the solution $\psi(\vec{x},t)$ in the left
half-space $x_3 < 0$ by $\psi_{\text{i}}(\vec{x},t)$ and in the right
half-space by $\psi_{\text{traf}}(\vec{x},t)$. Let
$\psi_{\text{i}}(\vec{x},t)$ be the $x_3 < 0$-part of a wave packet with $p_3
> 0$,
\begin{equation}\label{wavepack} \psi(\vec{x},t) =
\left(\frac{1}{2\pi\hbar}\right)^{3/2}\int_{{\mathbb
R}^3}d^3p\,\tilde{\psi}(\vec{p})\exp\left[\frac{i}{\hbar}\left(-\frac{|\vec{p}|^2}{2\mu}t
+ \vec{p}\cdot\vec{x}\right)\right]\ ,
\end{equation} where $\tilde{\psi}(\vec{p})$ is a rapidly decreasing function
(see \cite{RS}, p.\ 133) with $\tilde{\psi}(\vec{p}) = 0$ for all $p_3 \leq
0$, and let, for any fixed (finite) time, function
$\psi_{\text{traf}}(\vec{x},t)$ is rapidly decreasing.

At the points of the screen, the wave function is discontinuous. From the
left, the boundary values
$$
\lim_{x_3\rightarrow -0} \psi(\vec{x},t) = \lim_{x_3\rightarrow 0}
\psi_{\text{i}}(\vec{x},t)\ ,\quad \lim_{x_3\rightarrow -0} \frac{\partial
\psi}{\partial x_3}(\vec{x},t) = \lim_{x_3\rightarrow 0} \frac{\partial
\psi_{\text{i}}}{\partial x_3}(\vec{x},t)\ ,
$$
are determined by the solution $\psi_{\text{i}}(\vec{x},t)$. From the right,
\begin{equation}\label{bcscreen0} \lim_{x_3\rightarrow 0}
\psi_{\text{traf}}(\vec{x},t) = 0\ ,\quad \lim_{x_3\rightarrow 0}
\frac{\partial \psi_{\text{traf}}}{\partial x_3}(\vec{x},t) = 0
\end{equation} for $(x_1,x_2) \not\in D$ and
\begin{equation}\label{bcscreen1} \lim_{x_3\rightarrow 0}
\psi_{\text{traf}}(\vec{x},t) = \lim_{x_3\rightarrow -0} \psi(\vec{x},t)\
,\quad \lim_{x_3\rightarrow 0} \frac{\partial \psi_{\text{traf}}}{\partial
x_3}(\vec{x},t) = \lim_{x_3\rightarrow -0} \frac{\partial \psi}{\partial
x_3}(\vec{x},t)
\end{equation} for $(x_1,x_2) \in D$.
\end{assump} This expresses the notion that all particles arrive at the screen
from the left and those that hit the screen are absorbed by the screen and
cannot reappear.

The mathematical problem defined by Assumptions \ref{asgeomscr} and
\ref{asdynscr} can be solved by the same method as the diffraction problem in
optics can (see \cite{BW}, Section 8.3.1)\footnote{The author is indebted to
Pavel Kurasov for clarifying this point.} even if the wave equation is a
rather different kind of differential equation than the Schr\"{o}dinger
equation. Indeed, for a monochromatic wave,
$$
\psi(\vec{x},t) = \exp\left(-\frac{i}{\hbar}Et \right)\Psi(\vec{x})\ ,
$$
Eq.\ (\ref{schrodfree}) implies
$$
\bigtriangleup \Psi(\vec{x}) + k^2 \Psi(\vec{x}) = 0\ ,
$$
where
$$
k^2 = \frac{2\mu E}{\hbar^2}\ ,
$$
which coincides with Helmholtz equation (\cite{BW}, p.\ 375). The solution of
Helmholtz equation in the half-space $x_3 > 0$ given by Fresnel-Kirchhof
diffraction formula (\cite{BW}, p.\ 380) then leads to the general solution
$\psi_{\text{traf}}(\vec{x},t)$ (which is a Fourier integral of monochromatic
waves defined by $\tilde{\psi}(\vec{p})$ of Eq.\ (\ref{wavepack})) that
satisfies the required boundary conditions. Hence, the solution exists and is
unique.

We can define absorption, ${\mathrm P}_{\text{abs}}$, and transmission,
${\mathrm P}_{\text{tra}}$, probabilities for the screen as follows:
\begin{equation}\label{tranfprob} {\mathrm P}_{\text{tra}} =
\lim_{t\rightarrow\infty}\int_{{\mathbb R}^2}d^2x\,\int_0^\infty dx_3
|\psi_{\text{traf}}(\vec{x},t)|^2
\end{equation} and
$$
{\mathrm P}_{\text{abs}} = 1 - {\mathrm P}_{\text{tra}}\ .
$$
This is based on the idea that the initial rapidly decreasing wave packet will
leave the left half-space completely for $t\rightarrow\infty$.

Function $\psi_{\text{traf}}(\vec{x},t)$ is not normalised and its norm is
${\mathrm P}^2_{\text{tra}} < 1$. Hence, the model defines a dynamics that is
not unitary. This is clearly due to the incompleteness of the model: particles
that hit the screen are absorbed and this part of the process was ignored
above. Let us give a short account of the physics of absorption. The body
${\mathcal B}$ is assumed to be a perfect absorber so that $S$ does not leave
it. Thus, the screen is assumed to be {\em ideal}: every particle that arrives
at it is either absorbed or goes through the opening. Let the quantum model
$B_q$ of ${\mathcal B}$ be a macroscopic quantum system with Hilbert space
${\mathbf H}^B$ (a real screen is somewhat thicker than a plane, but we just
construct a model). The process of disappearance of a quantum system $S$ in a
macroscopic body $B_q$ can be decomposed into three steps. First, $S$ is
prepared in a state that has a separation status so that a further preparation
or registration (in which the screen participates) can be made. Second, such
$S$ enters $B_q$ and ditch most of its kinetic energy somewhere inside
$B_q$. Third, the energy passed to $B_q$ is dissipated and distributed
homogeneously through $B_q$ in a process aiming at thermodynamic
equilibrium. Then, system $S + B_q$ cannot be decomposed into well-defined
subsystems $S$ and $B_q$ any more, $S$ ceases to be an object and it does not
possess any individual state of its own after being absorbed if there are any
particle of the same type within $B_q$, as it has been explained in Section
4.1. $S$ loses its separation status. Even if, originally, no particle of the
same type as $S$ is within $B_q$, in the course of the experiment, $B_q$ will
be polluted by many of them.

It is important that the absorption process is (or can be in principle)
observable. For instance, the increase of the temperature of $B_q$ due to the
energy of the absorbed particles can be measured. That is, either a single
particle $S$ has enough kinetic energy to cause an observable temperature
change, or there is a cumulative effect of more absorbed particles. More
precisely, suppose that the energy $E^S$ of the absorbed particle is small,
\begin{equation}\label{limclscreen} E^S < \Delta E^{\mathcal B}\ ,
\end{equation} where $\Delta E^{\mathcal B}$ is the variance of the screen
energy in the initial state of the screen so that it would seem that the
absorption could not change the classical state of the screen. However, after
a sufficient number of absorptions, the total change of the energy will
surpass the limit (\ref{limclscreen}) so that the average change of the screen
energy due to one absorption is well defined. In any case, the initial and
final states of $B_q$ cannot be described by wave functions and they differ by
their classical properties from each other, e.g.\ by the temperature.

Let us now try to complete the model including the process of absorption by
writing the initial state as a linear combination of the absorbed and the
transmitted ones. We define a function $\psi_{\text{trai}}(\vec{x},t)$ for
$x_3 < 0$ as the solution of Schr\"{o}dinger equation (\ref{schrodfree})
satisfying the boundary conditions
\begin{equation}\label{bcscreen0-} \lim_{x_3\rightarrow 0}
\psi_{\text{trai}}(\vec{x},t) = 0\ ,\quad \lim_{x_3\rightarrow 0}
\frac{\partial \psi_{\text{trai}}}{\partial x_3}(\vec{x},t) = 0
\end{equation} for $(x_1,x_2) \not\in D$ and
\begin{equation}\label{bcscreen1-} \lim_{x_3\rightarrow 0}
\psi_{\text{trai}}(\vec{x},t) = \lim_{x_3\rightarrow 0}
\psi_{\text{i}}(\vec{x},t)\ ,\quad \lim_{x_3\rightarrow 0} \frac{\partial
\psi_{\text{trai}}}{\partial x_3}(\vec{x},t) = \lim_{x_3\rightarrow 0}
\frac{\partial \psi_{\text{i}}}{\partial x_3}(\vec{x},t)
\end{equation} for $(x_1,x_2) \in D$.

Then, the pair of functions $\psi_{\text{trai}}(\vec{x},t)$ and
$\psi_{\text{traf}}(\vec{x},t)$ define a $C^1$ solution to the
Schr\"{o}dinger equation in the whole space as if the screen did not
exist. Let us denote this function by $\sqrt{{\mathrm
P}_{\text{tra}}}\psi_{\text{tra}}(\vec{x},t)$. Then,
$\psi_{\text{tra}}(\vec{x},t)$ is a normalised solution running from the left
to the right and vanishing in the left-hand half-space for large times.

Finally, let us define function $\psi_{\text{abs}}(\vec{x},t)$ in the
left-hand half-space by
\begin{equation}\label{decomp1} \psi(\vec{x}) = c_{\text{tra}}
\psi_{\text{tra}}(\vec{x}) + c_{\text{abs}} \psi_{\text{abs}}(\vec{x})\ ,
\end{equation} where $c_{\text{tra}} = \sqrt{{\mathrm P}_{\text{tra}}}$,
$c_{\text{abs}} = \sqrt{1 - {\mathrm P}_{\text{tra}}}$ and
$\psi_{\text{tra}}(\vec{x},t)$ is a normalised wave function of the part that
will be left through and $\psi_{\text{abs}}(\vec{x})$ that that will be
absorbed by $B_q$. Indeed, the two wave functions $\psi_{\text{tra}}$ and
$\psi_{\text{abs}}$ must be orthogonal to each other because their large-time
evolution gives $\psi_{\text{abs}} = 0$ in the right-hand half space and
$\psi_{\text{tra}} = 0$ in the left-hand half space.

Decomposition (\ref{decomp1}) is determined by the nature of $B_q$: for a
polariser, these are the two orthogonal polarisation states, and for a simple
screen consisting of a wall with an opening, these can be calculated from the
geometry of $B_q$.

The initial state of $B_q$ is a high-entropy one (see Chapter 3). It is,
therefore, described by a state operator ${\mathsf T}_{\text{i}}$. Then the
initial state for the evolution of the composite is
$$
\bar{\mathsf T}_{\text{i}} = N^2_{\text{exch}} \bar{\mathsf
\Pi}_S(|\psi_{\text{i}}\rangle \langle\psi_{\text{i}}| \otimes {\mathsf
T}_{\text{i}})\bar{\mathsf \Pi}_S\ ,
$$
where $N^2_{\text{exch}} = tr\Bigl(\bar{\mathsf \Pi}_S(|\psi\rangle
\langle\psi| \otimes {\mathsf T}_{\text{i}})\bar{\mathsf \Pi}_S\Bigr)$ and
$\bar{\mathsf \Pi}_S$ is the symmetrization or anti-symme\-trization over all
particles indistinguishable from $S$ (see Section 4.1) within the composite
system $S + B_q$ (we leave open the question of whether they are fermions or
bosons---thus we make a more general theory than that of the previous
section). It is an operator on the Hilbert space ${\mathbf H} \otimes {\mathbf
H}^B$. Further steps are analogous to those for the absorption of the
registered system in the photo-emulsion ${\mathcal D}$ that has been analysed
in more details in the previous section and we can skip the details here.

Let the evolution of the composite $S + B_q$ be described by operator
$\bar{\mathsf U}$. It contains the absorption and dissipation process in
$B_q$. $\bar{\mathsf U}$ is a unitary operator on the Hilbert space ${\mathbf
H} \otimes {\mathbf H}^B$ that commutes with projection $\bar{\mathsf \Pi}_S$
(see Section 4.5) so that it leaves subspace $\bar{\mathsf \Pi}_S({\mathbf H}
\otimes {\mathbf H}^B)$ invariant and so defines a unitary operator on Hilbert
space $\bar{\mathsf \Pi}_S({\mathbf H} \otimes {\mathbf H}^B)$ of the
composite. It is independent of the choice of the initial state. After the
process is finished, we obtain
$$
\bar{\mathsf T}_{\text{f}} = N^2_{\text{exch}}\bar{\mathsf \Pi}_S\bar{\mathsf
U} (|\psi_{\text{i}}\rangle \langle\psi_{\text{i}}| \otimes {\mathsf
T}_{\text{i}})\bar{\mathsf U}^\dagger\bar{\mathsf \Pi}_S\ .
$$

Using decomposition (\ref{decomp1}), we can write
\begin{multline}\label{formscreenf} \bar{\mathsf T}_{\text{f}} =
N^2_{\text{exch}}c_{\text{abs}}c^*_{\text{abs}}\bar{\mathsf \Pi}_S\bar{\mathsf
U} (|\psi_{\text{absi}}\rangle \langle\psi_{\text{absi}}| \otimes {\mathsf
T}_{\text{i}})\bar{\mathsf U}^\dagger\bar{\mathsf \Pi}_S \\ +
N^2_{\text{exch}}c_{\text{tra}}c^*_{\text{tra}}\bar{\mathsf \Pi}_S\bar{\mathsf
U} (|\psi_{\text{trai}}\rangle \langle\psi_{\text{trai}}| \otimes {\mathsf
T}_{\text{i}})\bar{\mathsf U}^\dagger\bar{\mathsf \Pi}_S \\ +
N^2_{\text{exch}} c_{\text{tra}}c^*_{\text{abs}}\bar{\mathsf
\Pi}_S\bar{\mathsf U} (|\psi_{\text{trai}}\rangle \langle\psi_{\text{absi}}|
\otimes {\mathsf T}_{\text{i}})\bar{\mathsf U}^\dagger\bar{\mathsf \Pi}_S \\ +
N^2_{\text{exch}}c_{\text{abs}}c^*_{\text{tra}}\bar{\mathsf \Pi}_S\bar{\mathsf
U} (|\psi_{\text{absi}}\rangle \langle\psi_{\text{trai}}| \otimes {\mathsf
T}_{\text{i}})\bar{\mathsf U}^\dagger\bar{\mathsf \Pi}_S\ .
\end{multline} The first term describes the process that starts with state
$\psi_{\text{absi}}$. Thus, $S$ does not reappear at the end and the result is
an excited state $\bar{\mathsf T}'_{\text{f}}$ of the screen that has absorbed
$S$. The second term represents the evolution that starts with $S$ in the
state $\psi_{\text{trai}}$. Then the screen remains in its initial state
${\mathsf T}_{\text{i}}$ and $S$ reappears in state
$\psi_{\text{traf}}$. Hence,
$$
\bar{\mathsf T}_{\text{f}} = \bar{\mathsf T}_{\text{end}1} + \bar{\mathsf
T}_{\text{end}0}\ ,
$$
where
$$
\bar{\mathsf T}_{\text{end}1} = |c_{\text{abs}}|^2 \bar{\mathsf T}'_{\text{f}}
+ |c_{\text{tra}}|^2 |\psi_{\text{traf}}\rangle \langle \psi_{\text{traf}}|
\otimes {\mathsf T}_{\text{i}}\ .
$$
State $\bar{\mathsf T}_{\text{end}1}$ is a convex combination of two states
that differ from each other by their classical properties while
$$
tr(\bar{\mathsf T}_{\text{end}0}) = 0\ .
$$

We can now argue in analogy with the previous section: RCU interpretation
suggests together with the observation that only the first two terms describe
the true end state of the composite after each individual individual process
and the state is not just a convex combination but a proper mixture (see
Section 1.3):
\begin{equation}\label{screenf} \bar{\mathsf T}_{\text{true}f} = {\mathrm
P}_{\text{tra}} |\psi_{\text{traf}}\rangle \langle\psi_{\text{traf}}| \otimes
{\mathsf T}_{\text{i}}\ +_s\ {\mathrm P}_{\text{abs}} \bar{\mathsf
T}'_{\text{f}}\ .
\end{equation} The transition from $\bar{\mathsf T}_f$ to $\bar{\mathsf
T}_{\text{true}f}$ such a mixture is our definition of state reduction as in
Section 5.2.

Again, the state reduction is not a unitary transformation: First, the
non-diagonal terms in (\ref{formscreenf}) have been erased. Second, we have
also assumed that state $\psi_{\text{traf}}$ is the state of $S$ that has been
{\em prepared} by the screening. This means for us that it is a real state
with a separation status. Hence, operator $\bar{\mathsf \Pi}_S$ can be left
out in Formula (\ref{screenf}). This is, of course, another violation of
unitarity.

The disappearance of $S$ in $B_q$, as well as the disappearance of $S$ in the
photo-emulsion ${\mathcal D}$ described in the previous section, is a physical
process that have a definite time and place. This suggests that the state
reduction occurs at the time and the place of the possible absorption of the
particle in $B_q$ or ${\mathcal D}$. The possible absorption had to be viewed
as a part of the whole process even in the case that an individual particle is
not absorbed but goes through. Indeed, that an individual particle goes
through is only a result of the state reduction, which is a change from the
linear superposition of the transition and the absorption states.

\section{The structure of meters} Here, we extend some ideas of Section 5.2 on
Stern-Gerlach apparatus to all meters with the aim to improve the
understanding of registrations. Most theoretical descriptions of meters that
can be found in the literature are strongly idealised (see, e.g.,
\cite{BLM,WM}): the meter is a not further specified quantum system with a
``pointer'' observable. We are going to give a more elaborated picture and
distinguish between {\em fields, screens, ancillas} and {\em detectors} as
basic structural elements of meters.

Screens have been dealt with in Section 5.2. It is also more or less clear
what are fields: for example, in the Stern-Gerlach experiment, the beam is
split by an inhomogeneous magnetic field. In some optical experiments, various
crystals are used that make possible the split of different polarisations or
the split of a beam into two mutually entangled beams such as by the
down-conversion process in a crystal of KNbO$_3$ \cite{MW}. The corresponding
crystals can also be considered as fields. In any case, the crystals and
fields are macroscopic systems the (classical) state of which is not changed
by the interaction with the registered system.

In many modern experiments, in particular in non-demolition and weak
measurements, but not only in these, the following idea is employed. The
registered system $S$ interacts first with an auxiliary quantum system $A$
that is prepared in a suitable state. After $S$ and $A$ become entangled, $A$
is subject to further registration and, in this way, some information on $S$
is revealed. Subsequently, further measurements on $S$ can but need not be
made. The state of $S$ is influenced by the registration of $A$ just because
of its entanglement with $A$. Such auxiliary system $A$ is usually called {\em
ancilla} (see, e.g., \cite{peres}, p.\ 282).

Finally, important parts of meters are {\em detectors}. Indeed, even a
registration of an ancilla needs a detector. It seems that any registration on
microscopic systems has to use detectors in order to make features of
microscopic systems visible to humans. Detector is a large system that changes
its (classical) state during the interaction with the registered
system. ``Large'' need not be macroscopic but the involved number of particles
ought to be at least about $10^{10}$. For example, the photo-emulsion grain or
nanowire single photon detector (see, e.g., \cite{natarayan}) are large in
this sense. A criterion for a physical object to be large is that the object
has a classical model that gives a good approximation to some aspect of its
behaviour. For example, the object has well-defined thermodynamic states.

For example, in the so-called cryogenic detectors \cite{stefan}, $S$
interacts, e.g., with superheated superconducting granules by scattering off a
nucleus in a granule. The resulting phonons induce the phase transition from
the superconducting to the normally conducting phase. The detector can contain
very many granules (typically $10^9$) in order to enhance the probability of
such scattering if the interaction between $S$ (a weakly interacting massive
particle, neutrino) and the nuclei is very weak. Then, there is a solenoid
around the vessel with the granules creating a strong magnetic field. The
phase transition of only one granule leads to a change in magnetic current
through the solenoid giving a perceptible electronic signal.

Modern detectors are constructed so that their signal is electronic. For
example, to a scintillation film, a photomultiplier is attached (as in
\cite{tono}). We assume that there is a signal collected immediately after the
detector changes its classical state, which we call {\em primary}
signal. Primary signal may still be amplified and filtered by other electronic
apparatuses, which can transform it into the final signal of the detector. For
example, the light signal of a scintillation film in the interference
experiment of \cite{tono} is a primary signal. It is then transformed into an
electronic signal by a photocathode and the resulting electronic signal is
further amplified by a photomultiplier.

A detector contains {\em active volume} ${\mathcal D}$ and {\em signal
collector} ${\mathcal C}$ in thermodynamic state of metastable
equilibrium. The term ``thermodynamic equilibrium'' is correct even in the
cases in which mechanical or electrodynamic forces play a relevant role (such
as that of Geiger-Muller counter). Notice that the active volume is a physical
system, not just a volume of space. For example, the photo-emulsion or the set
of the superconducting granules are active volumes. Interaction of the
detected systems with $\mathcal D$ triggers a relaxation process leading to a
change of the classical state of the detector---the {\em detector signal}. For
some theory of detectors, see, e.g., \cite{leo,stefan}.

What is the difference between ours and the standard ideas on detectors? The
standard ideas are, e.g., stated in (Ref.\ \cite{peres} p.\ 17) with the help
of the Stern-Gerlach example:
\begin{quote} The microscopic object under investigation is the magnetic
moment $\mathbf \mu$ of an atom.... The macroscopic degree of freedom to which
it is coupled in this model is the centre of mass position $\mathbf r$... I
call this degree of freedom {\em macroscopic} because different final values
of $\mathbf r$ can be directly distinguished by macroscopic means, such as the
detector... From here on, the situation is simple and unambiguous, because we
have entered the macroscopic world: The type of detectors and the detail of
their functioning are deemed irrelevant.
\end{quote} The root of such notion of detectors may be found among some ideas
of the grounding fathers of quantum mechanics. For example, Ref.\
\cite{pauli}, p. 64, describes a measurement of energy eigenvalues with the
help of scattering similar to Stern-Gerlach experiment, and Pauli explicitly
states:
\begin{quote} We can consider the centre of mass as a 'special' measuring
apparatus\ldots
\end{quote}

In these statements, no distinction is made between ancillas and detectors:
indeed, the centre-of-mass position above can be considered as an
ancilla. However, such a distinction can be made and it ought to be made
because it improves our understanding of registrations. Thus, to improve the
understanding, we have slightly modified the current notions of detector and
ancilla. Our detectors are more specific than what is often assumed.

The foregoing analysis motivates the following trial hypothesis.
\begin{sh}\label{aspointerh} Any meter for microsystems must contain at least
one detector and every reading of the meter can be identified with a primary
signal from a detector. The state reduction required by realism and
observational evidence on measurements takes place in detectors and screens.
\end{sh} A similar hypothesis has been first formulated in \cite{hajicek2}. TH
\ref{aspointerh} makes the reading of meters less mysterious. Moreover, TH
\ref{asobjectcq} allows us to distinguish between the quantum and the
classical models of a detector and to consider the classical model as a meter
for the quantum model, so that von Neumann's chain begins and ends at the
detector.

\section{Two hypotheses on state reduction} Here, we study the form of state
reduction and the objective circumstances with which it is connected.
\begin{sh}\label{assh} Let ${\mathcal O}$ be an object (such as a detector)
with classical model $O_c$ and quantum model $O_q$. Let the standard unitary
evolution describing some process in which $O_q$ takes part results in an end
state of the form:
\begin{equation}\label{Tf} \bar{\mathsf T}_f = \sum_{k=1}^n {\mathrm P}_k
\bar{\mathsf T}_k + \bar{\mathsf T}_{\text{end}0}\ ,
\end{equation} where $\bar{\mathsf T}_k$ are states of $O_q$ such that each is
associated with a classical state of $O_c$ and these classical states are
different for different $k$'s. The coefficients satisfy ${\mathrm P}_k > 0$
for $k = 1,\ldots,n$ and $\sum_k {\mathrm P}_k = 1$. $\bar{\mathsf
T}_{\text{end}0}$ is a s.a.\ operator with trace 0. Then, the standard unitary
evolution must be corrected so that $\bar{\mathsf T}_f$ is replaced by
\begin{equation}\label{Tend} \bar{\mathsf T}_{\text{end}} = \sum_{k=}^n +_s
{\mathrm P}_k \bar{\mathsf T}_k\ ,
\end{equation} the proper mixture of states $\bar{\mathsf T}_k$.
\end{sh} TH \ref{assh} is applicable to those unitary evolutions that have an
end state of the form (\ref{Tf}). However, classical objects may have some
properties that make such a form to be a general case. For example, it may be
impossible for a quantum model of a classical object to be in a convex
combination of states, one of which is associated with a classical state and
the other not having classical properties or in a state equal to two different
convex compositions so that the two sets of classical states defined by the
two compositions are different from each other. This seems to follow from the
classical realism described at the beginning of Chapter 1.

To illustrate the difference to an ordinary convex decomposition, let us
consider an arbitrary normalised state vector $\Phi$ of some quantum
system. Such a state can be decomposed into two orthonormal vectors in an
infinite number of different ways, for example,
$$
\Phi = c_1\Phi_1 + c_2\Phi_2 = d_1\Psi_1 + d_2\Psi_2\ .
$$
Then
$$
|\Phi\rangle \langle \Phi| = |c_1|^2 |\Phi_1\rangle \langle \Phi_1| + |c_2|^2
|\Phi_2\rangle \langle \Phi_2| + c_1c^*_2|\Phi_1\rangle\langle \Phi_2| +
c_2c^*_1|\Phi_2\rangle\langle \Phi_1|
$$
and
$$
|\Phi\rangle \langle \Phi| = |d_1|^2 |\Psi_1\rangle \langle \Psi_1| + |d_2|^2
|\Psi_2\rangle \langle \Psi_2| + d_1d^*_2|\Psi_1\rangle\langle \Psi_2| +
d_2d^*_1|\Psi_2\rangle\langle \Psi_1|
$$
are two different decompositions of state $|\Phi\rangle \langle \Phi|$ that
have the form of (\ref{Tf}).

We leave the detailed questions of applicability of TH \ref{assh} open to
future investigations in the hope that the approach that it suggests is more
or less clear. In any case, TH \ref{assh} defines a rule that determines the
correction to unitary evolution {\em uniquely} in a large class of scattering
and registration processes (see \cite{hajicek4,hajicek5}).

Both detectors and screens, where the state reductions occur, are mezzo- or
macroscopic (for example, the emulsion grains can be considered as
mezzoscopic), but there are processes of interaction between microscopic and
macroscopic objects, the standard quantum description of which gives always a
unique classical end state of the macroscopic part. For example, the
scattering of neutrons by ferromagnetic crystals in which the crystal remains
in the same classical state during the process of scattering. In such
processes, TH \ref{assh} implies no state reductions. It is the structure of
the final quantum state that makes the difference: for a state reduction, the
standard quantum evolution had to give a convex combination of states that
differ in their classical properties.

What is the cause of the change $\bar{\mathsf T}_f$ into $\bar{\mathsf
T}_{\text{end}}$? For example, the detector that detects microsystem $S$
achieves the signal state so that $S$ interacts with its active volume
${\mathcal D}$ and the state of $S + D_q$ dissipates, which leads to a loss of
separation status of $S$. A similar process runs in a screen that absorbs
$S$. The dissipation is necessary to accomplish the loss. The dissipation
process does not have anything mysterious about it. It can be a usual
thermodynamic relaxation process in a macroscopic system or a similar process
of the statistical thermodynamics generalised to nano-systems (see, e.g.,
\cite{horodecki}). $S$ might be the registered object or an ancilla of the
original experiment. In all such cases, state $\bar{\mathsf T}_{\text{end}}$
originates in a process of relaxation triggered by $S$ in $D_q$ or $B_q$ and
accompanied by the loss of separation status of $S$. This motivates the
following hypothesis:
\begin{sh}\label{asavh} The cause of the state reduction postulated by TH
\ref{assh} is an uncontrollable disturbance due to a loss of separation
status.
\end{sh} The loss of separation status is an objective process and the
significance of TH \ref{asavh} is that it formulates an objective condition
for the applicability of an alternative kind of dynamics.

Actually, the assumption that a measuring process disturbs the measured system
in an uncontrollable way and that this is the cause of the state reduction is
not new (see, e.g., \cite{messiah}, Section 4.3.1). What we add to it is just
the role of separation-status loss.

The three Trial Hypotheses \ref{aspointerh}, \ref{assh} and \ref{asavh} form a
basis of our theory of state reduction. They generalise some empirical
experience, are rather specific and, therefore, testable. That is, they cannot
be disproved by purely logical argument but rather by an experimental
counterexample. For the same reason, they also show a specific direction in
which experiments ought to be proposed and analysed: if there is a state
reduction, does then a loss of separation status take part in the process?
What system loses its status? How the loss of the status can lead to state
reduction?

In fact, our theory remains rather vague with respect to the last question in
that it suggests no detailed model of the way from a separation status change
to a state reduction. Such a model would require some new physics and we
believe that hints of what this new physics could be will come from attempts
to answer the above questions by suitable experiments.

The new notions and hypotheses are studied on some examples and models
\cite{hajicek4}, where the old definition of separation status is used. A
reformulation of these examples based on the new definition seems to be more
or less straightforward.

\chapter{Summary of RCU interpretation} The project described by this paper is
to reformulate a popular version of the Copenhagen interpretation in more
rigorous terms and augment it with some new ideas that make it more self
consistent and comprehensible. The result is the so-called RCU
interpretation. The starting point was the Minimum Interpretation, i.e., the
notion that quantum mechanics was a set of rules to calculate probabilities of
registration values. To this, some further assumptions were added, called
Trial Hypotheses (TH). This was done step by step so that the motivation and
impact of each TH could be explained. In this way, however, the whole
resulting picture remained rather elusive. The purpose of this chapter is to
give a short survey of all changes done.

Our first TH is Completeness Hypothesis
\begin{quote} There are no unknown causes beyond the probabilities given by
the Born rule.
\end{quote} It might seem more cautious if one would leave this question open
but then some main principles of the standard quantum mechanics had to be
changed. The Completeness Hypothesis has important consequences for
interpretation of states.

The next aim is to provide a tool against the positivist and instrumentalist
standpoint to which quantum physicists and are motivated by Minimum
Interpretation. To accomplish this task, we lean strongly on the realist
version of the model approach to the philosophy of science, the Constructive
Realism \cite{giere}. This leads us to distinguish between a real object and
its models, in particular physical objects and their classical and quantum
models.

Our definition of quantum (real) object, TH \ref{shobject}, reads:
\begin{quote} A quantum object is defined by a preparations. The objects are
thus distinguished from each other by the properties that are determined by
their preparations. These include the structural properties describing a
system type, the prepared state and properties that are uniquely defined by
the state. Objects of non-relativistic quantum mechanics can be classified
into electrons, neutrons, nuclei, atoms, molecules and their composites.
\end{quote} This represents our requirement that real objects must have a
sufficient number of objective properties and our basic assumption that such
properties of quantum systems are those defined uniquely preparations. TH
\ref{shobject} strongly influenced almost all further development. In
particular, it motivates the interpretation of observable values in TH
\ref{OCR} (Outcomes Created by Registration):
\begin{quote} The outcome of an individual registration performed on a quantum
object $S$ in state ${\mathsf T}$ is in general only created during the
registration. It is an objective property of the whole registration process,
not of the registered system.
\end{quote} TH \ref{OCR} also helps explain some quantum paradoxes, such as
the wave-particle dualism or existence of a special quantum ``logic'', which
is no logic but an algebra of projection operators so that working with this
algebra requires the ordinary mathematical logic.

TH \ref{shobject} is further specified by TH \ref{shobject1}:
\begin{quote} Objective properties of a quantum system $S$ can be divided into
three classes: 1. structural properties of $S$, 2. a state of $S$ and the
properties determined uniquely by the state (such as expectation values of a
fixed observable), 3. the properties of the state of a system that has been
prepared so that it contains $S$ as a subsystem if such properties concern $S$
but are not determined by the state of $S$ itself (such as the way $S$ is
entangled with other systems).
\end{quote} Structural properties are those that are common to all systems of
the same type, such as mass and charge. The TH gives a list of objective
properties. It also includes the correlations between pairs of observables,
each belonging to one of two entangled systems.

We have derived some important consequences of TH \ref{shobject}: the
existence of proper mixtures and the fundamental distinction between proper
mixtures and all other quantum states. We have motivated the notion that the
popular distinction between pure states and mixtures is rather
misleading. Instead, leaning on the Completeness Hypothesis, we introduce the
distinction between ontic and epistemic states as the only physically
meaningful one by TH \ref{shontic}:
\begin{quote} Extremal states and improper mixtures are ontic, proper mixtures
are epistemic, quantum states.
\end{quote} Hence, the quantum states that are not proper mixtures give a
maximum information of quantum systems and have by themselves, no statistical
character.

The universality of quantum mechanics, TH \ref{rhmacro}, reads:
\begin{quote} Every object has a quantum model that accounts for all its known
physical properties.
\end{quote} It also postulates that quantum mechanics is applicable to
classical systems. Chapters 2 and 3 start a new approach to the problem of
construction of quantum models describing classical systems. The basis is TH
\ref{asobjectcq}:
\begin{quote} Let $S_c$ be a classical model and $S_q$ a quantum model of
object $S$. Then $S_c$ can be considered as an essential part of the
preparation device and, simultaneously, as an essential part of a meter, for
the quantum model $S_q$. The meter in question registers values distinguishing
the quantum states of $S_q$ that are associated with different classical
states of $S_c$.
\end{quote} It relates the classical model to the preparation and registration
apparatus for the quantum model. TH \ref{asobjectcq} also illustrates the
practical value and advantage of the model approach.

The relation between quantum and classical properties is postulated by TH
\ref{ashighS}:
\begin{quote} Let a real object ${\mathcal S}$ has a classical model $S_c$ and
a quantum model $S_q$. Then all properties of $S_c$ are selected properties of
some high-entropy states of $S_q$.
\end{quote} This is motivated by thermodynamics, where it is well-known to be
valid, and contradicts sharply the common prejudice that quantum coherent
states carry classical properties. For the special case of Newtonian
mechanics, the idea that we call Exner-Born Conjecture is stressed:
\begin{quote} States of Newtonian systems that are described by sharp points
of the phase space do not exist. Newtonian models that can approach the
reality better are non-trivial probability distribution function on the phase
space.
\end{quote} The conjecture opens the way to a unified theory for both
thermostatic and mechanical properties. The technical tool is the ME-packet
notion, the state maximizing entropy for given averages and variances, and its
significance is expressed by the ME-Packet Conjecture:
\begin{quote} For most mechanical objects ${\mathcal S}$, all measurable
predictions of Newtonian mechanics can be obtained from a classical model
$S_c$, that is an ME packet defined by Theorem \ref{propold19}.
\end{quote} It allows to choose specific high-entropy states of Newtonian
systems---ME packets---as those that are to be approximated by quantum models.

The classical and quantum mathematics of maximum entropy packets has been
developed. Its main result (Theorem \ref{thmclaslim}) is that the trajectories
of classical and quantum ME packets match each other the better, the higher
their entropy is. That is, the fuzziness improves the match between classical
and quantum theory. This confirms the feeling that quantum mechanics is more
accurate and finer than Newtonian mechanics. Newtonian mechanics also becames
conspicuously similar to phenomenological thermodynamics. The result is proved
for all polynomial potential functions. We hope to be able to prove this
statement for non-polynomial potentials such as Coulomb potentials associated
with Kepler orbits. Also, the idea ought yet to be generalised to classical
electro- and magnetostatic properties, as well as to the relativistic
classical electrodynamics.

The theory also suggests promising ideas of how the well-known conceptual
problems associated with classical properties (explained at the start of
Chapter 2) could be solved.

Next, we turn to measurement theory. Chapter 4 studies the well-known (c.f.\
\cite{peres}) but generally ignored disturbance of registrations of system $S$
by environmental particles of the same type. We have shown that it can only be
avoided if the registration is made by incomplete meters. The incomplete
apparatus must give probability zero to all values that are registered on
states of the environment. Moreover, the measured system must be prepared in a
state that lifts it from the sea of identical particles: it must have the
so-called separation status, Definition \ref{dfsseq}. Then, the environmental
particles that are indistinguishable from the measured system can be ignored
in the practice of registrations and in their theoretical treatment, as it is
usually done.

This leads to a new notion of preparation process, TH \ref{shprep}:
\begin{quote} Any preparation of a single microsystem must yield a state
having a separation status.
\end{quote} This correction concerns all previous Trial Hypotheses that use
the notion of preparation. For instance, it refines Trial Hypothesis
\ref{shobject} about objectivity of structural properties and prepared states
concerning the word ``prepared'' in it.

The definition of separation status proposed in Chapter 4 is different from
that of \cite{hajicek2,hajicek4} so that some problems of the old definition
are removed. Moreover, the new notion of separation status is more general and
simpler to use than the old one.

The relation between a quantum observable and its registration apparatus
becomes then more complicated than is usually assumed. The observables
themselves are defined as in the standard quantum theory (as self-adjoint
operators). In this way, the simplicity and elegance of the standard
mathematics (such as $C^*$-algebras) of quantum observables is
preserved. However, the role of meters was described by Trial Hypothesis
\ref{shobsmeter}:
\begin{quote} For any observable ${\mathsf O}$ of a system $S$, there is a
meter that can register ${\mathsf O}$. There is no meter that can register
${\mathsf O}$ on any state.
\end{quote} This means that each observable represents a whole set of
apparatuses, each registering only the part of it that is associated with a
subset of states. We do not want to deny that some measurements are better
described by positive-operator valued measures (POVM) than by self-adjoint
operators but just do not call POVM ``observables''. The relation between
observables and apparatuses described by TH \ref{shobsmeter} is also valid for
that of POVM and apparatuses.

The new theory is logically consistent with the Minimum Interpretation of
exchange symmetry, agrees with the praxis of real measurements and our
understanding of registration apparatuses as well as that of preparation
processes is improved.

A reformulation of the quantum theory of measurement theory is given in
Chapter 5. Section 5.4 gives a narrower definition to the notion of detector
(already proposed in \cite{hajicek2}) than what is usually assumed. The basic
assumption is TH \ref{aspointerh}:
\begin{quote} Any meter for microsystems must contain at least one detector
and every reading of the meter can be identified with a primary signal from a
detector. The state reduction required by realism and observational evidence
on measurements takes place in detectors and screens.
\end{quote} The form of state reduction is uniquely specified by TH
\ref{assh}:
\begin{quote} Let ${\mathcal O}$ be an object (such as a detector) with
classical model $O_c$ and quantum model $O_q$. Let the standard unitary
evolution describing some process in which $O_q$ takes part results in an end
state of the form:
$$
\bar{\mathsf T}_f = \sum_{k=1}^n {\mathrm P}_k \bar{\mathsf T}_k +
\bar{\mathsf T}_{\text{end}0}\ ,
$$
where $\bar{\mathsf T}_k$ are states of $O_q$ such that each is associated
with a classical state of $O_c$ and these classical states are different for
different $k$'s. The coefficients satisfy ${\mathrm P}_k > 0$ for $k =
1,\ldots,n$ and $\sum_k {\mathrm P}_k = 1$. $\bar{\mathsf T}_{\text{end}0}$ is
a s.a.\ operator with trace 0. Then, the standard unitary evolution must be
corrected so that $\bar{\mathsf T}_f$ is replaced by
$$
\bar{\mathsf T}_{\text{end}} = \sum_{k=}^n +_s {\mathrm P}_k \bar{\mathsf
T}_k\ ,
$$
the proper mixture of states $\bar{\mathsf T}_k$.
\end{quote} There are two innovations in TH \ref{assh}: first, it specifies
the form of state reductions uniquely and second, it postulates state
reduction as a transformation between high-entropy states of large
systems. Thus, it needs more sophisticated mathematical methods, which are
introduced in Sections 5.2 and 5.3. It is also necessary to stress that our
theory leads to some doubts on the standard theory of quantum measurement as
it is explained e.g.\ in \cite{WM}. The basic notion of the standard theory is
that of ``state transformer'': a transformation between the initial state of
the measured system $S$ and the state of $S$ after the registration by
detector $M$. We have shown that the registered system is lost after the
registration, that is it cannot be identified with any well-defined subsystem
of the composed system $S + M$.

A suggestion that the cause of state reduction may be associated with a loss
of separation status is TH \ref{asavh}:
\begin{quote} The cause of the state reduction postulated by TH \ref{assh} is
an uncontrollable disturbance due to a loss of separation status.
\end{quote} It is a more precise form of the original suggestion in
\cite{hajicek2}. It promotes the state reduction to a physical process,
specifies the objective conditions of its occurrence and shows where and when
it takes place. However, it does not give a detailed account of the process.

Many applications of our measurement theory are described in
\cite{hajicek4}. It is also shown there that the theory has measurable
consequences and a direction of possible further development is suggested.

\end{document}